\pdfoutput=1
\documentclass[11pt]{article}
\usepackage{jheppub}

\usepackage{amsmath}
\usepackage{amssymb}
\usepackage{graphicx,color}
\usepackage{amsthm}
\usepackage{mathrsfs}

\usepackage{tikz}
\usetikzlibrary{matrix,arrows}

\usepackage{ifpdf}

\usepackage{subfigure}

\usepackage{url}

\usepackage[boxsize=0.5em,aligntableaux=center]{ytableau}

\newcommand{\bC}{\mathbb{C}}
\newcommand{\bF}{\mathbb{F}}

\newcommand{\bL}{\mathbb{L}}
\newcommand{\bP}{\mathbb{P}}
\newcommand{\bR}{\mathbb{R}}
\newcommand{\bRP}{\mathbb{RP}}
\newcommand{\bZ}{\mathbb{Z}}

\newcommand{\cC}{\mathcal{C}}
\newcommand{\cE}{\mathscr{E}}
\newcommand{\cF}{\mathscr{F}}

\newcommand{\cL}{\mathcal{L}}
\newcommand{\cM}{\mathcal{M}}
\newcommand{\cN}{\mathcal{N}}
\newcommand{\cO}{\mathcal{O}}
\newcommand{\cP}{\mathcal{P}}

\newcommand{\cS}{\mathcal{S}}

\newcommand{\tN}{{\tilde{N}}}

\newcommand{\ov}{\overline}

\newcommand{\sL}{\mathsf{L}}
\newcommand{\sP}{\mathsf{P}}

\newcommand{\dsz}[2]{\bigl\langle#1,#2\bigr\rangle}

\newcommand{\symm}{\ydiagram{2}}
\newcommand{\asymm}{\ydiagram{1,1}}
\newcommand{\fund}{\ydiagram{1}}
\newcommand{\ssymm}{\scalebox{0.7}{$\symm$}}
\newcommand{\sasymm}{\scalebox{0.7}{$\asymm$}}

\DeclareMathOperator{\Hom}{Hom}
\DeclareMathOperator{\ch}{ch}
\DeclareMathOperator{\Td}{Td}
\DeclareMathOperator{\Tr}{Tr}
\DeclareMathOperator{\Ext}{Ext}
\DeclareMathOperator{\diag}{diag}
\DeclareMathOperator{\re}{Re}
\DeclareMathOperator{\im}{Im}
\DeclareMathOperator{\cone}{Cone}
\DeclareMathOperator{\meijerG}{G\!}

\newcommand{\Sp}{U\!Sp}
\newcommand{\SO}{SO}
\newcommand{\SU}{SU}
\newcommand{\SL}{SL}
\newcommand{\U}{U}
\newcommand{\NS}{N\!S5}

\renewcommand{\arraystretch}{1.2}

\newcommand{\alt}[2]{\texorpdfstring{#1}{#2}}

\def\Label#1{\label{#1}%
  \smash{\hbox to0pt{\raise1ex\hbox{\tiny[#1]}\hss}}}
\def\noLabels{\let\Label=\label}
\def\nobbibitem{\let\bbibitem=\bibitem}
 \def\noBibitem{\let\Bibitem=\bibitem}

\newcommand{\be}{\begin{equation}}
\newcommand{\ee}{\end{equation}}
\newcommand{\ba}{\begin{eqnarray}}
\newcommand{\ea}{\end{eqnarray}}

\newcommand{\lp}{\left(}
\newcommand{\rp}{\right)}
\newcommand{\ls}{\left[}
\newcommand{\rs}{\right]}

\newcommand{\w}{\wedge}
\newcommand\varpm{\mathbin{\vcenter{\hbox{%
  \oalign{\hfil$\scriptstyle+$\hfil\cr
          \noalign{\kern-.3ex}
          $\scriptscriptstyle({-})$\cr}%
}}}}
\newcommand\varmp{\mathbin{\vcenter{\hbox{%
  \oalign{$\scriptstyle({+})$\cr
          \noalign{\kern-.3ex}
          \hfil$\scriptscriptstyle-$\hfil\cr}%
}}}}

\title{\centering New $\cN=1$ dualities from orientifold transitions\\
\Large{--- Part II: String Theory ---}}

\author[a]{I\~naki Garc\'ia-Etxebarria,}
\author[b]{Ben Heidenreich,}
\author[c]{and Timm Wrase}

\affiliation[a]{Theory Group, Physics
  Department, CERN, CH-1211, Geneva 23, Switzerland}
\affiliation[b]{Department of Physics, Cornell University, Ithaca, NY
  14853, USA}
\affiliation[c]{Stanford Institute for Theoretical Physics, Stanford University, Stanford, CA 94305, USA}

\emailAdd{inaki@cern.ch}
\emailAdd{bjh77@cornell.edu}
\emailAdd{timm.wrase@stanford.edu}

\abstract{We present a string theoretical description, given in terms
  of branes and orientifolds wrapping vanishing cycles, of the dual
  pairs of gauge theories analyzed in \cite{gauge}. Based on the
  resulting construction we argue that the duality that we observe in
  field theory is inherited from S-duality of type IIB string theory. We analyze in detail
  the complex cone over the zeroth del Pezzo surface and discuss an
  infinite family of orbifolds of flat space. For the del Pezzo case
  we describe the system in terms of large volume objects, and show
  that in this language the duality can be understood from the
  strongly coupled behavior of the $O7^+$ plane, which we analyze
  using simple F-theory considerations. For all cases we also give a
  different argument based on the existence of appropriate torsional
  components of the 3-form flux lattice. Along the way we clarify some
  aspects of the description of orientifolds in the derived category
  of coherent sheaves, and in particular we discuss the important role
  played by exotic orientifolds --- ordinary orientifolds composed
  with auto-equivalences of the category --- when describing
  orientifolds of ordinary quiver gauge theories.}

\begin{document}
\noLabels % uncomment for final production
%\nobbibitem % uncomment for final production
%\noBibitem % uncomment for final production

\makeatletter
\let\old@fpheader\@fpheader
\renewcommand{\@fpheader}{\old@fpheader\hfill
CERN-PH-TH/2013-153,
SU/ITP-13/11}
\makeatother

\maketitle

\section{Introduction}
\label{sec:Intro}

In a companion paper \cite{gauge} we have argued for existence of a
duality between the following $\cN=1$ field theories in four
dimensions. The first theory is given by
\begin{align} \label{eqn:dP0thyA}
  \begin{array}{c|cc|ccc}
     & \SO(N-4) & \SU(N)  & \SU(3) & \U(1)_R & \bZ_{3}\\ \hline
     A^i & \fund & \fund & \fund & \frac23+\frac2N & \omega_{3 N} \\
     B^i & 1 & \ov{\asymm} & \fund & \frac23 - \frac4N & \omega_{3 N}^{-2}
  \end{array}
\end{align}
with $\omega_n \equiv e^{2\pi i/n}$ and the superpotential
\begin{align}
  W = \frac{1}{2} \epsilon_{ijk} \Tr A^i A^j B^k\, .
\end{align}
The second theory is given by
\begin{align} \label{eqn:dP0thyB}
  \begin{array}{c|cc|ccc}
     & \Sp(\tilde N + 4) & \SU(\tilde N) & \SU(3) & \U(1)_R & \bZ_{3}\\ \hline
     \tilde{A}^i & \fund & \fund & \fund & \frac23 - \frac{2}{\tilde{N}} & \omega_{3 \tilde{N}} \\
    \tilde{B}^i & 1 & \ov{\symm} & \fund & \frac23 + \frac{4}{\tilde{N}} & \omega_{3 \tilde{N}}^{-2}
  \end{array}
\end{align}
with the superpotential
\begin{align}
  \tilde{W} = \frac{1}{2} \epsilon_{ijk} \Tr \tilde{A}^i \tilde{A}^j
  \tilde{B}^k\, .
\end{align}
In \cite{gauge}, we have argued that the $\Sp$ theory is dual to the $\SO$ theory when $\tilde N = N-3$ for odd $N$, where in our conventions
$\tilde{N}$ has to be even for $\Sp(\tilde{N}+4)$ to be defined. (In~\S\ref{sec:torsion}, we argue that the $\SO$ theory is self-dual for even $N$.)

Although both theories describe the worldvolume gauge theory on D3 branes probing orientifolds of the $\bC^3/\bZ_3$ singularity,
the arguments for the duality presented in \cite{gauge} were formulated mainly in field
theoretic terms, verifying the agreement of several protected quantities between the two theories. One may well wonder if a careful study of the corresponding branes in string theory could
shed light on the physical origin and nature of the duality.

We show in this paper that this is indeed the case. In particular, we present two converging lines of argument leading to the main claim of our paper: \emph{the duality found in \cite{gauge} is a strong/weak duality, directly inherited from S-duality in type IIB string theory}. As such, its closest known analogues are the electromagnetic dualities relating $\cN=4$ $\SO$ and $\Sp$ gauge theories.

Our paper employs two complementary arguments to establish our main claim. The first approach, outlined in~\S\ref{sec:torsion}, focuses on topological aspects of the gravity dual.
Following~\cite{Witten:1998xy}, we argue that there are four possible choices of NSNS and RR 2-form discrete torsion, splitting into a singlet and a triplet of $SL(2,\bZ)$. As in~\cite{Witten:1998xy}, the different torsion values naturally correspond to the different possible gauge theories. The action of $SL(2,\bZ)$ on the discrete torsion triplet reproduces the duality found in field theory; in particular, the dual theories are related by S-duality ($\tau \to -1/\tau$), and at most one can be weakly coupled for a given value of the string coupling, leading to a strong/weak duality which descends from ten-dimensional S-duality.

As a non-trivial check of this argument, in~\S\ref{sec:infinitefamilies} we apply the same reasoning to other orbifold singularities and show that they admit the same choices of discrete torsion. We then write down the corresponding field theories and show that they have matching anomalies, as expected for S-dual theories. In particular, we carry out this program for an infinite family of orbifold singularities, resulting in an infinite family of new dualities with an increasing number of gauge group factors.

In the second line of argument, developed in~\S\ref{sec:DC-preliminaries} and~\S\ref{sec:interpretation}, we
reformulate the system in terms of large volume objects,
i.e.\ $D7$ branes and $O7$ planes. We then connect the discussion in~\S\ref{sec:torsion} to the large volume perspective, giving a
direct brane interpretation of the different torsion assignments.
We show that the behavior of the resulting brane system
under S-duality of type IIB reproduces the duality structure found in
field theory. Critical to this statement is the behavior of the $O7$
plane at strong coupling, which we analyze in~\S\ref{sec:O7-S-duality}. Along the way we discuss in detail
some interesting points in the dictionary relating orientifolds at the
quiver point and large volume which are important for our
considerations.

Based on these arguments it seems natural to conjecture, as in~\cite{gauge}, that
$\bC^3/\bZ_3$ is just the simplest member of an infinite class of
toric geometries giving rise to $\cN=1$ S-dual pairs, including but not limited to the infinite family of orbifolds discussed in~\S\ref{sec:infinitefamilies}. To illustrate how our ideas can be generalized to these other cases, we also discuss a Seiberg dual (non-toric) phase of $dP_0$ in~\S\ref{sec:Z3-phaseII}. We defer consideration of non-orbifold examples, such as those introduced in~\cite{gauge}, to an upcoming work~\cite{strings-all}, where we also discuss an infinite family of dual gauge theories obtained from $D3$ branes probing orientifolds of the real cone over $Y^{2p,2p-1}$.

We close in~\S\ref{sec:conclusions} with our conclusions and a
review of the main questions that our analysis does not address. In appendix~\ref{sec:SD-mirror} we discuss some aspects of the mirror to $\bC^3/\bZ_3$
that complement and clarify the analysis in \S\ref{sec:Z3-phaseII}.

\medskip

While this paper was in preparation \cite{Bianchi:2013gka} was published which has some overlap with~\S\ref{sec:infinitefamilies}.

\section{Discrete torsion and S-duality}
\label{sec:torsion}

In this section we generalize the argument~\cite{Witten:1998xy} that
$O3$ planes fall into $\SL(2,\bZ)$ multiplets classified by their
discrete torsion\footnote{A more general treatment of discrete torsion
  than that needed here is given for orbifolds in
  \cite{Vafa:1986wx,Vafa:1994rv,Douglas:1998xa,Douglas:1999hq,Sharpe:2000ki,Sharpe:2000wu}
  and for orientifolds in \cite{Sharpe:2009hr}.}  to the case of
fractional $O7$ planes at an orbifold singularity.

We first review the argument for an $O3$ plane in a flat background, resulting in an $\cN=4$ $\SO$ or $\Sp$ gauge theory in the presence of $D3$ branes. The electromagnetic dualities which arise in these theories~\cite{Goddard:1976qe,Montonen:1977sn,Osborn:1979tq} can be understood by considering the action of the $\SL(2,\bZ)$ self-duality of type IIB string theory on the $O3$ plane.

The gravity duals of the different possible gauge theories are
distinguished by $B_2$ and $C_2$ discrete torsion on a cycle
surrounding the point in $\bR^6$ where the $O3$ is
located~\cite{Witten:1998xy}. To explain the geometric origin of this
discrete torsion, note that $B_2$ is not a globally-valued two-form,
but a connection on a
gerbe~\cite{Hitchin:1999fh,Kapustin:1999di}.\footnote{This is the
  viewpoint we adopt in this paper. A complete treatment of $C_2$ and
  $B_2$ in general orientifold backgrounds is more subtle
  \cite{Kapustin:1999di,Braun:2002qa,Distler:2009ri,Distler:2010an},
  and may be required to analyze more involved
  singularities.}  The underlying gerbe is classified by a cohomology
class $[H]\in H^3(M,\bZ)$, where the de Rham cohomology class of the
curvature three-form $H_3$ (locally $d B_2$) is the image of $[H]$
under the natural map $H^3(M,\bZ) \to H^3(M,\bR)$ which takes $\bZ \to
\bR$ and $\bZ_n \to 0$ for each factor of the cohomology group. The
same considerations apply to $C_2$, the integral class $[F]$, and the
curvature three-form $F_3$.

Before orientifolding the cycle surrounding the $D3$ branes is an
$S^5$, which has $H^3(S^5,\bZ) = 0$ and thus does not admit
a nontrivial gerbe. After orientifolding this
becomes a $\bRP^5$. Since $B_2$ and $C_2$ are odd under the worldsheet
part of the orientifold projection the associated gerbes are classified by the twisted cohomology group $H^3(\bRP^5, \widetilde{\bZ}) =
\bZ_2$. Hence the $B_2$ gerbe is classified by the ``discrete torsion'' $\theta_{NSNS} \in \bZ_2$, and likewise the $C_2$ gerbe by the discrete torsion $\theta_{RR} \in \bZ_2$, for a total of four possible topologically distinct configurations. Denoting the trivial and nontrivial elements of $H^3(\bRP^5, \widetilde{\bZ})$ as $\{0,\frac{1}{2}\}$ respectively, the four choices are
$(\theta_{RR},\theta_{NSNS})=\left\{(0,0), (\frac{1}{2}, 0), (0,
  \frac{1}{2}), (\frac{1}{2}, \frac{1}{2})\right\}$.

The action of $\SL(2,\bZ)$ on these torsion classes follows directly from its action on NSNS and RR fluxes. In particular, the action of the generator $T\in SL(2,\bZ)$ is:
\begin{align}
  T \begin{pmatrix}\theta_{RR}\\ \theta_{NSNS}\end{pmatrix}
  = \begin{pmatrix}1 & 1\\0&
    1\end{pmatrix}\begin{pmatrix}\theta_{RR}\\ \theta_{NSNS}\end{pmatrix}
  = \begin{pmatrix}\theta_{RR}+\theta_{NSNS}\\ \theta_{NSNS}\end{pmatrix}\, .
\end{align}
whereas the action of the $S\in SL(2,\bZ)$ S-duality generator is
\begin{align}
  S\begin{pmatrix}\theta_{RR}\\ \theta_{NSNS}\end{pmatrix}
  = \begin{pmatrix}0 & -1\\1 &
    0\end{pmatrix}\begin{pmatrix}\theta_{RR}\\ \theta_{NSNS}\end{pmatrix}
  = \begin{pmatrix}
    \theta_{NSNS}\\\theta_{RR}
    \end{pmatrix}\, ,
\end{align}
where we have used the $\bZ_2$ nature of the cohomology class to set
$\theta_{NSNS}=-\theta_{NSNS}$. Combining the two generators, we conclude that $(\theta_{RR},\theta_{NSNS})=(0,0)$ is an $\SL(2,\bZ)$ singlet whereas the three remaining choices with non-trivial torsion form an $\SL(2,\bZ)$ triplet.

Vanishing discrete torsion must therefore correspond to the $\mathcal{N}=4$ $\SO(N)$ gauge theory with even $N$, as this is the only case with a full $\SL(2,\bZ)$ self-duality.
Introducing discrete torsion for $B_2$
 leads to an extra sign in the worldsheet path integral for unoriented worldsheets
and changes the gauge group to $\Sp(\tilde{N})$.
The case with both $B_2$ and $C_2$ discrete torsion is related to this one by a shift of $C_0$. Up to an overall normalization $C_0$ is the theta angle of the gauge theory, so this case corresponds to the same $\Sp(\tilde{N})$ gauge theory at a different theta angle. By a process of elimination, we conclude that the remaining choice with only $C_2$ discrete torsion must correspond to the $\mathcal{N}=4$ $\SO(N)$ gauge theory with odd $N$, which is S-dual to the $\Sp(\tilde{N})$ gauge theory. $D3$ charge is $\SL(2,\bZ)$ invariant, which implies that $\tilde{N}=N-1$ under this S-duality, as expected from the Montonen-Olive duality relating $\SO(2k+1)$ with $\Sp(2k)$.

These identifications can be subjected to various consistency checks
via the AdS/CFT correspondence \cite{Witten:1998xy}. Here we confine
our attention to $D3$ branes wrapping the torsion three-cycle
generating $H_3(S^5/\bZ_2,\bZ) = \bZ_2$, corresponding to a particle
in four-dimensions. For $\SO(N)$ with even $N$ the wrapped brane is
dual to a single-trace operator: the Pfaffian of $N/2$ adjoint
scalars, a ``baryon'' of $\SO(N)$. The product of two Pfaffians can be
rewritten as the product of $N/2$ mesons, but a single Pfaffian cannot
be reduced to a product of mesons, since it is charged under the
$\bZ_2$ outer automorphism group of $\SO(N)$, perfectly reproducing
the $\bZ_2$ stability of the wrapped $D3$ brane.

For the other gauge groups there is no corresponding Pfaffian
operator. To understand this fact in the gravity dual, note that the
existence of a $\U(1)$ gauge bundle on a $D$-brane wrapping a cycle
$\Sigma$ embedded via the map $i\colon \Sigma\hookrightarrow \bRP^5$
imposes a restriction on the pullback of the $B_2$
gerbe~\cite{Witten:1998xy}:
\begin{equation} \label{eqn:FreedWitten}
i^*([H]) = W
\end{equation}
where $W \in H^3(\Sigma,\widetilde{\bZ})$ is a torsion class equal in the
absence of orientifolds to the third integral Stiefel-Whitney
class~\cite{Freed:1999vc}. We follow the prescription
of~\cite{Witten:1998xy} for computing $W$ in the presence of
orientifolds; in particular, this implies that $W$ vanishes whenever
$\Sigma$ admits a spin structure.

For the wrapped $D3$ brane considered above, $\Sigma$ has the topology of $\bR\bP^3$,
which is spin (as is any orientable three-manifold). Therefore $W=0$ according to the prescription
of~\cite{Witten:1998xy} and so a singly wrapped brane requires
$\theta_{NSNS}=0$. For a $D3$ brane an analogous condition also
restricts the $C_2$ gerbe $[F]$, so we must also require
$\theta_{RR}=0$. Thus, topologically stable wrapped $D3$ branes only
exist for the case of trivial discrete torsion, in perfect agreement
with field theory expectations.

\medskip

These arguments generalize readily to the case of $D3$ branes probing an orbifold singularity, for which the near-horizon geometry
is given by $AdS_5 \times X$ with $X$ a 5-dimensional lens space. Recall that the lens space
$L_n(a_1,a_2,a_3)$ is defined as the quotient of $S^5$ by the $\bZ_n$
action
\begin{align} \label{eqn:smoothorbifold}
  (z^1,z^2,z^3) \to (\omega_n^{a_1} z^1, \omega_n^{a_2} z^2, \omega_n^{a_3} z^3)\, ,
\end{align}
where $\omega_n\equiv\exp(2\pi i/n)$ and we have taken the natural
embedding of $S^5$ in $\bC^3$, with $\bC^3$ parameterized by
$(z^1,z^2,z^3)$. We confine our attention to the cases with unbroken supersymmetry and a smooth horizon, which requires $\sum_i a_i = 0\bmod n$ and $\gcd(a_i,n)=1$ respectively. For instance, the infinite family of orbifolds we study in~\S\ref{sec:OrbifoldFamily} have horizons $L_n(1,1,n-2)$, which is supersymmetric and smooth for odd $n$ and reduces to the horizon of $\bC^3/\bZ_3$ for $n=3$. More generally smoothness and supersymmetry together require odd $n$.\footnote{Unbroken $\mathcal{N}=2$ supersymmetry requires $a_3=0$ up to a permutation of the labels, precluding a smooth horizon except in the trivial $n=1$ case with maximal supersymmetry. Thus, all of the examples we study will have $\mathcal{N}=1$ supersymmetry.}

We choose the orientifold involution
\begin{align}
  z^i \to - z^i
\end{align}
This involution acts freely on the horizon for odd $n$, and therefore
corresponds to $O3$ planes or fractional $O7$ planes at the orbifold
singularity $z^i=0$. In particular, since $\bZ_n \times \bZ_2 \cong
\bZ_{2 n}$ for odd $n$, the orientifolded horizon is the smooth lens
space $L_{2n}(n+2 a_1,n+2 a_2,n+2 a_3)$, where $\gcd(n+2 a_i,2n)=1$
follows directly from $\gcd(a_i,n)=1$ for odd $n$.

We denote the orientifold quotient of the horizon as $X=Y/\bZ_2$. To classify discrete torsion in this geometry, we compute $H^3(X,\widetilde\bZ)$.
Since $X$ is orientable, $H^3(X,\widetilde\bZ)\cong H_2(X,\widetilde\bZ)$ by Poincare duality, whereas the latter group is more easily computed.
As we discuss below, $H_2(X,\widetilde \bZ)$ has a natural physical
interpretation as the group classifying possible domain walls changing
the rank and type of the orientifold theory.

The computation is facilitated by the existence of a long exact
sequence relating the twisted homology groups to ordinary
homology (see section 3.H in \cite{AT}
for a derivation):
\begin{equation}
  \label{eq:local-coeffs-from-global}
  \begin{gathered}
    \begin{tikzpicture}[descr/.style={fill=white,inner sep=1.5pt}]
      \matrix (m) [
      matrix of math nodes,
      row sep=1em,
      column sep=2.5em,
      text height=1.5ex, text depth=0.25ex
      ]
      { \ldots & H_i(X,\widetilde \bZ) & H_i(Y, \bZ) & H_i(X,\bZ) & \\
        & H_{i-1}(X,\widetilde \bZ) & H_{i-1}(Y, \bZ) & H_{i-1}(X,\bZ) & \ldots \\
      };

      \path[overlay,->, font=\scriptsize,>=latex]
      (m-1-1) edge (m-1-2)
      (m-1-2) edge (m-1-3)
      (m-1-3) edge node[above]{$p^i_\ast$} (m-1-4)
      (m-1-4) edge[out=355,in=175] (m-2-2)
      (m-2-2) edge (m-2-3)
      (m-2-3) edge node[above]{$p^{i-1}_\ast$} (m-2-4)
      (m-2-4) edge (m-2-5);
    \end{tikzpicture}
  \end{gathered}
\end{equation}
where the map $p^i_\ast$ is the
induced map on homology coming from the double covering $p\colon Y\to X$.
Since both $X$ and $Y$ are lens spaces their homology groups are
well known (see for instance example 2.43 in \cite{AT}). For $L_k(a,b,c)$ the homology groups are
$H_\bullet(L_k(a,b,c),\bZ)=\{\bZ, \bZ_k, 0, \bZ_k, 0, \bZ\}$. The maps $p^5_\ast$ and $p^0_\ast$ take $\bZ \to 2 \bZ$ and $\bZ \to \bZ$, respectively.
The action of $p^1_\ast$ and $p^3_\ast$ can be deduced by considering representative one and three-cycles $z^2=z^3=0$ and $z^3=0$ respectively, which gives $\bZ_n \to \bZ_n \subset \bZ_{2n}$ in both cases. The remaining cases are trivial, so $p^i_\ast$ is injective for all $i$ and the long exact sequence splits into short exact sequences:
\begin{equation}
0 \longrightarrow H_i(Y, \bZ) \overset{p^i_\ast}{\longrightarrow} H_i(X,\bZ) \longrightarrow H_{i-1}(X,\widetilde \bZ) \longrightarrow 0
\end{equation}
From this it is straightforward to compute $H_\bullet(X,\widetilde\bZ)=\{\bZ_2, 0, \bZ_2, 0, \bZ_2, 0\}$, and in particular $H_2(X,\widetilde\bZ) = \bZ_2$. The corresponding twisted two-cycle is an $\bRP^2$ given by the orientifold quotient of the $S^2 \subset S^5/\bZ_n$ defined by $\im(z^i)=0$.

By analogy with the $\cN=4$ examples, we expect three different gauge
theories corresponding to the possible choices of NSNS and RR discrete
torsion, where the two cases with NSNS discrete torsion give the same
gauge theory at different theta angles. Indeed, for $\bC^3/\bZ_3$,
there are three theories:\ the $\SO(N-4)\times\SU(N)$ theory
(\ref{eqn:dP0thyA}) with even and odd $N$ as well as the
$\Sp(\tN+4)\times\SU(\tN)$ theory~(\ref{eqn:dP0thyB}), where $\tN$ is
necessarily even. The field theory duality between the odd-$N$ $\SO$
theory and the $\Sp$ established in~\cite{gauge} suggests that they
form the expected $SL(2,\bZ)$ triplet of theories (which reduces to 2
perturbatively distinct theories, as in the $\cN=4$ case) and the
even-$N$ $\SO$ theory corresponds to vanishing discrete torsion and
has an $\SL(2,\bZ)$ self-duality. As before, we expect that the
introduction of NSNS torsion will change the sign of the orientifold
projection, and therefore the cases with NSNS torsion should
correspond to the $\Sp$ theory, whereas the case with only RR discrete
torsion should correspond to the $\SO$ theory with odd $N$. We thus
find that at the level of discrete torsion the duality structure is
compatible with the action of S-duality in IIB.

As a further check that we have identified the correct field theory duals of the different possible choices of discrete torsion, we consider D3 branes wrapping a torsion three-cycle, as above. For $\bC^3/\bZ_3$, these correspond to ``baryons'' charged under a $\bZ_6$ discrete symmetry. Indeed, for the $\SO$ theory with even $N$, there is a candidate $\bZ_6$ discrete symmetry as follows:
\begin{align} \label{eqn:dP0thyAevenN}
  \begin{array}{c|cc|ccc}
     & \SO(N-4) & \SU(N)  & \SU(3) & \U(1)_R & \cC \bZ_{6}\\ \hline
     A^i & \fund & \fund & \fund & \frac23+\frac2N & \omega_{6 N} \\
     B^i & 1 & \ov{\asymm} & \fund & \frac23 - \frac4N & \omega_{6 N}^{-2}
  \end{array}
\end{align}
where $\cC$ denotes the generator of the $\bZ_2$ outer automorphism
group of $\SO(N-4)$, so that $\cC \bZ_6$ is an (anomalous) flavor
symmetry, whereas $\bZ_6$ is anomaly-free but does not commute with
$\SO(N-4)$. The corresponding minimal baryon is the Pfaffian of $B$,
which in close analogy with the $\mathcal{N}=4$ case we conjecture to
be dual to the wrapped D3 brane.

For odd $N$ and for the $\Sp$ theory, there is no $\bZ_6$ discrete symmetry, but only a $\bZ_3$ discrete symmetry as in~(\ref{eqn:dP0thyA}, \ref{eqn:dP0thyB}), with corresponding minimal baryons $B^N$ and $\tilde{A}^\tN$ respectively. As before, this is explained in the gravity dual by the topological condition~(\ref{eqn:FreedWitten}), which requires a D3 brane to wrap the torsion three-cycle an even number of times in the presence of discrete torsion.

\medskip

As in~\cite{Witten:1998xy} (see also~\cite{Hyakutake:2000mr}), we can relate the different possible choices of discrete torsion with domain walls in $AdS_5$. We briefly review how the argument can be generalized to $\bC^3/\bZ_3$ and other orbifolds, as this result will be needed in~\S\ref{sec:microscopic-torsion}.

Domain walls in $AdS_5$ are three-branes, and can arise both from $D3$ branes at a point on $S^5/\bZ_6$ and from $D5$ or $NS5$ branes wrapping the torsion cycle $\bR\bP^2$ found above, where the dual gauge theory will in general change when crossing the wall. For unwrapped $D3$ branes, it is easy to see that the domain wall changes the $D3$ charge by one unit, and hence takes $N \to N+2$ without altering the discrete torsion. We now argue that domain walls coming from wrapped five-branes will change the discrete torsion.

To do so, consider the torsion cycles $S^3/\bZ_{2n} \subset X$ defined by $z^3 = 0$ and $\bRP^2\subset X$ defined by $\im(z^{1,2}) = 0$ and $\im(z^3+i z^1) = 0$. The latter is Poincare dual to the nontrivial element of $H^3(X,\tilde{\bZ})$, whereas it intersects the $S^3/\bZ_{2n}$ transversely at a point, itself Poincare dual on $S^3/\bZ_{2n}$ to the nontrivial element of $H^3(S^3/\bZ_{2n},\tilde{\bZ})$. We conclude that the pullback of the $B_2$ gerbe is trivial on $S^3/\bZ_{2n}$ if and only if it is trivial on $X=S^5/\bZ_{2n}$ and likewise for the $C_2$ gerbe.

We now consider a $D5$ brane wrapping the $\bR\bP^2$ at a radial distance $r_0$ from the orbifold singularity. We construct a four-chain $\Sigma$ given by $S^3/\bZ_{2n}$ times the interval $r \in [r_-,r_+]$ for $r_-<r_0<r_+$, which therefore intersects the $D5$ brane transerversely at a point. Cutting out a small ball $B$ surrounding the point of intersection, the $C_2$ gerbe is well-defined on the remainder $\Sigma - B$, where $\partial (\Sigma - B)$ is given by two copies of $S^3/\bZ_{2n}$ at $r_+$ and $r_-$ as well as $\partial B \cong S^3$. We now wish to relate $[F]$ on the three boundaries using the fact that the gerbe is well-defined in the interior of $\Sigma - B$.

We start by solving a simpler problem: suppose that $\omega$ is a closed three-form defined globally on a four-manifold $M$ with boundary $\partial M$. Stokes' theorem implies that the pullback of $\omega$ is trivial in $H^3(\partial M, \bR)$. Using the long exact sequence $\ldots \longrightarrow H^3(M,\bR) \overset{\times 2^{\ast}}{\longrightarrow} H^3(M,\bR) \longrightarrow H^3(M,\bZ_2) \longrightarrow 0$ associated to the short exact sequence $\bR \overset{\times 2}{\longrightarrow} \bR \longrightarrow \bZ_2$, we obtain a surjective map $H^3(M,\bR) \longrightarrow H^3(M,\bZ_2)$ which commutes with the pullback map. Thus, we conclude that if $[\omega]$ is an element of $H^3(M,\bZ_2)$, then its pullback is trivial within $H^3(\partial M,\bZ_2)$.

\begin{figure}
  \begin{center}
    \includegraphics[width=0.4\textwidth]{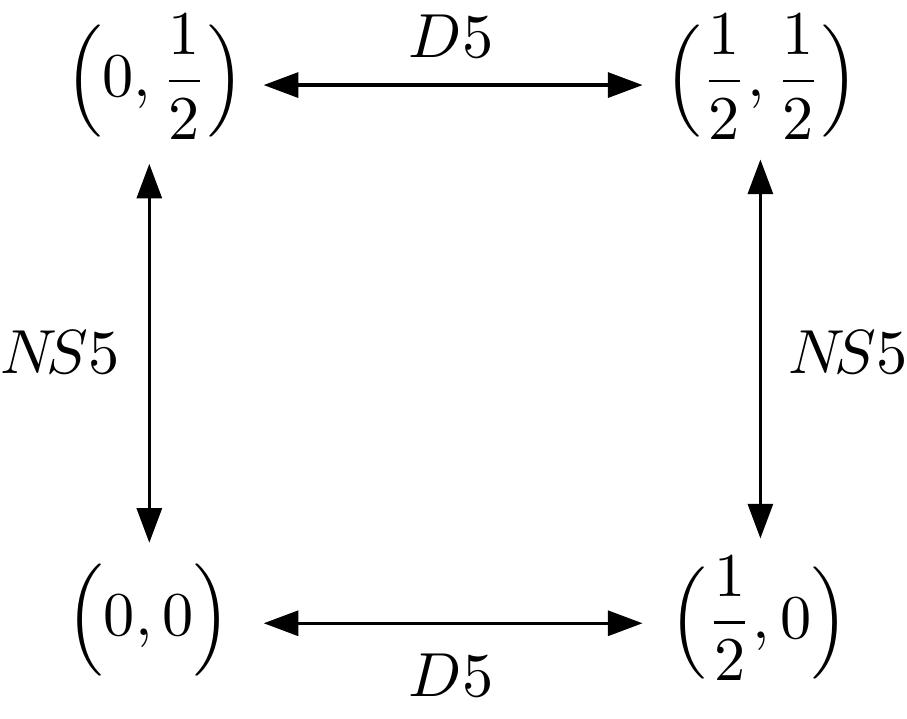}
  \end{center}
  \caption{The effect of wrapped five-brane domain walls on discrete torsion.}
  \label{fig:domainwalls}
\end{figure}

However, the long exact sequence associated to the short exact sequence $\tilde{\bZ} \overset{\times 2}{\longrightarrow} \tilde{\bZ} \longrightarrow \bZ_2$ also induces a surjective map
$f: H^3(M,\tilde{\bZ}) \longrightarrow H^3(M,\bZ_2)$ which once again commutes with the pullback map. For $S^3/\bZ_{2n}$ $f$ is an isomorphism, whereas for $\partial B \cong S^3$, $f$ maps $\bZ \to \bZ/2\bZ \cong \bZ_2$. Since $f([F])$ is well-defined on $\Sigma - B$, we conclude that the product of the classes of $f([F])$ pulled back to the three boundaries is trivial. Thus, for a single $D5$ brane (or any odd number) wrapping the $\bR\bP^2$, the $C_2$ discrete torsion jumps upon crossing the domain wall between $r_+$ and $r_-$. The same argument applies to wrapped $\NS$ branes and by extension to wrapped $(p,q)$ five branes. We summarize the situation in figure~\ref{fig:domainwalls}.

The geometric arguments presented in this section apply to other isolated orbifold singularities besides $\bC^3/\bZ_3$, suggesting that in each case there should be three different gauge theories corresponding to the different choices of discrete torsion. Heuristically, since $n$ is odd the parent quiver theory has an odd number of nodes, and applying the rules outlined in appendix A of~\cite{gauge} will lead to (at least) one $\SO$ or $\Sp$ gauge group factor. Thus, the pattern of discrete torsion can plausibly be explained in a manner closely analogous to the examples already discussed. Less trivially, we expect new gauge theory dualities relating the $\SO$ and $\Sp$ theories. The appearance of these dualities is a highly nontrivial check on the discrete torsion classification presented above, and is the subject of the next section.

\section{An infinite family of dual gauge theories}
\label{sec:infinitefamilies}

In this section we discuss the gauge theory dualities which arise from orientifolds of other $\cN=1$ orbifold singularities. In particular, based on the arguments given in the previous section, we expect a dual pair of gauge theories and a self-dual gauge theory for every isolated orbifold singularity. As a nontrivial check of this conjecture, we write down the possible gauge theories for the orientifolds of an infinite family of orbifold singularities and show that two of three possibilities have matching anomalies, consistent with an S-duality relating these theories.

We focus our attention on the $z^i \to - z^i$ orientifolds
of the orbifolds $\bC^3/\bZ_{n=2k+1}$\footnote{Throughout the discussion
we use both $n=2k+1$ and $k$ to simplify the presentation.} with the $\bZ_n$ action
\be
z^{1,2} \rightarrow \omega_n z^{1,2} \;\;,\;\; z^3 \rightarrow \omega_n^{-2} z^3 \,,
\ee
where $\omega_n\equiv\exp(2\pi i/n)$. The resulting geometry is toric,\footnote{See e.g.~\cite{Cox:2010tv,Garcia:2010tn} for a review of toric geometry.} where the $\U(1)^3$ toric isometry group is enhanced to $\SU(2)\times\U(1)\times\U(1)_R$ with $\SU(2)$ acting on the doublet $(z^1,z^2)$, except that in the special case of $\bC^3/\bZ_3$ (studied in~\cite{gauge}) the isometry group is further enhanced to $\SU(3)\times\U(1)_R$ with $\SU(3)$ acting on the triplet $(z^1,z^2,z^3)$.

\begin{figure}
\begin{center}
\subfigure[Toric diagram with points labeled.\label{fig:labeledToric}]{\includegraphics[width=0.32\textwidth]{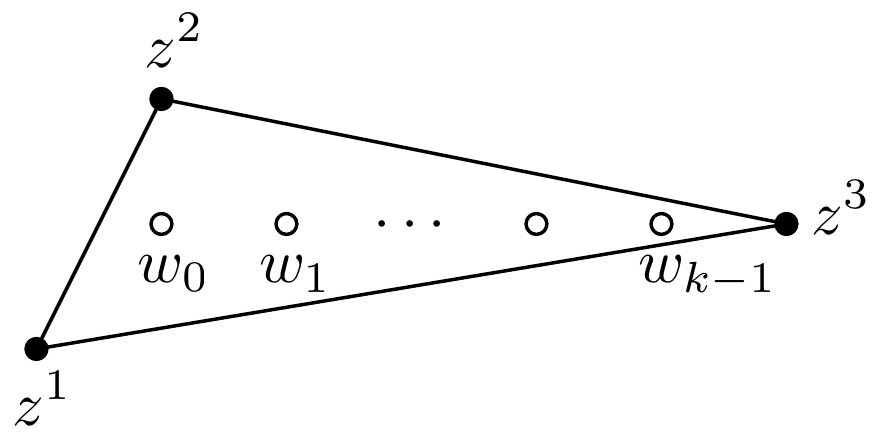}}\hspace{1mm}
\subfigure[Triangulated toric diagram.\label{fig:triangulatedToric}]{\includegraphics[width=0.32\textwidth]{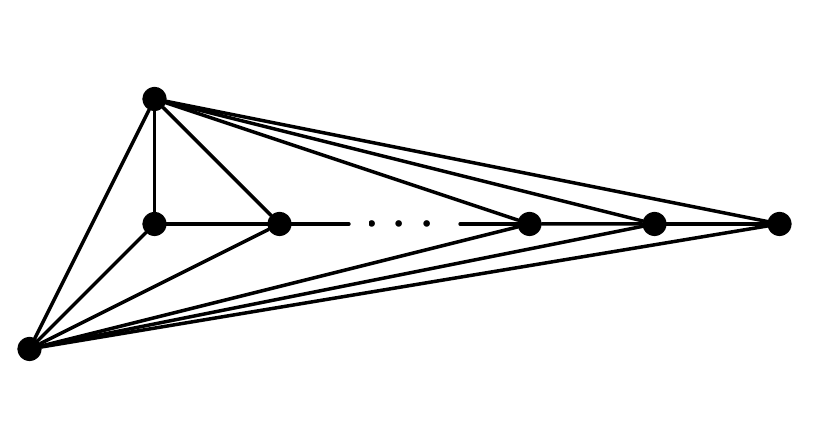}}\hspace{1mm}
\subfigure[Fan for the exceptional divisor $w_i=0$.\label{fig:exceptionalFan}]{\includegraphics[width=0.32\textwidth]{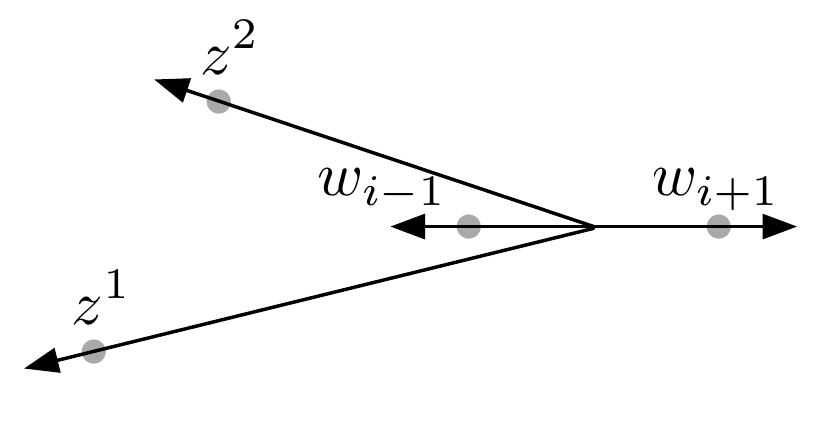}}
\end{center}
\caption{\subref{fig:labeledToric}~The toric diagram for the $\bC^3/\bZ_{2k+1}$ orbifold discussed in the text. The black vertices $z^1,z^2,z^3$ are fields in the corresponding gauged linear sigma model (GLSM) (see e.g.~\cite{Cox:1993fz, Kennaway:2007tq}), whereas each internal point corresponds to an exceptional divisor. \subref{fig:triangulatedToric}~Triangulating the toric diagram corresponds to fully resolving the geometry, where each internal point $w_j$ is now a GLSM field. \subref{fig:exceptionalFan}~The fan for the exceptional divisor $w_j=0$, with a ray for each edge connected to $w_j$ in the triangulated toric diagram. For $j>0$, this is the fan for the Hirzebruch surface $\bF_{2j+1}$, whereas for $j=0$ the left-facing ray is absent, and we obtain the fan for $\bP^2$.\label{fig:ZnToric}}
\end{figure}

The toric diagram\footnote{The fan of a toric Calabi-Yau threefold can be drawn with all of its primitive generators $\vec{n}_i \in \bZ^3$ in the $(\cdot,\cdot,1)$ plane~\cite{Kennaway:2007tq}. Having done so, the toric diagram is the intersection of the fan with this plane in $\bR^3$, and contains the same information as the fan: one, two, and three-dimensional cones in the fan correspond to vertices, edges and faces in the toric diagram.} corresponding to this singularity is shown in figure~\ref{fig:labeledToric}.
Fully resolving the singularity corresponds to triangulating the toric diagram, as in figure~\ref{fig:triangulatedToric}, giving a GLSM description with fields $z^i$ and $w_j$, $0\le j<k$. Acting on the resolved geometry, one can show that the involution $z^i \to -z^i$ has fixed planes $w_j = 0$ for $k-j$ odd, as well as a separate fixed point $z^1 = z^2 = w_0 = 0$ for even $k$. Thus, for odd $k$ there are $\frac{k+1}{2}$ $O7$ planes wrapping exceptional divisors $\bP^2$ and $\bF_{2 j+1}$ for $0<j<k$ even (see figure~\ref{fig:exceptionalFan}) where $\bF_m$ denotes the $m$th Hirzebruch surface, whereas for even $k$ there are $k/2$ $O7$ planes wrapping exceptional divisors $\bF_{2 j+1}$ for $0<j<k$ odd as well as an $O3$ plane at a point on the $\bP^2$.

To obtain the worldvolume gauge theory of $D3$ branes probing these orientifolds, we work within the framework of dimer models~\cite{Hanany:2005ve,Franco:2005rj}.
We refer the reader to~\cite{Franco:2007ii} for the state of the art on orientifolds of
generic dimer models, and~\cite{Franco:2010jv} for a treatment in a
dimer model language of a family of orientifolds similar to the one
considered here and in~\cite{strings-all}. Brane/orientifold systems
very similar to some of the ones we consider here have
appeared in the literature many times before, most often studied in
the CFT language. See for example~\cite{Bianchi:1991eu,Gimon:1996rq,Berkooz:1996iz,Angelantonj:1996uy,Aldazabal:1997wi,Antoniadis:1998ep,Camara:2007dy,Bianchi:2009bg}
for some relevant work on orientifolds and in particular orientifolds
of orbifolds.\footnote{We would like to highlight in particular~\cite{Angelantonj:1996uy}, where S-duality of a type I configuration
  T-dual to our main example, the orientifolded $\bC^3/\bZ_3$ quiver,
  was studied. (See also~\cite{Berkooz:1996iz,Angelantonj:1996uy,Aldazabal:1997wi,Antoniadis:1998ep,Camara:2007dy}
  for other heterotic/type I S-dual orbifold pairs.)  An important
  physical difference of these works with respect to the configuration
  studied here is that the heterotic/type I S-duality of string theory
  in ten dimensions generically gives rise to a weak/weak duality in four dimensions, while
  S-duality in our singular configurations naturally gives strong/weak
  dualities in the four-dimensional field theory.}

\subsection{An infinite family S-dual gauge theories}
\label{sec:OrbifoldFamily}

In this subsection we derive the worldvolume gauge theories for $D3$ branes probing the infinite family of orientifold singularities considered above.
As anticipated in~\S\ref{sec:torsion}, for each singularity we obtain three different gauge theories, two of which are expected to be S-dual. We demonstrate that the prospective S-dual gauge theories have matching anomalies. Other orbifold singularities not belonging to this infinite family are briefly considered in~\S\ref{sec:other-orbifolds-generalization}.

\begin{figure}
  \begin{center}
    \includegraphics[width=.7\textwidth]{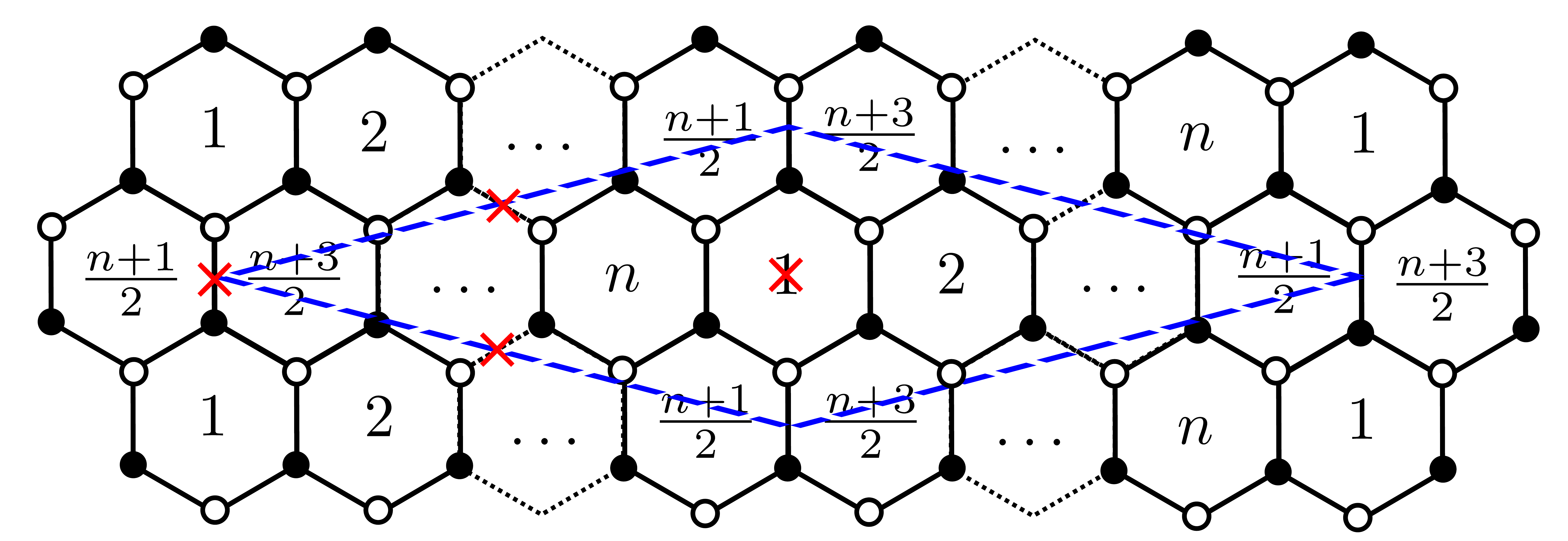}
  \end{center}
  \caption{The brane tiling for the $\bC^3/\bZ_n$ orbifold singularity described in the text. The red crosses indicate the orientifold fixed points and the dashed blue line outlines the unit cell.}
  \label{fig:Zntiling}
\end{figure}

The brane tiling corresponding to the $\bC^3/\bZ_n$ orbifold singularity described above is shown in figure~\ref{fig:Zntiling}.
The tiling is invariant under reflection about a horizontal line
through the middle row, which gives rise to a global
$\SU(2)$ symmetry in the corresponding gauge theory.

We focus on involutions of the dimer with isolated fixed points, since only these involutions can leave the global symmetries completely unbroken~\cite{strings-all}.
Since there are $2n$ nodes in the unit cell and
$n=2k+1$ is odd, the sign rule in~\cite{Franco:2007ii} requires the
product of the four signs associated to the four orientifold fixed
points to be odd. Due to the reflection symmetry there are six
inequivalent choices. Labelling the fixed point signs counter-clockwise from the leftmost fixed point in figure~\ref{fig:Zntiling}, these are $(\mp,\pm,\pm,\pm)$, $(\pm,\mp,\pm,\pm)$, and $(\pm,\pm,\mp,\pm)$. Following the meson sign rules given in~\cite{Franco:2007ii}, we find the corresponding geometric involutions $z^3 \to - z^3$, $z^1 \to - z^1$ and $z^i \to - z^i$, respectively. Thus, the cases $(\mp,\pm,\pm,\pm)$, $(\pm,\mp,\pm,\pm)$ correspond to non-compact $O7$ planes, and lead to field theories with gauge anomalies in the absence of flavor branes.

The remaining two cases $(\pm,\pm,\mp,\pm)$ correspond to the desired $z^i \to -z^i$ involution, and lead to anomaly-free gauge theories for certain choices of the gauge group ranks. For $(-,-,+,-)$, we obtain the gauge group $\prod_{a=1}^{k} \SU(N_a) \times \SO(N_{k+1})$ with the ranks fixed by anomaly
cancellation to be $N_a = N-4 \lfloor \tfrac{a}{2} \rfloor$ for some $N$. The
charge table for this theory is
\begin{center}
  \begin{tabular}{c|ccccccc|c}
     & $\SU(N_1)$ & $\SU(N_2)$ & $\SU(N_3)$ & \ldots & $\SU(N_{k-1})$ & $\SU (N_k)$ & $\SO(N_{k+1})$ & $\SU(2)$ \\ \hline
    $X^i$ & $\ov{\asymm}$ & 1 & 1 & \ldots & 1 & 1 & 1 & $\fund$\\
    $Y$ & $\fund$ & $\fund$ & 1 & \ldots & 1 & 1 & 1 & 1\\
    $A^i_{(1)}$ & $\fund$ & $\ov{\fund}$ & 1 & \ldots & 1 & 1 & 1 & $\fund$\\
    $B_{(1)}$ & $\ov{\fund}$ & 1 & $\fund$ & \ldots & 1 & 1 & 1 & 1\\
    $A^i_{(2)}$ & 1 & $\fund$ & $\ov{\fund}$ & \ldots & 1 & 1 & 1 & $\fund$\\
    \vdots & \vdots & \vdots & \vdots & \vdots & \vdots & \vdots & \vdots & \vdots\\
    $A^i_{(k-1)}$ & 1 & 1 & 1 & \ldots & $\fund$ & $\ov{\fund}$ & 1 & $\fund$ \\
    $B_{(k-1)}$ & 1 & 1 & 1 & \ldots & $\ov{\fund}$ & 1 & $\fund$ & 1\\
    $A^i_{(k)}$ & 1 & 1 & 1 & \ldots & 1 & $\fund$ & $\fund$ & $\fund$ \\
    $Z$ & 1 & 1 & 1 & \ldots & 1 & $\ov{\asymm}$ & 1 & 1
  \end{tabular}
\end{center}
where we have omitted the abelian global symmetries, which are discussed below.

The remaining choice $(+,+,-,+)$ is related to the $\SO$ theory described above by the negative rank duality discussed in
appendix B of \cite{gauge}. Therefore this theory has gauge group $\prod_{a=1}^{k} \SU(\tilde{N}_a) \times \Sp (\tilde{N}_{k+1})$ with $\tilde{N}_a = \tilde{N}+4 \lfloor \tfrac{a}{2} \rfloor$ and charge table
\begin{center}
  \begin{tabular}{c|ccccccc|c}
     & $\SU(\tilde{N}_1)$ & $\SU(\tilde{N}_2)$ & $\SU(\tilde{N}_3)$ & \ldots & $\SU(\tilde{N}_{k-1})$ & $\SU (\tilde{N}_k)$ & $\Sp(\tilde{N}_{k+1})$ & $\SU(2)$ \\ \hline
    $\tilde{X}^i$ & $\overline{\symm}$ & 1 & 1 & \ldots & 1 & 1 & 1 & $\fund$\\
    $\tilde{Y}$ & $\fund$ & $\fund$ & 1 & \ldots & 1 & 1 & 1 & 1\\
    $\tilde{A}^i_{(1)}$ & $\fund$ & $\ov{\fund}$ & 1 & \ldots & 1 & 1 & 1 & $\fund$\\
    $\tilde{B}_{(1)}$ & $\ov{\fund}$ & 1 & $\fund$ & \ldots & 1 & 1 & 1 & 1\\
    $\tilde{A}^i_{(2)}$ & 1 & $\fund$ & $\ov{\fund}$ & \ldots & 1 & 1 & 1 & $\fund$\\
    \vdots & \vdots & \vdots & \vdots & \vdots & \vdots & \vdots & \vdots & \vdots\\
    $\tilde{A}^i_{(k-1)}$ & 1 & 1 & 1 & \ldots & $\fund$ & $\ov{\fund}$ & 1 & $\fund$ \\
    $\tilde{B}_{(k-1)}$ & 1 & 1 & 1 & \ldots & $\ov{\fund}$ & 1 & $\fund$ & 1\\
    $\tilde{A}^i_{(k)}$ & 1 & 1 & 1 & \ldots & 1 & $\fund$ & $\fund$ & $\fund$ \\
    $\tilde{Z}$ & 1 & 1 & 1 & \ldots & 1 & $\overline{\symm}$ & 1 & 1
  \end{tabular}
\end{center}
where we once more omit the abelian global symmetries.

\begin{figure}
  \begin{center}
    \includegraphics[width=\textwidth]{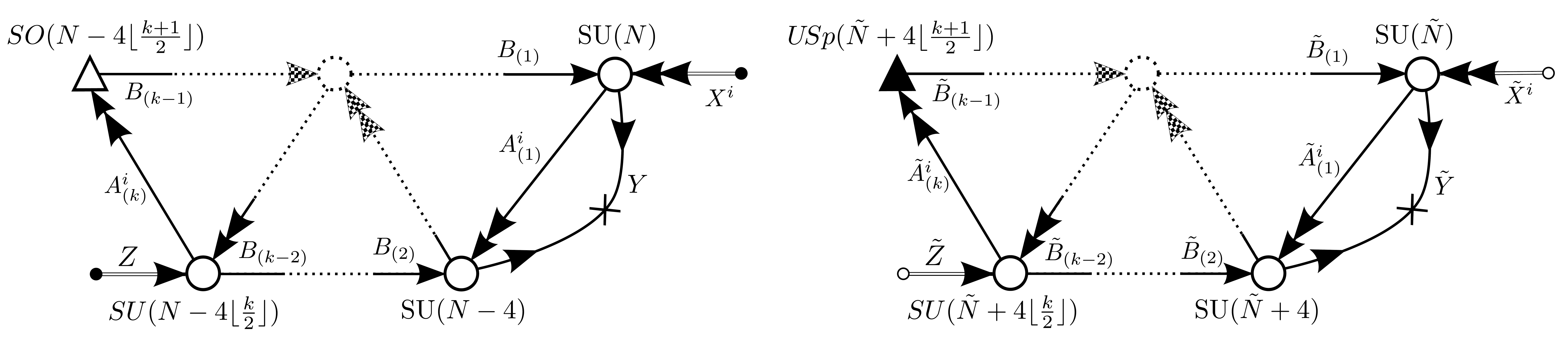}
  \end{center}
  \caption{The quivers for the two gauge theories arising from orientifolding $D3$-branes probing the $\bC^3/\bZ_{n=2k+1}$ singularity. There are $k$ $\SU$ factors. The $X^i$ and $Z$ matter fields of the $\SO$ theory transform as $\ov{\asymm}$ while the matter fields $\tilde{X}^i$ and $\tilde{Z}$ of the $\Sp$ model transform as $\overline{\symm}$.}
  \label{fig:Znquivers}
\end{figure}

The superpotential for the $\SO$ theory is
\be
W = \epsilon_{ij} \Tr \lp X^i A^j_{(1)} Y + \sum_{a=1}^{k-1} A^i_{(a)} A^j_{(a+1)} B_{(a)} +  A^i_{(k)} A^j_{(k)} Z \rp\,,
\ee
and for the $\Sp$ theory one has similarly
\be
\tilde{W} = \epsilon_{ij} \Tr \lp \tilde{X}^i \tilde{A}^j_{(1)} \tilde{Y} + \sum_{a=1}^{k-1} \tilde{A}^i_{(a)} \tilde{A}^j_{(a+1)} \tilde{B}_{(a)} +  \tilde{A}^i_{(k)} \tilde{A}^j_{(k)} \tilde{Z} \rp\,.
\ee
The quiver diagrams for the two gauge theories are shown in figure~\ref{fig:Znquivers}.

In addition to the continuous $\SU(2)\times\U(1)\times\U(1)_R$ global symmetry group, there are sometimes additional discrete symmetries. In particular, an extra $\bZ_n$ symmetry appears whenever $N$ or $\tilde{N}$ is a multiple of $n$, whereas an extra $\bZ_2$ symmetry is present in the $\SO$ theory for even $N$. The latter arises from a combination of the $\bZ_2$ outer automorphism group of $\SO(2 m)$ with a discrete flavor symmetry. Thus, a duality (if it exists) must relate the $\Sp$ theory and the odd-$N$ $\SO$ theory with $N-\tilde{N}$ an odd multiple of $n$.

The fields carry the following charges under the $\U(1) \times \U(1)_R
\times \bZ_n$ symmetries\footnote{The discrete symmetry
  group is $\bZ_{n}$ rather than $\bZ_{nN}$ since the generator to the
  $n$-th power is gauge equivalent to the identity. The generator of
  the discrete $\bZ_n$ symmetry to the $k N$-th power is gauge
  equivalent to a global $\U(1)$ transformation, which means that the
  global discrete $\bZ_n$ symmetry is contained in the $\U(1)$ unless
  $N$ is a multiple of $n$. The R-charges are assigned so that the global
  anomalies take a relatively simple form. We discuss the correct
  R-charges obtained from $a$-maximization below.}\,\,\footnote{For simplicity, we omit the extra $\bZ_2$ symmetry which appears for even $N$ in the $\SO$ theory.}
\begin{center}
\renewcommand\arraystretch{1.6}
  \begin{tabular}{c|c|c|c}
    & $X^i$ & $Y$ & $A^i_{(a)}$, $a$ odd \\ \hline
    $\U(1)$ & $1$  & $-2+\frac{2(n-3)}{N-4}$ & $1+\frac{(n-2a+1)(a-1)}{N-2(a-1)}-\frac{(n-2a-1)(a+1)}{N-2(a+1)}$ \\ \hline
    $\U(1)_R $ & $\frac23 -\frac{n\mp1}{N}$ & $\frac23+\frac{n\mp1}{2N}+\frac{(n-3)5+3\pm3}{6(N-4)}$ & $\frac23+\frac{(n-2a+1)(4a-1)+3 \mp3}{6(N-2(a-1))}-\frac{(n-2a-1)(4a+1)+3\pm3}{6(N-2(a+1))}$\\\hline
    $\bZ_n$ & $\omega_{nN}^{-2}$ & $\omega_{nN} \cdot \omega_{n(N-4)}^3$ & $\omega_{n(N-2(a-1))}^{2a-1} \cdot \ls \omega_{n(N-2(a+1))}^{-2a-1} \rs^{1-\delta_{a,k}}$ \\
  \end{tabular}
\end{center}

\begin{center}
\renewcommand\arraystretch{1.6}
  \begin{tabular}{c|c|c}
    & $B_{(a)}$, $a$ odd & $A^i_{(a)}$, $a$ even\\ \hline
    $\U(1)$ & $-2-\frac{(n-2a+1)(a-1)}{N-2(a-1)}+\frac{(n-2a-3)(a+1)}{N-2(a+1)}$ & $1+\frac{2a}{N-2a}$ \\ \hline
    $\U(1)_R $ & $\frac23-\frac{(n-2a+1)(4a-1)+3 \mp3}{6(N-2(a-1))}+\frac{(n-2a-3)(4a+7)+3 \mp3}{6(N-2(a+1))}$ & $\frac23+\frac{10a-3(n\mp1)}{3(N-2a)}$\\\hline
    $\bZ_n$ & $\omega_{n(N-2(a-1))}^{1-2a} \cdot \ls \omega_{n(N-2(a+1))}^{2a+3}\rs^{1-\delta_{a,k-1}}$ & $\omega_{n(N-2a)}^{2a-1} \cdot \ls \omega_{n(N-2a)}^{-2a-1}\rs^{1-\delta_{a,k}}$ %$\omega_{n(N-2a)}^{-2} $
  \end{tabular}
\renewcommand\arraystretch{1}
\end{center}

\begin{center}
\renewcommand\arraystretch{1.6}
  \begin{tabular}{c|c|c}
    & $B_{(a)}$, $a$ even & $Z$\\ \hline
    $\U(1)$ & $-2-\frac{(n-2a+1)a}{N-2a}+\frac{(n-2a-3)(a+2)}{N-2(a+2)}$ &  $-2-\frac{2(n-2\pm1)}{N-n+2 \mp1}$\\ \hline
    $\U(1)_R $ & $\frac23-\frac{(n-2a+1)(4a-3)+3 \pm3}{6(N-2a)}+\frac{(n-2a-3)(4a+5)+3\pm3}{6(N-2(a+2))}$ & $\frac23-\frac{2(2n-1\mp1)}{3(N-n+2\mp1)}$\\\hline
    $\bZ_n$ & $\omega_{n(N-2a)}^{1-2a} \cdot \ls \omega_{n(N-2(a+2))}^{2a+3}\rs^{1-\delta_{a,k-1}}$ & $\omega_{n(N-n+2\mp 1)}^{-2(n-2)}$
  \end{tabular}
\renewcommand\arraystretch{1}
\end{center}
where $\omega_m \equiv e^{2\pi i/m}$ and the upper/lower sign is for $k=(n-1)/2$ even/odd. The $\U(1) \times \U(1)_R \times \bZ_n$ charges for the $\Sp$ theory are obtained by replacing $N \rightarrow -\tilde{N}$, as dictated by the negative rank duality relating the two theories.

Calculating the global anomalies that are relevant for anomaly
matching in dual theories~\cite{Csaki:1997aw} one finds that the
$\SU(2)^2 \, \bZ_n$ anomaly as well as the gravitational $\bZ_n$
and $\U(1)$ anomalies vanish and the other anomalies are given by
\begin{center}
\renewcommand\arraystretch{1.1}
  \begin{tabular}{c|c|c}
    & $\SO$ theory & $\Sp$ theory\\ \hline
    $\SU(2)^3$ & $\frac12 N(N-n) $ mod 2 & $\frac12 \tilde{N}(\tilde{N}+n) $ mod 2\\ \hline
    $\SU(2)^2 \, \U(1)$ & $\frac12 n N (N-n)$ & $\frac12 n \tilde{N} (\tilde{N}+n)$\\ \hline
    $\SU(2)^2 \, \U(1)_R$ & $\frac16 (\pm3-n(2+n^2+N(N-n)))$ & $\frac16 (\pm3-n(2+n^2+\tilde{N}(\tilde{N}+n)))$ \\ \hline
    $\U(1)^3$ & $-3n N(N-n)$ & $-3n \tilde{N}(\tilde{N}+n)$ \\\hline
    $\U(1)^2 \, \U(1)_R$ & $-n (n^2\mp n+ N(N-n))$ & $-n (n^2 \mp n+ \tilde{N}(\tilde{N}+n))$ \\\hline
    $\U(1) \, \U(1)_R^2$ & $-\frac13 (n\pm3)n(n\mp1)$ & $-\frac13 (n\pm3)n(n\mp1)$ \\\hline
    $\U(1)_R^3$ & $\begin{array}{c}
                     \frac{1}{18}(9\pm90-n(99+6n(n\mp1) \\
                     -8 N(N-n)))
                   \end{array}$
     & $\begin{array}{c}
                     \frac{1}{18}(9\pm90-n(99+6n(n\mp1) \\
                     -8 \tilde{N}(\tilde{N}+n)))
                   \end{array}$ \\\hline
    $\U(1)_R$ & $-\frac12 \lp 5 n -1 \mp 4\rp$ & $-\frac12 \lp 5n -1 \mp 4\rp$
  \end{tabular}
\renewcommand\arraystretch{1}
\end{center}
where again the upper/lower sign is for $k$ even/odd. We conclude that the two
theories have matching anomalies for $\tilde{N}=N-n$. This is a highly non-trivial check of our previous assertion that these theories should be S-dual, and is in perfect agreement with the arguments given in~\S\ref{sec:torsion}.

As an aside we note that the two dual theories have the same number of chiral multiplets and vector multiplets. In particular for $\tilde{N}=N-n$ we find for the $\Sp$ theory
\ba
&\text{number of vector multiplets:     } & \frac16 (3\mp9 +n(n^2+5+ 3N(N-n))\\
&\text{number of chiral multiplets:     } & \frac{1}{2} \left(6\pm3 + n \left(n^2-10+3N(N-n)\right)\right)
\ea
which agrees with the $\SO$ theory where as usual the upper/lower sign is for $k$ even/odd. There is no obvious reason for the dual theories to be related in this fashion, and indeed the relation does not persist for nonorbifold singularities~\cite{gauge, strings-all}.

For completeness, we describe $a$-maximization for these theories.
To find the superconformal R-charge we define a trial R-symmetry under
which the fields carry the charge $q_R + b\, q_{\U(1)}$, where $q_R$
and $q_{\U(1)}$ are the charges given above. Since the gravitational
anomaly for the $U(1)$ flavor symmetry vanishes,
$a$-maximization \cite{Intriligator:2003jj} reduces to maximization
with respect to $b$ of the $\U(1)_R^3$ anomaly for the trial
R-symmetry. We find for the $\SO$ theory
\begin{align}
b
=-\frac{n(n\mp1)+\!N(N-n)-\sqrt{n^2(n\mp1)^2+\!N(N-n)(n\mp3)(n\mp1)+\!N^2(N-n)^2}}{3N(N-n)}\,,
\end{align}
where again the upper/lower signs are for $k=(n-1)/2$ even/odd. For
the $\Sp$ theory $b$ is given by replacing $N \rightarrow
-\tilde{N}$. Note that the vanishing of the gravitational $\U(1)$
anomaly implies that the $\U(1)_R^2\, \U(1)$ anomaly vanishes
after $a$-maximization. Using the above formula, the central charge and other anomalies involving the
R-symmetry can easily be obtained. The results are
rather lengthy, so we refrain from spelling them out explicitly.

\subsection {Generalization to other orbifolds}
\label{sec:other-orbifolds-generalization}

A general supersymmetric $\bC^3/\bZ_n$ orbifold with an isolated singularity takes the form
\begin{equation}
z^j \rightarrow e^{\frac{2\pi i a_j}{n}} z^j
\end{equation}
where $\sum a_i = 0 \bmod n$ and $\gcd(a_i,n)=1$, so that $(a_1,a_2,a_3) = (1,\ell,-\ell-1)$ for $\gcd(\ell,n)=\gcd(\ell+1,n)=1$. The first nontrivial example not belonging to the infinite family discussed above is the $\bZ_7$ orbifold $(1,2,4)$. Choosing the same $z^i \to - z^i$ involution
one obtains the gauge theories found in~\cite{Kakushadze:1998tr}. The gauge groups are $\SO(N+4) \times \SU(N)^3$ and $\Sp(\tN-4) \times \SU(\tN)^3$, where the explicit charge table is given in Table III of \cite{Kakushadze:1998tr}. Both theories have a global $\U(1)^2 \times \U(1)_R$ symmetry, and their anomalies match for $\tN = N+1$, as expected based on our arguments in~\S\ref{sec:torsion}.

We leave it to the interested reader to work out the details of other $\bZ_n$ orbifolds with isolated singularities, which are expected to behave similarly to the cases studied here. For simplicity, we omit discussion of non-isolated singularities --- such as for $\bZ_n$ orbifolds with even $n$, $\bZ_m \times \bZ_n$ orbifolds with $\gcd(m,n)>1$, and nonabelian orbifolds --- and proceed to discuss a different physical viewpoint on the duality.

\section{The large volume picture}
\label{sec:DC-preliminaries}

While the arguments presented in~\S\ref{sec:torsion} provide a clear
link between the duality relating the $\bC^3/\bZ_3$ $\SO$ and $\Sp$
gauge theories and ten-dimensional S-duality, the interpretation of
the duality in terms of branes is initially less obvious. Whereas the
$\cN = 4$ case involved $O3$ planes, which transform into each other
under $\SL(2,\bZ)$, the $\bC^3/\bZ_3$ orientifold we consider
corresponds to an $O7$ plane wrapping the $\bP^2$ exceptional
divisor. Since $O7$ planes do not transform simply under $\SL(2,\bZ)$
(in particular, the S-dual of an $O7$ plane is not an $O7$ plane), the
$\cN = 4$ story requires substantial modification to correctly
describe the ``microscopics'' of how the duality acts on the
fractional branes. The primary goal of the following sections is to
develop this story. Along the way, we will also provide an explanation
for the rank relation $\tN = N - 3$ in terms of $D3$ charge
conservation.

In order to systematically study D-branes in type IIB string theory,
it is convenient to work in the framework of the derived category of
coherent sheaves (we refer the reader to
\cite{workingMath,Sharpe:1999qz,Douglas:2000gi,Sharpe:2003dr,Aspinwall:2004jr}
for excellent reviews and some of the original works on this topic in
the physics literature). For completeness we review certain parts of
this description below, highlighting those aspects that will be most
important in our analysis. Much of the following formalism is well
understood in the absence of orientifolds, see for example
\cite{Douglas:2000qw,Cachazo:2001sg,Wijnholt:2002qz,Herzog:2003dj,Aspinwall:2004vm,Wijnholt:2005mp,Hanany:2006nm}
for some early works. The action of orientifolds on the derived
category of coherent sheaves has been discussed in
\cite{Diaconescu:2006id}; we follow the formalism and notation in that
paper, extending it to include non-trivial $B_2$ fields and
auto-equivalences of the category.

\subsection{Preliminaries on derived categories and orientifolds}
\label{sec:dc-general}

In the language of the derived category, branes are described by a
complex of sheaves $\mathfrak{E}$ in an ambient space $X$. We will be interested in
branes wrapping a complex surface $\cS$ in a Calabi-Yau manifold
$X$, with embedding map $i:\cS\hookrightarrow X$, and supporting a
sheaf $\cE$, possibly with some non-trivial integer shift in the
grading. In other words, we do not need to deal with general
complexes, but only objects of the form $\mathfrak{E}=i_*\cE[k]$, with
$\cE$ an ordinary sheaf on $\cS$ and $k$ the position of
$\cE$ in the complex.\footnote{For convenience of notation, we will
  often denote the brane described by the sheaf $i_*\cE[k]$ simply by
  $\cE$. Whether we are talking of the brane or its associated bundle
  should always be clear from the context.}

The branes $\cE$ corresponding to the fractional branes at the
singularity can be constructed by (left) \emph{mutation} of a basic
set of projective objects $\sP_i$
\cite{BondalHelixes,Cachazo:2001sg,Feng:2002kk,Herzog:2003dj,Aspinwall:2004vm}.
On a del Pezzo surface $\cS$ the $\sP_i$ objects can be easily
constructed as line bundles on $\cS$, and there are systematic
algorithms for finding such a collection \cite{Hanany:2006nm}. We will
give various examples below. The left mutation of a brane $\cF_i$
through $\cF_j$, denoted $\sL_{\cF_j}(\cF_i)$, is defined as
\cite{Aspinwall:2004vm}:
\begin{align}
  \label{eq:left-mutation}
  \sL_{\cF_j}(\cF_i) = \cone\bigl(\Hom(\cF_j, \cF_i)\otimes\cF_j \to
    \cF_i\bigr)[-1]\, .
\end{align}
We refer the reader to \cite{Aspinwall:2004jr} for a review of the
cone construction. If one is only interested in the Chern characters
of the branes, then~\eqref{eq:left-mutation} simplifies to:
\begin{align}
  \ch(\sL_{\cF_j}(\cF_i)) = \ch(\cF_i) - \chi(\cF_j,\cF_i)\ch(\cF_j) \,,
\end{align}
with
\begin{align}
  \chi(\cF_j,\cF_i) = \int_\cS \ch(\cF_j^\vee)\wedge \ch(\cF_i)\wedge
  \Td(T_\cS)\, .
\end{align}
Using these definitions, a basis of fractional branes can be
constructed by taking:
\begin{align}
  \begin{split}
    \cE_1 & = \sP_1 \, ,\\
    \cE_2 & = \sL_{\sP_1}\sP_2 \, ,\\
    & \cdots\\
    \cE_{n-1} & = \sL_{\sP_1}\sL_{\sP_2}\cdots\sL_{\sP_{n-2}}\sP_{n-1}\, ,\\
    \cE_n & =
    \sL_{\sP_1}\sL_{\sP_2}\cdots\sL_{\sP_{n-1}}\sP_{n}\, .
  \end{split}
\end{align}

\medskip

Now that we know how to construct the fractional branes at the
singularity in terms of geometric objects, we want to define a
suitable orientifold action. Consider first the case with vanishing
$B_2$ field. In the case that the orientifold wraps $\cS$ itself, the
action on the fractional branes is given by \cite{Diaconescu:2006id}
\begin{align}
  \label{eq:orientifold-action}
  i_*\cE[k] \longrightarrow i_*(\cE^\vee\otimes K_\cS)[2-k] \, ,
\end{align}
with $K_\cS$ the anti-canonical class of $\cS$. This agrees with the
usual large volume action on $D7$s, which is generally considered only
at the level of Chern classes. In this
case~\eqref{eq:orientifold-action} maps $D7$s to $D7s$ and $\ov{D7}$s
to $\ov{D7}$s. Furthermore, the worldvolume flux $F$ on the brane is
related to $\cE$ by $F=\cE\otimes K_\cS^{-1/2}$
\cite{Katz:2002gh}. Acting on $F$, \eqref{eq:orientifold-action} then
gives $F\to F^\vee$, in agreement with the usual prescription.

Incorporating a $B_2$ field in $H^2(\cS,\bR)$ is relatively straightforward. Usually one introduces a quantity
$\mathcal{F}=c_1(\cF)-B_2$, with $c_1(\cF)$ the field strength for the
connection in the bundle $\cF$, in terms of which the orientifold acts
as $\mathcal{F}\to -\mathcal{F}$.

However, there is an equivalent alternative
viewpoint that fits better with the derived-category description of
branes. Notice that since orientifolds map $B_2\to -B_2$, only
half-integrally quantized $B_2$ fields are allowed, and we can view $2B_2$ as the field strength of a line bundle $\cL_{2B_2}$.
The orientifold therefore has
a double effect: it acts on the D-branes as in~\eqref{eq:orientifold-action} while also reflecting the real
part of the K\"ahler moduli space. We can trivially undo the action on
K\"ahler moduli space (so we can compare branes and their images at
the same point in moduli space) by shifting $-B_2\to -B_2+2B_2 = B_2$. Due to the
invariance of the theory under joint integral shifts of $B_2$ and the
bundles on the branes, this is equivalent to tensoring the sheaf $\cE$
on the brane by $\cL_{2B_2}$. Thus, in the presence of $B_2\in
H^2(\cS,\bR)$, \eqref{eq:orientifold-action} gets amended
to
\begin{align}
  \label{eq:orientifold-action-B}
  i_*\cE[k] \longrightarrow i_*(\cE^\vee\otimes K_\cS\otimes \cL_{2B_2})[2-k] \, .
\end{align}
This action can also be understood purely in
terms of the derived category, forgetting about the physical origin of
$\cL_{2B_2}$. From this viewpoint, \eqref{eq:orientifold-action-B} generalizes
the ordinary action of the orientifold by twisting the elements of the
category with a line bundle. Twisting all the elements of the category
by the same line bundle is an autoequivalence of the category, so we
have our first example of an orientifold that combines the ordinary large-volume orientifold action with an auto-equivalence of the
derived category. We discuss a variation of this idea below, which
turns out to be useful in understanding the orientifolds of
various quiver configurations.

\medskip

While the description of the branes in terms of the derived category
of coherent sheaves is relatively simple (at least in comparison with
the objects in the mirror description, the Fukaya category, see
\cite{Aspinwall:2004jr} for a review), there is an important
complication: as we move in K\"ahler moduli space the supersymmetry
preserved by the D-branes changes, in a way highly influenced by
world-sheet instanton corrections. The most convenient way to deal
with this issue is by considering the central charge $Z(\cE,t)$, with
$\cE$ our brane of interest and $t=B_2+iJ$ our position in
complexified K\"ahler moduli space.\footnote{For the sake of brevity,
  we will often drop the dependence on $t$ from the notation. Also,
  despite the fact that in $\cN=1$ compactifications the holomorphic
  field involving $J$ is $C_4-\tfrac{i}{2} J \w J$, we will keep referring to
  $t$ as parameterizing the complexified K\"ahler moduli space, since
  this is the natural variable entering the central charge formulas
  for BPS D-branes in the theory.} At large volume, where we can
ignore $\alpha'$ corrections, a brane $\cE$ wrapping $\cS$ has central
charge
\begin{align}
  Z(\cE[k], t) = (-1)^k \int_\cS e^{-t}  \ch(\cE)\sqrt{\frac{\Td(T_\cS)}{\Td(N_\cS)}}\, .
\end{align}
It will be convenient to introduce a charge vector for the brane given
by:
\begin{align}
  \label{eq:charge-vector}
  \Gamma(\cE[k]) = (-1)^k[S]\, \ch(\cE)\sqrt{\frac{\Td(T_\cS)}{\Td(N_\cS)}}
\end{align}
in terms of which the large volume central charge is given by:
\begin{align}
  \label{eq:central-charge}
  Z(\cE[k], t) = \int_X e^{-t} \, \Gamma(\cE[k])\, .
\end{align}

One can incorporate $\alpha'$ corrections to the central charge by
going to the mirror of the configuration in question. In what follows, we quote
the relevant results as needed, referring the interested reader to
\cite{Cox:2000vi,Hori:2003ic,Aspinwall:2004jr} for surveys of the techniques required to derive these results and further references.

\medskip

Once we have the set of branes and the orientifold action we can
compute the spectrum of light states. The precise calculation requires
computation of $\Ext$ groups \cite{Katz:2002gh}. In the particular
case that we will be considering --- $\Ext$ groups between sheaves
$A$, $B$ supported on a surface $\cS$ in the Calabi-Yau $X$ --- there
is a one-term spectral sequence \cite{SeidelThomas,Katz:2002gh} giving
the $\Ext$ groups in $X$ in terms of $\Ext$ groups in $\cS$:
\begin{align}
  \begin{split}
    \Ext^k_X(i_*A, i_*B)  & = \sum_{p+q=k}\Ext^p_\cS(A, B\wedge N_\cS^q)\\
    & = \Ext^k_\cS(A,B) \oplus \Ext^{k-1}_\cS(A, B\otimes K_\cS)\, ,
  \end{split}
\end{align}
where in the second line we have used the fact that $\cS$ is a divisor
in a Calabi-Yau, so $N_\cS=K_\cS$. Using Serre duality on $\cS$ we can
rewrite the final expression as:
\begin{align}
  \label{eq:ext-restriction}
  \Ext^k_X(i_*A, i_*B) = \Ext_\cS^k(A,B)\oplus
  \Ext_\cS^{3-k}(B,A)^\vee\, .
\end{align}
If one were interested only in the dimensions of the $\Ext$ groups
 the dual sign in the second term could be ignored, but we will keep it as
it nicely encodes some of the flavor structure of the quiver, as demonstrated in some of the examples below.

For our purposes it is usually sufficient to compute only
the chiral index of states between the branes. This is defined as:
\begin{align}
  \dsz{\cE_i}{\cE_j} = \sum (-1)^k \Ext^k_X(\cE_i,\cE_j)\, ,
\end{align}
with $X$ the Calabi-Yau threefold. As is often the case with an index, this
expression can be expressed as an integral of forms on $X$. In
particular, it is given by the Dirac-Schwinger-Zwanziger (DSZ) product
of the corresponding charge vectors:
\begin{align}
  \label{eq:dsz-general}
  \dsz{\cE_i}{\cE_j} = \sum_k (-1)^k \int_X \Gamma^{(2k)}(\cE_i)\,\,\wedge\,\,
  \Gamma^{(6-2k)}(\cE_j)\, ,
\end{align}
with $\Gamma^{(m)}$ denoting the part of $\Gamma$ of degree m. This
product is clearly antisymmetric, and in fact it is the mirror to the
usual intersection product in IIA. Since we have branes wrapping a
complex surface $\cS$, the charge vector takes the form:
\begin{align}
  \Gamma(\cE_i) = [\cS]\wedge \left(\omega^{(i)}_0 + \omega^{(i)}_2 +
  \omega^{(i)}_4\right)\, ,
\end{align}
with $\omega_n\in H^n(\cS,\bR)$. Eq.~\eqref{eq:dsz-general} then
simplifies to:
\begin{align}
  \label{eq:dsz-product}
  \dsz{\cE_i}{\cE_j} = \int_\cS c_1(T_\cS)\wedge
  \left(\omega_0^{(i)}\wedge \omega_2^{(j)} - \omega_0^{(j)}\wedge
    \omega_2^{(i)}\right) \,,
\end{align}
where we have used adjunction: $c_1([\cS]|_\cS) = c_1(N_\cS) =
-c_1(T_\cS)$, since $X$ is a Calabi-Yau manifold.

\begin{table}
  \begin{center}
    \begin{tabular}{c | c}
      Number of chiral multiplets & Representation\\
      \hline
      $\dsz{\cE_i}{\cE_j}$ & $(\fund_i,\ov\fund_j)$\\
      $\dsz{\cE_i}{\cE_j'}$ & $(\fund_i,\fund_j)$\\
      $\frac{1}{2}\dsz{\cE_i}{\cE_i'} +
        \frac{1}{8}\dsz{e^{-B_2}\cE_i}{O7^\pm}$ & $\symm_{\,i}$\\
      $\frac{1}{2}\dsz{\cE_i}{\cE_i'} -
        \frac{1}{8}\dsz{e^{-B_2}\cE_i}{O7^\pm}$ & $\asymm_{\,i}$\\
    \end{tabular}
  \end{center}
  \caption{Spectrum of chiral multiplets charged under brane $\cE_i$
    in the presence of a $O7^\pm$ plane. $\cE_j$ denotes a generic brane
    intersecting $\cE_i$, and primes denote image branes. Any
    resulting negative signs should be interpreted as conjugate
    representations. Notice that $\cE_i$ refers to the basic
    fractional branes, we count multiplicities separately.}
  \label{table:orientifold-spectrum}
\end{table}

Orientifold planes will also contribute to the D-brane charges.
The charge
vector for an $O7$ plane is given by:
\begin{align}
  \label{eq:orientifold-charges}
  \Gamma(O7^\pm) = \pm8 [\cS]\wedge \sqrt{\frac{\hat L(T_\cS/4)}{\hat
      L(N_\cS/4)}} = \pm[\cS]\wedge\left(8 - \frac{1}{6} c_2(T_\cS)\right)\,,
\end{align}
with $\hat L$ the Hirzebruch genus $\hat
L(E)=1+\frac{1}{3}(c_1^2(E)-2c_2(E))+\ldots$, where we have omitted
terms of degree 6 or higher, since they vanish on $\cS$. In the
presence of such an orientifold, the spectrum gets truncated
to invariant states, given by bifundamentals and (anti-)symmetric
representations. The precise matter content in the presence of the
orientifold plane can be read off from the mirror formulas in IIA
\cite{Blumenhagen:2000wh,Ibanez:2001nd,Cvetic:2001tj,Cvetic:2001nr,Marchesano:2007de}. Given
branes $\cE_i,\cE_j$ with orientifold images $\cE_i',\cE_j'$
respectively, one obtains the spectrum in
table~\ref{table:orientifold-spectrum}.

As in \cite{Cvetic:2001nr},
this spectrum can essentially be derived from tadpole/anomaly
cancellation and linearity of the DSZ product. Tadpole cancellation
requires that the 4-form and 2-form parts of the charge vectors
satisfy:
\begin{align}
  \label{eq:general-tadpole-cancellation}
  \Gamma(O7^\pm) + \sum e^{-B_2}\Gamma(\cE_i) = 0\, ,
\end{align}
where the sum is over all branes in our configuration, including
images under the orientifold involution and multiplicities for
non-abelian stacks. We have added the $B_2$ field explicitly, since we
did not include it in our definition of the charge
vector~\eqref{eq:charge-vector}, but it enters in the definition of
the Chern-Simons charge. Consider now a fractional brane $\cE_i$ not
invariant under the orientifold involution, and let us put a stack of
$N_i(\cE_i+\cE_i')$ branes on our singularity, with gauge group
$\U(N_i)$. Taking the DSZ product
of~\eqref{eq:general-tadpole-cancellation} with $e^{-B_2}\Gamma(\cE_i)$
one gets (with a slight abuse of notation):
\begin{align}
  \label{eq:dsz-constraint}
  \sum_{j\neq i,i'}\dsz{\cE_i}{\cE_j} + \dsz{\cE_i}{N_i\cE_i'} +
  \dsz{e^{-B_2}\cE_i}{O7^\pm} = 0\, ,
\end{align}
where we have used the fact that the DSZ product is an index, so it does not
change by deforming both sides by the same $B_2$ field, and in
particular the $B_2$ field can be ignored if it appears in both sides of
the DSZ product. (The 6-form part of the charges (i.e. the $D3$ charge)
was not constrained by~\eqref{eq:general-tadpole-cancellation}, but
since we have no Calabi-Yau filling branes in our background the
$D9$-$D3$ contribution drops out of~\eqref{eq:dsz-constraint}
anyway.) Notice that the first term in~\eqref{eq:dsz-constraint} is
just the field theory anomaly coming from the chiral fields in the
fundamental representation of $\U(N_i)$, in conventions where each
chiral fundamental field contributes 1 unit to the anomaly. The second
and third terms must then equal the net anomaly coming from two-index
tensors:
\begin{align}
  n_{\ssymm}(N_i+4) + n_{\sasymm}(N_i-4) = N_i\dsz{\cE_i}{\cE_i'} +
  \dsz{e^{-B_2}\cE_i}{O7}\, .
\end{align}
Imposing that the relation is satisfied for any $N_i$ we obtain the
relations in table~\ref{table:orientifold-spectrum}.

Finally, given a generic brane $\cE_i$, one has $\Ext^0(\cE_i,\cE_i)$
gauge bosons from the brane to itself. In the absence of orientifolds, this gives rise to a $\U(N_i)$
gauge stack, with $N_i^2 = \dim \Ext^0(\cE_i,\cE_i)$. If the brane is
invariant under the orientifold projection, the involution projects $\U(N_i)$ to either $\Sp(N_i)$ or
$\SO(N_i)$. If the brane is not invariant, but is mapped to an
image brane instead, then the original $\U(N_i)\times \U(N_i)$ gauge
group gets projected down to $\U(N_i)$.

\subsection{Large volume description of the orientifolded
  \alt{$\bC^3/\bZ_3$}{C3/Z3} quiver}

Let us put what we just described into practice. The theory for branes
at a $\bC^3/\bZ_3$ is conventionally described by the exceptional
collection
\begin{align}
  \label{eq:usual-collection}
  \cC = \{\cO[0], \Omega(1)[1], \cO(-1)[2]\}
\end{align}
with $\Omega$ the cotangent bundle on $\bP^2$. This collection can be
obtained by mutation of the basic set of projective objects:
\begin{align}
  \label{eq:usual-collection-projective}
  \sP = \{\cO, \cO(1), \cO(2)\}\, .
\end{align}
In our case it will be convenient to tensor all the elements in the
collection with $\cO(-1)$, and thus we will be dealing with the following
collection instead:
\begin{align}
  \label{eq:collection}
  \cC = \{\cO(-1)[0], \Omega[1], \cO(-2)[2]\}\, .
\end{align}
The reason for tensoring with $\cO(-1)$ is simple: since we want to
orientifold, the branes that we identify under the involution should have the same
mass, but it is not hard to see using the explicit expressions for the
central charge given below that the elements
of~\eqref{eq:usual-collection} have different central charges, and
thus different masses. Tensoring the whole basis by a line bundle does
not change the quiver structure, but it changes the central charge
and hence fixes the problem. To wit, if we have a $D7$ brane wrapping
$\cS$ with charge vector
\begin{align}
  \Gamma(D7) = [\cS]\wedge(Q_0 + Q_2\ell + Q_4 \ell^2)
\end{align}
with $\ell$ the hyperplane in $\cS=\bP^2$, then its exact central
charge is given by \cite{Aspinwall:2004jr}
\begin{align}
  \label{eq:dp0-general-Z}
  Z(D7) = \left(\frac{Q_0}{2} - Q_2 \right)\Phi_1  +
  \left(\frac{1}{2}Q_0\right)\Phi_2 + \frac{1}{12}Q_0 + Q_4\, ,
\end{align}
where $\Phi_i$ are the quantum periods. We will discuss these periods
in more detail in section~\ref{sec:Z3-phaseII}, but for our current
purposes we will only need the fact that they vanish at the quiver
point, where
\begin{align}
  \label{eq:Z-at-quiver-point}
  Z^\bullet(D7) = \frac{1}{12}Q_0 + Q_4 \, .
\end{align}
This implies in particular that:
\begin{align}
  \begin{split}
    Z^\bullet(\cO(n)) & = \frac{1}{12} + \left(\frac{1}{2}n^2 +
      \frac{3}{2}n + \frac{5}{4}\right)\\
    Z^\bullet(\cO(n-1)) & = \frac{1}{12} + \left(\frac{1}{2}n^2 +
      \frac{1}{2}n + \frac{1}{4}\right)\, .
  \end{split}
\end{align}
Imposing that both central charges are equal gives $n=-1$, as
claimed. Tensoring the collection by $\cO(-1)$ can also be
achieved by a change in conventions, see \cite{Aspinwall:2004jr} for
an example.

\begin{figure}
  \begin{center}
    \includegraphics[width=0.5\textwidth]{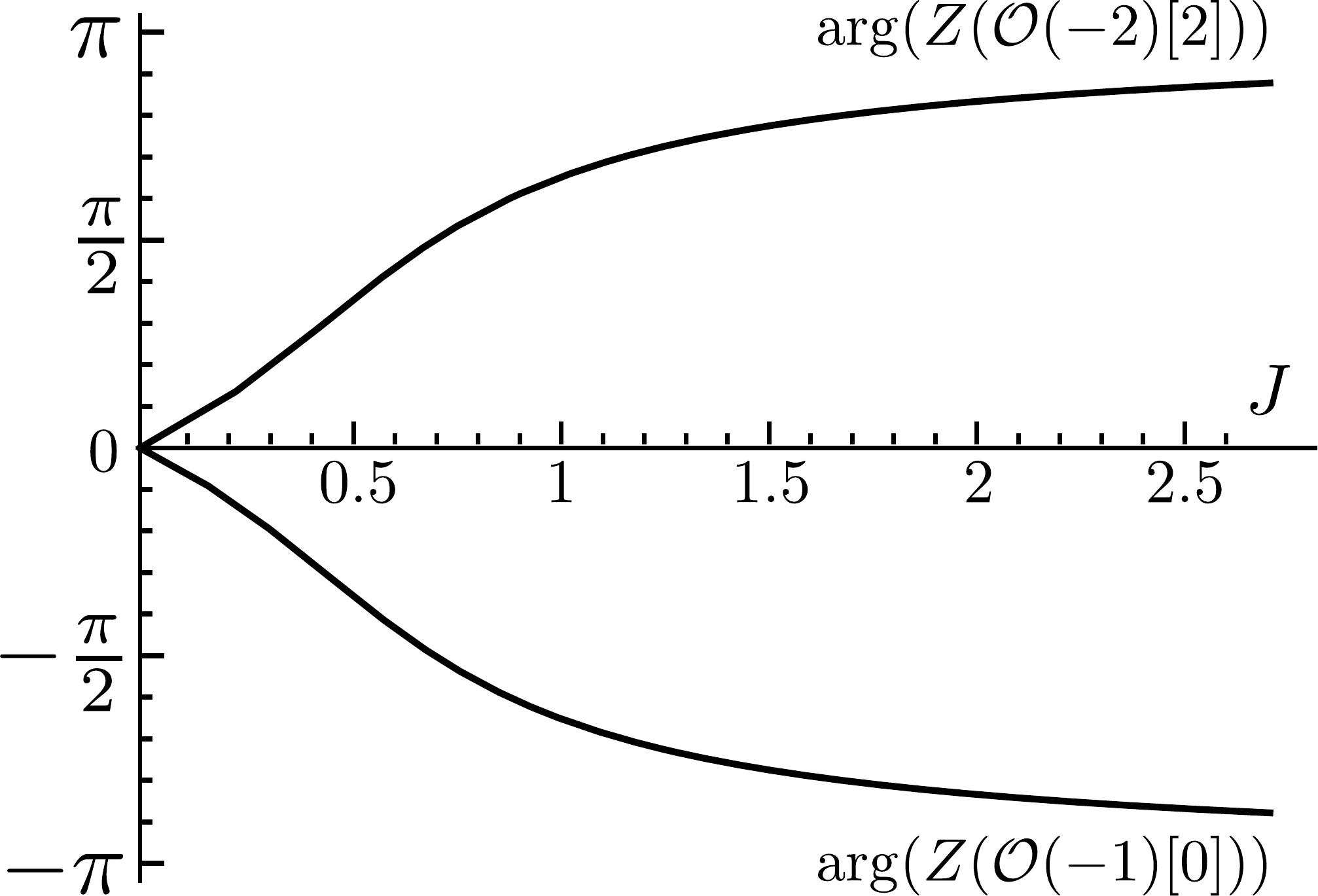}
  \end{center}
  \caption{Phase of the central charges for the $\bC^3/\bZ_3$
    fractional branes as we go from the quiver point at $J=B_2=0$
    towards large volume ($J=+\infty$) along the $B_2=0$ line. The
    central charge of $\Omega[1]$ stays real and positive, so we do
    not show it. The phases of $\cO(-2)[2]$ and $\cO(-1)[0]$ asymptote
    to $\pm\pi$.\label{fig:phases}}
\end{figure}

The charge vectors for the fractional branes
in~\eqref{eq:collection} are given by:
\begin{align}
  \label{eq:induced-charges}
  \begin{split}
    \Gamma(\cO(-1)[0]) = [\cS]\wedge\ch(\cO(-1))
    \sqrt{\frac{\Td(T_\cS)}{\Td(N_\cS)}} & = [\cS]\wedge\left(1+\frac{1}{2}\ell +
      \frac{1}{4}\ell^2\right)\\
    \Gamma(\Omega[1]) = -[\cS]\wedge\ch(\Omega) \sqrt{\frac{\Td(T_\cS)}{\Td(N_\cS)}}
    & = [\cS]\wedge\left(-2 + \frac{1}{2}\ell^2\right)\\
    \Gamma(\cO(-2)[2]) = [\cS]\wedge\ch(\cO(-2))
    \sqrt{\frac{\Td(T_\cS)}{\Td(N_\cS)}} & = [\cS]\wedge\left(1-\frac{1}{2}\ell +
      \frac{1}{4}\ell^2\right)\, .
  \end{split}
\end{align}
Plugging these expressions in~\eqref{eq:Z-at-quiver-point} we easily
see that
\begin{align}
  \label{eq:dP0-phaseI-Z}
  Z^\bullet(\cO(-1)[-1])=Z^\bullet(\Omega[0])=Z^\bullet(\cO(-2)[1])=\frac{1}{3}\,  ,
\end{align}
which is what one expects from the $\bZ_3$ symmetry permuting the
fractional branes. For illustration we show the behavior of the
central charges as we go to large volume in figure~\ref{fig:phases},
where we have used the explicit form~\eqref{eq:C3/Z3-periods} of the
periods $\Phi_i$ and the mirror map \cite{Cox:2000vi}, which in our
case is just $B_2+iJ=\Phi_1(z)$.

Using~\eqref{eq:orientifold-action} it is straightforward to check
that~\eqref{eq:collection} maps to itself under the orientifold
involution. In the case of the line bundles this is easy to see:
\begin{align}
  i_*\cO(-1)[0] \longrightarrow i_*(\cO(-1)^\vee\wedge K_\cS)[2] =
  i_*(\cO(1)\wedge \cO(-3))[2] = i_*\cO(-2)[2]\, .
\end{align}
The case of $\Omega[1]$ is slightly more complicated, but follows from
the fact that $\Omega^\vee\otimes K_\cS=\Omega$, so the brane maps to
itself:
\begin{align}
  i_*\Omega[1] \longrightarrow i_*(\Omega^\vee\otimes K_\cS)[2-1] =
  i_*\Omega[1]\, .
\end{align}

We can in fact derive the quiver in full detail. By
\eqref{eq:induced-charges}, we know that $\Omega[1]$ has $D7$ charge
equal to $-2$, so we can cancel the $D7$ brane tadpole by adding 4
$\Omega[1]$ branes. Since $\Omega[1]$ maps to itself, and it has $\dim
\Ext^{0}(\Omega[1],\Omega[1])=1$, we see that under the $O7^+$ action
the resulting stack is $\Sp(4)$.\footnote{Determining the orientifold
  projection is not straightforward. It can be done in general by
  using the methods of \cite{Brunner:2008bi,Gao:2010ava}, or more
  simply in our example by matching with the CFT computation, since
  determining the symmetric/antisymmetric representation is
  straightforward.}  We can add $k$ pairs of regular $D3$ branes, and
thus enhance the symmetry to $\Sp(2k+4)\times \U(2k)$. The
intersection products are easily computed
using~\eqref{eq:induced-charges}, \eqref{eq:dsz-product} and the fact
that $c_1(T_\cS)=c_1(\cO(3))=3\ell$:
\begin{align}
  \begin{split}
    \dsz{\Omega[1]}{\cO(-1)[0]} & =
    \int_{\bP^2}(3\ell) \w \left(-2\cdot\frac{1}{2}\ell\right) = -3\\
    \dsz{\cO(-1)[0]}{\cO(-2)[2]} & = -3\\
    \dsz{\cO(-1)[0]}{O7^+} & = -12\, .
  \end{split}
\end{align}
By applying the rules in table~\ref{table:orientifold-spectrum} we
thus find the following matter content:
\begin{align}
  \label{eq:P2-Sp-quiver}
  \begin{array}{cc|c}
    \Sp(2k+4) & \U(2k) & \SU(3)\\
    \hline
    \ov\fund & \fund & \fund\\
    1 & \ov\symm & \fund
  \end{array}
\end{align}
with the $\SU(3)$ being a global symmetry, which at this level simply
encodes the multiplicity of the matter fields. We will see in some
simpler examples below how this group of global symmetries can also be
understood geometrically, but we avoid the discussion of this
particular case since it is slightly more technical.

\paragraph{Central charge for the orientifold.}
Let us assume that the mass of the orientifold can be computed exactly
(including all $\alpha'$ corrections) in a manner similar to that of a
$D7$ brane. We will analyze the $O7^+$ plane, and assume that it has
the same argument for the central charge as the $O7^-$ plane
(i.e. they preserve the same supersymmetries). Using the same notation
as above, we have from \eqref{eq:orientifold-charges} that
$Q_0=8$, $Q_2=0$, $Q_4=-\chi(\bP^2)/6=-\frac{1}{2}$. The central
charge at the quiver point is thus:
\begin{align}
  Z^\bullet(O7^+) = \frac{8}{12} - \frac{1}{2} = \frac{1}{6} > 0\, .
\end{align}
The central charge at large volume, on the other hand, is given by:
\begin{align}
  \begin{split}
    Z^{lv}(O7^+) & = 8\int_S e^{-B_2-iJ}\sqrt{\frac{\hat L(TS/4)}{\hat
        L(NS/4)}}\\
    & \xrightarrow{vol(S)\to\infty} - 8\int_S J^2 < 0\, .
  \end{split}
\end{align}
As it is clear, the central charge of the orientifold changes sign in
going from large to small volume, and in particular, since it is a
real quantity, it passes through 0 (the vanishing point is located at
$J\approx 0.0694$). Close to the quiver point, the orientifold has a
phase of the central charge \emph{opposite} to the one at large
volume. So we learn that the supersymmetry preserved by the $O7^\pm$
orientifold at the quiver point is opposite to the one preserved at
large volume, and furthermore it is the same supersymmetry preserved
by the fractional branes~\eqref{eq:collection}.

\subsection{Quantum symmetries and {\ae}rientifolds}
\label{sec:aerientifolds}

As we have just seen, the orientifold
action~\eqref{eq:orientifold-action} left the fractional brane
$\Omega[1]$ invariant, while it exchanged $\cO(-1)[0]$ and
$\cO(-2)[2]$. This beautifully reproduces the quiver structure
obtained via CFT or orientifolded dimer model techniques. However, a longer look to the
quiver may leave one puzzled: from the point of view of the
quiver gauge theory there is nothing special about the node associated
with the $\Omega[1]$ brane, one could have taken an involution of the
quiver leaving invariant any of the other two nodes.\footnote{See for
  example \cite{Wijnholt:2007vn} for some instances in which this
  perspective was taken.} This is in fact true in the full string theoretic description, and in the present language
follows from the fact that the derived category has
auto-equivalences, as we now explore.

We will not go into details of the mathematical meaning of
auto-equivalences (we refer the reader to \cite{Aspinwall:2004jr} for
a detailed review), but we can think of an auto-equivalence of the
derived category as a re-labeling of the D-branes in such a way that
the physics is unaffected. We have already encountered a simple
auto-equivalence of the derived category in
section~\ref{sec:dc-general}, where we discussed integer shifts of the
$B_2$ field. Another familiar context in which this phenomenon arises is
that of monodromy around a conifold point. Consider for example a
point in moduli space where a brane $\cF$ becomes massless. As we
circle once around this point in moduli space the charges of the
branes shift as \cite{Candelas:1990rm,Hori:2000ck}:
\begin{align}
  \Gamma(\cE) \to \Gamma(\cE) - \dsz{\cE}{\cF}\Gamma(\cF)\, .
\end{align}
As we see, the charges of most branes will change, but this cannot
induce a change in the physics of the background or the set of stable
branes, since we end up at the same point in moduli space. The
operation must then amount to a relabeling of the D-brane charges.

In our particular context the conifold points in moduli space are
precisely those where the fractional branes in the quiver point become
massless. In particular, we will present the moduli space in such a
way that it is $\cO(-1)[0]$ that becomes massless. The induced action
on the charge vectors as we go around the point where $\cO(-1)[0]$
becomes massless is:
\begin{align}
  \label{eq:O(-1)-monodromy}
  \Gamma(\cE) \to \Gamma(\cE) - \dsz{\cE}{\cO(-1)}\Gamma(\cO(-1))\, .
\end{align}
Since this is a linear transformation, it is convenient to rewrite
this as a matrix action on the charge vector:
\begin{align}
   \begin{pmatrix}
     \Gamma^{(2)}\\\Gamma^{(4)}\\\Gamma^{(6)}
   \end{pmatrix}\to
   \begin{pmatrix}
     -\frac{1}{2} & 3 & 0\\
     -\frac{3}{4} & \frac{5}{2} & 0\\
     -\frac{3}{8} & \frac{3}{4} & 1
   \end{pmatrix}
   \begin{pmatrix}
     \Gamma^{(2)}\\\Gamma^{(4)}\\\Gamma^{(6)}
   \end{pmatrix}\equiv \cM_C    \begin{pmatrix}
     \Gamma^{(2)}\\\Gamma^{(4)}\\\Gamma^{(6)}
   \end{pmatrix}
\end{align}
where as usual $\Gamma^{(2i)}$ denotes the $2i$-form part of $\Gamma$.

A similar phenomenon that appears in our context is monodromy around
the large volume point, which shifts the $B_2$ field by one unit, or
equivalently it acts on the charges as:
\begin{align}
  \Gamma(\cE) \to \Gamma(\cE)\wedge \ch(\cO(-1))\, .
\end{align}
Again writing this monodromy in matrix form, we have:
\begin{align}
   \begin{pmatrix}
     \Gamma^{(2)}\\\Gamma^{(4)}\\\Gamma^{(6)}
   \end{pmatrix}\to
   \begin{pmatrix}
     1  & 0 & 0\\
     -1 & 1 & 0\\
     \frac{1}{2} & -1 & 1
   \end{pmatrix}
   \begin{pmatrix}
     \Gamma^{(2)}\\\Gamma^{(4)}\\\Gamma^{(6)}
   \end{pmatrix}\equiv \cM_{LV}    \begin{pmatrix}
     \Gamma^{(2)}\\\Gamma^{(4)}\\\Gamma^{(6)}
   \end{pmatrix}\, .
\end{align}
The moduli space of $\bC_3/\bZ_3$ is a $\bP^1$ with three marked
points around which monodromy occurs: the large volume point, the
conifold point, and the quiver point. The total monodromy around all
three points must then vanish, and in this way we can easily obtain
the monodromy around the quiver point:
\begin{align}
  \label{eq:P2-quantum-symmetry}
  \cM_Q = \bigl(\cM_{LV}\cM_C\bigr)^{-1} = \begin{pmatrix}
    -\frac{1}{2} & -3 & 0\\
    \frac{1}{4} & -\frac{1}{2} & 0\\
    \frac{1}{8} & \frac{1}{4} & 1
    \end{pmatrix}\, .
\end{align}
It is straightforward to show that $\cM_Q^3=1$, and furthermore,
from the charges~\eqref{eq:induced-charges}:
\begin{align}
  \begin{split}
    \cM_Q\Gamma(\cO(-1)[0]) & = \Gamma(\Omega[1])\, ,\\
    \cM_Q\Gamma(\Omega[1]) & = \Gamma(\cO(-2)[2])\, ,\\
    \cM_Q\Gamma(\cO(-2)[2]) &= \Gamma(\cO(-1)[0])\, .
  \end{split}
\end{align}
We therefore identify $\cM_Q$ with the quantum symmetry rotating the
quiver.\footnote{It is possible to identify the quantum symmetry at
the level of the category itself, and not just at the level of
charges; we refer the reader to \cite{Aspinwall:2004jr} for details.}

We are now in a position to resolve the issue that we presented at the
beginning of this section. Denoting the ordinary orientifold
involution~\eqref{eq:orientifold-action} by $\cP$ (which acts on the
charges as $\diag(1,-1,1)$), one can construct a new class of
orientifolds by composing with the auto-equivalences of the category
just described. We call the resulting object an
\emph{{\ae}rientifold}, in order to distinguish it from the ordinary
large volume orientifold given by $\cP$, although we emphasize that
{\ae}rientifolds are just as natural from the quiver point of
view. For example, the {\ae}rientifold leaving the $\cO(-1)[0]$ node
invariant would be defined by $\cP'=\cM_Q^{-1}\cP\cM_Q$, and the one
leaving $\cO(-2)[2]$ invariant would be $\cP''=\cM_Q\cP\cM_Q^{-1}$. At
the level of charges we have that:
\begin{align}
  \cP' = \begin{pmatrix}
    -\frac{1}{2} & 3 & 0 \\
    \frac{1}{4} & \frac{1}{2} & 0 \\
    \frac{1}{8} & -\frac{1}{4} & 1
  \end{pmatrix}\, .
\end{align}
We see that from the quiver point of view it is very natural to dress
the ordinary large volume action of the orientifold with
auto-equivalences of the category, and such dressings appear very
naturally when orbifolding the quiver for certain
singularities.\footnote{The idea of dressing the large volume action
  by a quantum symmetry is not entirely new, see for example
  \cite{Brunner:2004zd,Diaconescu:2006id,Brunner:2008bi}, although the
  dressing considered in those papers is of a different nature of the
  one considered here, which is in some sense physically trivial (but
  still very useful when thinking about orientifolded quivers in large
  volume language).}

\subsection{Microscopic description of the discrete torsion}
\label{sec:microscopic-torsion}

We  now connect the classification of the different
orientifolds based on discrete torsion advocated in
section~\ref{sec:torsion} with the large volume picture discussed in this section. To do so, we make use of the fact (explained in~\S\ref{sec:torsion}) that a $D5$ brane wrapped on $\bR\bP^2 \subset S^5/\bZ_6$ induces a change in the $C_2$ discrete torsion when crossing the brane. Thus, allowing the wrapped $D5$ brane to collapse onto the singularity (restoring supersymmetry) should alter the configuration of fractional branes in a way which corresponds to changing the $C_2$ discrete torsion.

Consider the resolved geometry, i.e. $\cO(-3)\hookrightarrow\bP^2$. Contracting the $D5$ brane onto $\bP^2$ should induce some brane charge which is visible in the large volume description.
This charge should be $\bZ_2$ valued, stable only in the presence of an orientifold, and
associated with a 5-brane.
There is a natural
candidate fulfilling these conditions, given by a generalization of
the non-BPS $D7$ brane of type I string theory, which we
now briefly review.\footnote{Since we want to identify topological
  charges we will work in the classical (geometric) regime in this
  section, and in particular we will find that the different brane
  configurations are related by adding non-BPS objects. Similarly to
  what happens in~\cite{Witten:1998xy,Hyakutake:2000mr}, if we go to
  the singular locus and let the system relax it will find a BPS
  vacuum, in our case due to the familiar $\alpha'$ corrections to the
  central charges.}

It is well known that the stable states in type I string theory are
classified by elements of $KO(X)$, where $X$ is the spacetime manifold
\cite{Minasian:1997mm,Sen:1998rg,Sen:1998ii,Sen:1998sm,Bergman:1998xv,Sen:1998tt,Srednicki:1998mq,Witten:1998cd}
(see also \cite{Sen:1999mg,Schwarz:1999vu,Olsen:1999xx,Evslin:2006cj}
for nice reviews). For $X=\bR^{10}$, the classification of branes
reduces to computing the non-trivial homotopy groups
$\pi_{i}(O(32))$. In particular, due to the fact that
$\pi_1(O(32))=\bZ_2$
there is a topologically stable
7-brane in type I with $\bZ_2$-valued charge. This object is non-BPS
in type I, and it has some tachyonic modes with respect to the
background $D9$ branes~\cite{Frau:1999qs}.\footnote{In particular, it can decay into topologically non-trivial flux on the $D9$ branes, see~\cite{LoaizaBrito:2001ux}.} Of most interest to us is
that this brane admits an alternative description in terms of a
$D7$-$\ov{D7}$ pair in a type IIB orientifold description. The orientifold involution $\Omega$ of type I
removes the tachyon between the $D7$ and the $\ov{D7}$
\cite{Witten:1998cd,Frau:1999qs}, and renders the object stable
(modulo the tachyon with respect to the background $D9$ branes).

A first principles computation for the case at hand would require a
generalization of the $KO$ group to the wrapped orientifold, which
seems to be an involved technical problem. (We refer the reader to
\cite{Distler:2009ri,Gao:2010ava,Distler:2010an} for some recent work
on the definition of the proper K-theory in the contexts of interest
to us.)  Luckily, the observation in \cite{Witten:1998cd,Frau:1999qs}
that the non-BPS $D7$ can be constructed from a $D7$-$\ov{D7}$ pair
identified by the orientifold involution generalizes much more easily, if
somewhat more heuristically.

For the $\bC^3/\bZ_3$ $\SO$ theories there is an $O7^-$ plane wrapping the $\bP^2$ rather than the space-filling $O9^-$ plane of type I, so a
natural (in some sense T-dual) generalization of the $\bZ_2$-stable $D7$ brane of type I would be a $\bZ_2$-stable
$D5$ brane wrapping a divisor of the $\bP^2$. Recall that at the quiver
locus a single $D3$ (in covering space conventions) decomposes into a
$\Omega[1]+\cO(-1)+\cO(-2)[2]$ system. In particular, the $\Omega[1]$
has no induced $D5$ charge, so we will ignore it in what follows.
The other two branes have charge vectors given
by~\eqref{eq:induced-charges}, reproduced below for
convenience:
\begin{align*}
  \begin{split}
    \Gamma(\cO(-1)[0]) & = [\cS]\wedge\left(1+\frac{1}{2}\ell +
      \frac{1}{4}\ell^2\right)\\
    \Gamma(\cO(-2)[2]) & = [\cS]\wedge\left(1-\frac{1}{2}\ell +
      \frac{1}{4}\ell^2\right)\, .
  \end{split}
\end{align*}
Notice the appearance (at the level of the charges) of the $D5$-$\ov{D5}$ pair that we expected
would generalize the $D7$-$\ov{D7}$ stable object of type I.
It is therefore
natural to conjecture that a discrete $\bZ_2$ charge remains in
the system after tachyon condensation.\footnote{Since we have
  $D7$-branes in the background the $D5$ branes will decay into
  flux. The topological structure of the resulting flux in some
  particular examples is described in \cite{LoaizaBrito:2001ux}; we
  expect a similar structure to remain in our case.}  Since adding a
single stuck $D3$ in the covering space is precisely the change that
one would associate with wrapping a $D5$ on $\bR\bP^2$ (i.e.\ introducing some
discrete torsion for $C_2$), it must be the case that retracting
the $D5$ wrapping $\bR\bP^2$ to the exceptional locus $\bP^2$ induces this
stable $\bZ_2$-valued $D5$ charge.

For the $\Sp$ theory, we have an $O7^+$ plane wrapping the $\bP^2$ instead. Since there is no charge in $KSp(\bR^{10})$
that supports a $D7$ charge, we expect by analogy that there is no stable $D5$-$\ov{D5}$ pair, and thus wrapping a $D5$ brane on $\bR\bP^2$ does not change the gauge group, in agreement with the arguments of~\S\ref{sec:torsion}. Nonetheless, we expect a change in the theta angle of the gauge theory, though the mechanism for this change is not clear in the K-theory picture.

Finally, by allowing a wrapped $\NS$ brane to collapse onto the $\bP^2$, we expect the $O7^-$ plane to change into an $O7^+$ plane and vice versa. This is reminiscent of the general story for $Op$ planes in a flat background given in~\cite{Hyakutake:2000mr}, but a less heuristic justification remains elusive.

\section{Interpretation as an orientifold transition at strong coupling}
\label{sec:interpretation}

We have just seen how the system at the quiver point can be described
in terms of large volume objects. In this section we use
this picture to argue that the duality that we observe in field
theory is inherited from IIB S-duality. We first consider
the strongly coupled behavior of $O7^+$ planes in flat space, which we
analyze in~\S\ref{sec:O7-S-duality}. Once this is understood,
one can compactify the flat-space configuration, and the behavior at the quiver locus
can then be found by taking the continuation to small volume. Since the chiral structure of the quiver is topological in nature, it is not affected by the continuation to small volume, and we are able to reproduce the S-dual gauge theory expected from the field theoretic arguments of~\cite{gauge}. We work this out in detail for the $\bC^3/\bZ_3$ example in~\S\ref{sec:orientifold-transition}.

\subsection{S-duality for \alt{$O7^+$}{O7+} planes}
\label{sec:O7-S-duality}

As discussed in the introduction, our main claims in this paper are
that the field theories we analyze are related by a strong/weak
duality, and that this duality is inherited from S-duality of IIB
string theory. If this is the case, the structure of the dual pairs
should be compatible with the properties under S-duality of the
orientifolds and branes that engineer the field theory. There is no
issue with taking the $D7$ branes to strong coupling, but the
orientifold plane is more subtle. The strongly coupled limit of $Op$
planes with $p<6$ has already been extensively discussed in the
literature
\cite{Dasgupta:1995zm,Witten:1995em,Witten:1997kz,Uranga:1998uj,Witten:1998xy,Hori:1998iv,Gimon:1998be,Sethi:1998zk,Berkooz:1998sn,Hanany:1999jy,Uranga:1999ib,Hanany:2000fq}. Unfortunately,
the large volume picture of our system requires the introduction of
$O7^\pm$ planes, and the strongly coupled limit of these is less well
understood (some relevant papers are
\cite{Landsteiner:1997ei,Witten:1997bs,Bershadsky:1998vn}).

In this section we present evidence for a proposed
description of the strongly coupled limit of the $O7^+$ plane in flat
space as a bound state of an $O7^-$ plane with extra 7-branes, which
seems to be behind the duality between $\Sp$ theories and $\SO$
theories with odd rank. (We will comment at the end of the section on
what happens in the self-dual case.) Our proposal is the
following: at strong coupling, IIB string theory in the presence of an
$O7^+$ can be alternatively described as a weakly coupled IIB theory
in the presence of a bound state of an $O7^-$, 4 $(1,0)$ 7-branes
(i.e. ordinary $D7$s), and 4 $(0,1)$ 7-branes.

\begin{figure}
  \begin{center}
    \includegraphics[width=0.8\textwidth]{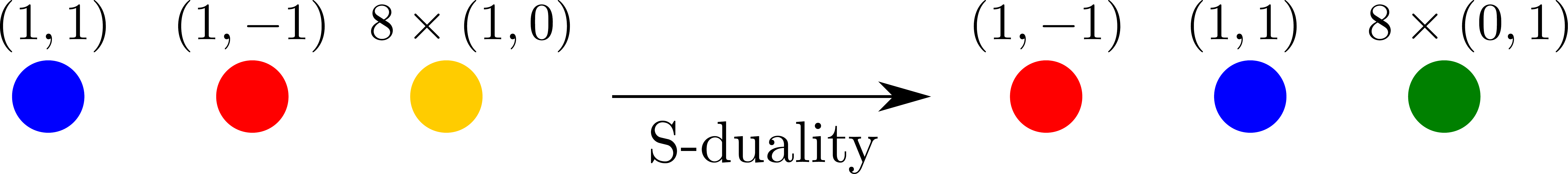}
  \end{center}
  \caption{S-duality for the standard components of a $D_8$
    singularity.\label{fig:O7-S-duality}}
\end{figure}

\begin{figure}
  \begin{center}
    \includegraphics[width=\textwidth]{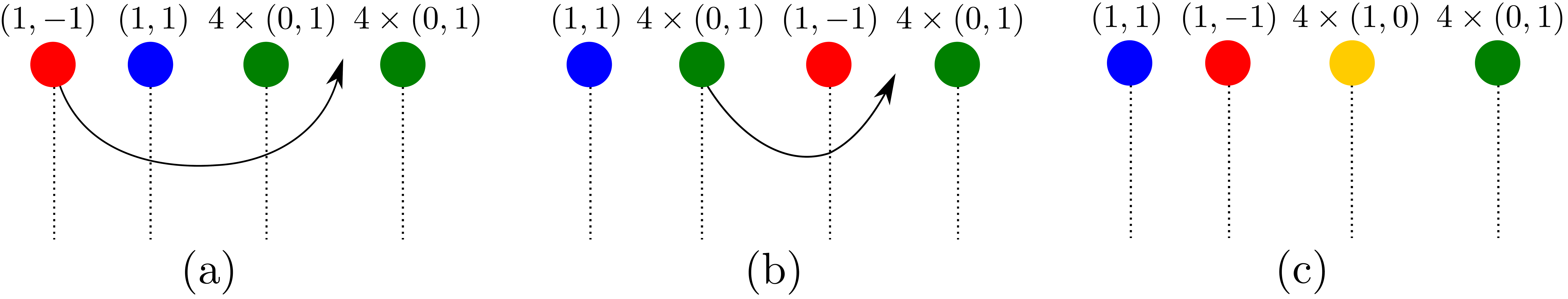}
  \end{center}
  \caption{Monodromy converting the S-dual in
    figure~\ref{fig:O7-S-duality} to our proposed dual. We have chosen
    the branch cuts to run downwards from the 7-branes, as indicated
    by the dotted lines.\label{fig:O7-monodromy}}
\end{figure}

This somewhat curious dual spectrum of branes can be motivated as
follows. Geometrically, the monodromy corresponding to an $O7^+$ plane
is that of a $D_8$ singularity.\footnote{As discussed in
  \cite{Witten:1997bs}, this is correct at the level of monodromies,
  but the actual realization in M-theory seems to be associated to a
  non-Weierstrass fiber of type $D_4\times D_4$ with $D_8$ monodromy.}
Such a monodromy can be engineered by locating 8 mobile $D7$ branes on
top of a $O7^-$ plane. By describing as usual the $O7^-$ plane as a
$(1,1)$ 7-brane together with a $(1,-1)$ 7-brane \cite{Sen:1997gv}, we
have a description of the $D_8$ singularity as 10 coincident $(p,q)$
7-branes.\footnote{The two components of an $O7^-$ plane by itself
  (with no $(1,0)$ branes on top) are separated due to $D(-1)$
  instanton effects by a distance of order $e^{-1/g_s}$, and thus the
  lift of a $O7^-$ is smooth. Adding the 8 extra $(1,0)$ branes
  removes this separation, and the total configuration is indeed
  singular, with $D_8$ singularity. This is easily seen using a probe
  argument, see \cite{Banks:1996nj} for the original probe argument
  and \cite{Witten:1997bs} for an explicit analysis of our case.}
We apply S-duality to each of the 7 branes in the standard way,
sending $(p,q)\to(q,-p)$. The original configuration and its dual
are shown (slightly resolved for clarity) in figure~\ref{fig:O7-S-duality}.
 We can now connect the resulting S-dual
system of $(p,q)$ 7-branes to our proposed dual by simple monodromy of
branes, as shown in figure~\ref{fig:O7-monodromy}.

We describe this connection in detail. We take the convention
that the branch cut for the monodromy associated to a $(p,q)$ 7-brane
runs downwards from the brane. Crossing this branch cut
counter-clockwise induces the monodromy:
\begin{align}
  \cM_{(p,q)} = \begin{pmatrix}
    1 - pq & p^2 \\
    -q^2 & 1+pq
    \end{pmatrix}\, .
\end{align}
We divide the stack of 8 $(0,1)$ 7-branes into two stacks of 4 branes
each. We now perform the rearrangement shown in
figure~\ref{fig:O7-monodromy}a, taking the $(1,-1)$ 7-brane to the
right of the $(1,1)$ 7-brane and the leftmost stack of $(0,1)$
7-branes. As it moves to its new position it crosses the $(1,1)$
branch cut counter-clockwise, and then the four $(0,1)$ branch
cuts. Its new $(p,q)$ labels are thus given by
\begin{align}
  \begin{pmatrix}p' \\ q'\end{pmatrix} =
  \cM_{(0,1)}^4\cM_{(1,1)}\begin{pmatrix}1 \\ -1 \end{pmatrix} =
  \begin{pmatrix}0 & 1 \\ -1 & -2\end{pmatrix}\begin{pmatrix}1 \\
    -1 \end{pmatrix} = \begin{pmatrix} -1 \\ 1 \end{pmatrix}\, .
\end{align}
The overall sign of the $(p,q)$ charge is not physical, therefore the 7-brane charge is unaltered following this operation. The second step
is depicted in figure~\ref{fig:O7-monodromy}b. We take the leftmost
group of $(0,1)$ branes to the right of the $(1,-1)$ brane. In doing
this we cross the $(1,-1)$ branch cut counter-clockwise, and thus the
$(p,q)$ labels of the $(0,1)$ stack become
\begin{align}
  \begin{pmatrix}p' \\ q'\end{pmatrix} =
  \cM_{(1,-1)}\begin{pmatrix}0 \\ 1 \end{pmatrix} =
  \begin{pmatrix}2 & 1 \\ -1 & 0\end{pmatrix}\begin{pmatrix}0 \\
    1 \end{pmatrix} = \begin{pmatrix} 1 \\ 0 \end{pmatrix}\, .
\end{align}
This gives the collection that we proposed, shown in figure~\ref{fig:O7-monodromy}c.

The above argument was made purely at the level of monodromies. This
cannot be the whole story, since eight $D7$ branes atop an $O7^-$
plane should naively yield an $\SO(16)$ gauge group, distinct from the
trivial gauge group expected for an $O7^+$ plane. This was already
observed in \cite{Witten:1997bs}, and given an explanation in the
context of $K3$ compactifications of F-theory in
\cite{Bershadsky:1998vn}. A general analysis from the type II
perspective was then presented in \cite{Brunner:2008bi,Gao:2010ava}
(see also \cite{Distler:2009ri}). From the type II perspective, the
type of the orientifold in our configuration will be determined by the
sign of the crosscap diagram around the orientifold. As explained
in~\cite{Gao:2010ava}, this sign is determined by a parallel section
--- known as the ``crosscap'' section --- of a line bundle with flat
connection. The line bundle and connection are derived from the
``twist'' line bundle and connection, whose curvature is
$B_2+\sigma^\ast B_2$ where $\sigma$ is the orientifold
involution. Thus, the orientifold type is indirectly related to the
$B_2$ discrete torsion $[H] \in H^3(X,\tilde{\bZ})$.

While in general the crosscap section is not completely determined by $[H]$, for the case of the $\bC^3/\bZ_3$ orientifold we argued in~\S\ref{sec:torsion} that trivial (non-trivial) $[H]$ should correspond to an $O7^-$ ($O7^+$) wrapping the exceptional divisor. This should follow from the general discussion of~\cite{Brunner:2008bi,Gao:2010ava}; it would be interesting to work this out in detail.

%, which is
%precisely the value of the torsional NSNS flux we have computed
%above. We claim that in this particular example the right assignment
%is that for non-vanishing NSNS torsion the brane configuration will
%perturbatively behave as an $O7^+$, while a vanishing NSNS torsion
%indicates the presence of an $O7^-$. This assignment should follow
%from the general discussion in \cite{Brunner:2008bi,Gao:2010ava}, it
%should be interesting to work out the details.

\subsection{The orientifold transition for \alt{$\bC^3/\bZ_3$}{C3/Z3}}
\label{sec:orientifold-transition}

Given the proposal for the strongly coupled behavior of the $O7^+$
above, let us try to obtain the field theory duality conjectured in
\cite{gauge} from the brane description of the system.

The most important change with respect to the flat space case is that
tadpole cancellation requires the introduction of some (anti-)branes
on top of the orientifold. Consistent configurations are of the form
$O7^++4\Omega[1]+2 k\, D3s$, with gauge group $\Sp(2k+4)\times
\U(2k)$. Under S-duality, the $O7^+$ becomes an $O7^-$ with some
7-branes on top. At the quiver point these 7-branes will decay into
the standard basis of fractional branes. We conclude that S-duality
acts on the wrapped $O7^+$ as follows:
\begin{align}
  \label{eq:P2-orientifold-transition}
  O7^+ \xrightarrow{\text{S-duality}} O7^- + 4\bigl(\cO(-1)[0] +
  \cO(-2)[2]\bigr) + 4\bigl(\widehat{\cO(-1)[0]} +
  \widehat{\cO(-2)[2]}\bigr) + n\, D3s\, .
\end{align}
Here $\widehat{\cE}$ indicates the S-dual of the brane $\cE$, and we
have allowed for the inclusion of $n$ $D3$ branes to take into account
lower charges induced by curvatures and fluxes. This integer can be
determined by imposing $D3$ charge conservation, since $D3$ branes are
self-dual under $\SL(2,\bZ)$. Using the expressions for the
charges~\eqref{eq:induced-charges} and~\eqref{eq:orientifold-charges}
we obtain $n=-5$. Note that the discussion in the previous
section was in terms of mobile 7-branes, so in terms of fractional
branes we need to consider $\cO(-1)[0]$ and its image $\cO(-2)[2]$
together.

We can now treat the whole system. Starting with
$O7^++4\Omega[1]+2k\, D3s$, S-duality gives:
\begin{align}
    O7^- + 4\bigl(\cO(-1)[0] +
    \cO(-2)[2]\bigr) + 4\bigl(\widehat{\cO(-1)[0]} +
    \widehat{\cO(-2)[2]} + \widehat{\Omega[1]}\bigr) + (2k-5)\, D3s\,.
\end{align}
Using the fact that the three fractional branes add up to a
regular $D3$, which is $\SL(2,\bZ)$ invariant, we can rewrite this
configuration as:
\begin{align}
    O7^- + 4\bigl(\cO(-1)[0] +
    \cO(-2)[2]\bigr) +  (2k-1)\, D3s\,.
\end{align}
Taking into account the change in orientifold projection and the
discussion in section~\ref{sec:DC-preliminaries}, we therefore find that
the full matter content of the theory after the transition it is given
by:
\begin{align}
  \label{eq:P2-SO-quiver}
  \begin{array}{cc|c}
    \SO(2k-1) & \U(2k+3) & \SU(3)\\
    \hline
    \ov\fund & \fund & \fund\\
    1 & \ov\asymm & \fund
  \end{array}
\end{align}
This is precisely the conjectured field theory dual of theory
\eqref{eq:P2-Sp-quiver}, where $\tN = 2 k = N-3$.

\medskip

The main features of this example will generalize to a number of
further examples, so let us highlight the primary consequences of the
orientifold transition. First of all, we find that under the
transition the sign of the orientifold projection changes. This immediately
implies that $SO$ and $\Sp$ groups get exchanged, while $SU$ groups
stay invariant. Similarly, symmetric and antisymmetric representations
get exchanged. This agrees perfectly with the features of the duality
that we are proposing.

The orientifold transition picture also naturally explains the change
in rank of the field theory: it is simply the manifestation of $D3$
charge being conserved. In any given example one can easily calculate
the change in rank one needs in order to conserve $D3$ charge, and in
all examples that we have checked this change is exactly what is needed for
agreement with anomaly matching in the field theory.

\medskip

The above discussion applies to the case where $[H]$ is changed by S-duality, leading to an orientifold transition. In the case where $[H]$ does not change under S-duality, i.e.\ where $[H]=[F]$, we expect a self-duality rather than an orientifold transition. In particular, both $[H]=[F]=0$ and $[H]=[F]=1/2$, corresponding to the $\SO$ theory for even $N$ and the ``$\widetilde{\Sp}$'' theory, should be self-dual under $S\in\SL(2,\bZ)$. We now describe these self-dualities at the level of the fractional branes.

In the $\SO$ case, we have the fractional branes $O7^-
+ 4\bigl(\cO(-1)[0] + \cO(-2)[2]\bigr) + (N-4) D3s$, but the $O7^- +
4\bigl(\cO(-1)[0] + \cO(-2)[2]\bigr)$ is self-dual (at the level of
monodromies) using the same argument as in~\S\ref{sec:O7-S-duality},
where we merely ignore the rightmost stack of branes in
figure~\ref{fig:O7-monodromy}; thus, the entire configuration of
fractional branes is self-dual. In the $\Sp$ case, we start with
the fractional branes $O7^+ + 4 \Omega[1] + \tilde{N} D3s$ and dualize
the 7-branes as in the transition above, except that due to the
non-vanishing $[H]$ after the duality we treat the resulting
7-brane cluster as the components of an $O7^+$ plane. This gives $O7^+
+4 \Omega[1] + n D3s$ for some $n$, where $D3$ charge conservation
requires $n = \tilde{N}$. Thus, this configuration is also self-dual, in perfect agreement with the results of~\S\ref{sec:torsion}.

\section{Phase II of \alt{$\bC^3/\bZ_3$}{C3/Z3}}
\label{sec:Z3-phaseII}

We will now compare the orientifold transition picture we just
discussed with the predictions of field theory in a related but
illustrative example, the theory Seiberg dual
\cite{Seiberg:1994pq,Intriligator:1995id} to that of branes at
$\bC^3/\bZ_3$, leaving the discussion of more involved singularities
for the upcoming \cite{strings-all}.

\subsection{Field theory}

\begin{figure}
  \begin{center}
    \includegraphics[width=0.3\textwidth]{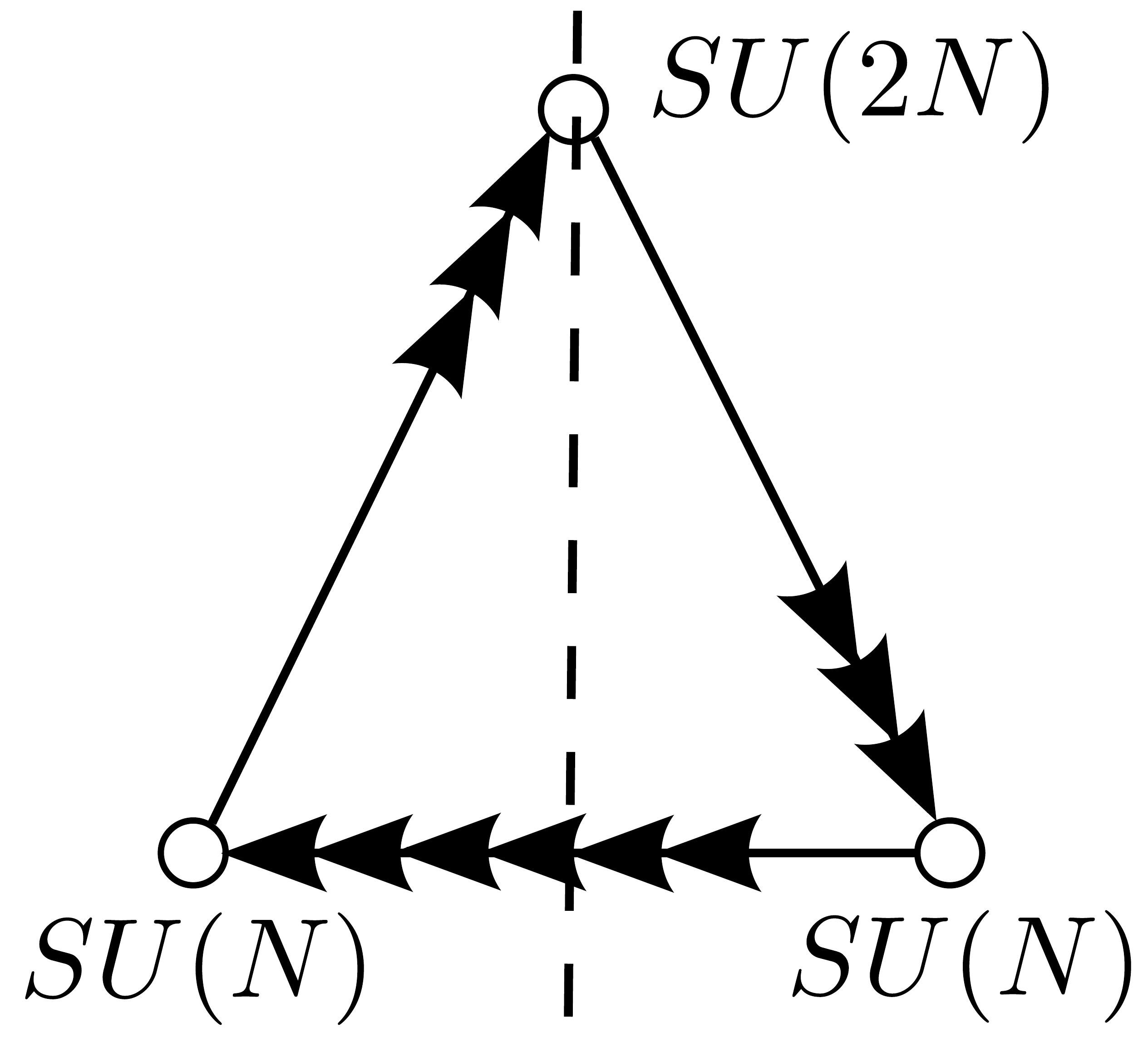}
  \end{center}
  \caption{Seiberg dual of the $\bC^3/\bZ_3$ orbifold theory. We have
    indicated the involution we want to study by the dashed line.}
  \label{fig:dp0phase2}
\end{figure}
Let us do a Seiberg duality on the top node of the quiver shown in
figure \ref{fig:dp0phase2}. This leads to an $\SU(2N) \times \SU(N)^2$
gauge theory that, following the procedure given in
\cite{Franco:2007ii,gauge}, can be orientifolded as indicated by the
dashed line in figure \ref{fig:dp0phase2}. There are two anomaly-free possibilities:
\begin{center}
  \begin{tabular}{c|cc|ccc}
     & $\SO(2(N+4))$ & $\SU(N)$ & $\SU(3)$ & $\U(1)_R$ & $\bZ_3$ \\ \hline
    $A^i$ & $\fund$ & $\fund$ & $\fund$ & $\frac13-\frac2N$ &$\omega_{3N}$ \\
    $B^{ij}$ & 1 & $\overline{\symm}$ & $\ov{\symm}$ & $\frac43 + \frac4N$ &$\omega_{3N}^{-2}$
  \end{tabular}
\end{center}
with superpotential
\begin{align}
W = \frac{1}{2} \Tr A_i A_j B^{ij}
\end{align}
and
\begin{center}
  \begin{tabular}{c|cc|ccc}
     & $\Sp(2(\tilde{N}-4))$ & $\SU(\tilde{N})$ & $\SU(3)$ & $\U(1)_R$ & $\bZ_3$\\ \hline
    $\tilde{A}^i$ & $\fund$ & $\fund$ & $\fund$ & $\frac13 + \frac{2}{\tilde{N}}$ &$\omega_{3 \tilde{N}}$\\
    $\tilde{B}^{ij}$ & 1 & $\ov{\asymm}$ & $\ov{\symm}$ & $\frac43 - \frac{4}{\tilde{N}}$ &$\omega_{3 \tilde{N}}^{-2}$
  \end{tabular}
\end{center}
with superpotential
\begin{align}
\tilde{W} = \frac{1}{2} \Tr \tilde{A}_i
\tilde{A}_j \tilde{B}^{ij},
\end{align}
where $\omega_n \equiv e^{2\pi i/n}$ and $\tilde{N}$ is even, since
for odd $\tilde{N}$ the $\Sp(2(\tilde{N}-4))$ gauge group has a Witten
anomaly. The discrete symmetry group is $\bZ_3$ since
the third power of the generator given above is contained in the gauge
group. For $N$ ($\tilde{N}$) not a multiple of 3 one can show that this
$\bZ_3$ is gauge equivalent to the center of the global $\SU(3)$. Thus
the global symmetry groups match only if $\tilde{N}$ and $N$ differ by
a multiple of 3.\footnote{For even $N$, the $\SO$ theory has an extra
  discrete $\bZ_2$ symmetry (cf. \eqref{eqn:dP0thyAevenN}) under which $A^i$ and $B^i$ carry charge
  $\omega_{6N}$ and $\omega_{6N}^{-2}$, respectively. The $\bZ_2$
  outer automorphism group of $\SO(2(N+4))$ is anomalous. Therefore,
  as expected from Seiberg duality, the global symmetry group of the
  $\SO$ theories for phase I and phase II match for even $N$. The
  $\bZ_2$ discrete anomalies satisfy the matching conditions given in
  \cite{Csaki:1997aw}.}

The global symmetry groups and anomalies for these two theories are
exactly as for the orientifold theories of phase I (see section
\ref{sec:Intro}) and are
\begin{center}
\begin{tabular}{ccc}
$\SO(2(N+4)) \times \SU(N)$ theory: && $\Sp(2(\tilde{N}-4)) \times \SU(\tilde{N})$ theory:\\
\begin{tabular}{|c|c|}
\hline
 $\SU(3)^3$ & $\frac{3}{2} (N-3) N$ \\
\hline
 $\SU(3)^2 \times \U(1)_R$ & $-\frac{1}{2} (N-3) N -6$ \\
\hline
 $\U(1)_R^3$ & $\frac{4}{3} (N-3) N-33$ \\
\hline
 $\U(1)_R$ & $-9$ \\
\hline
$\SU(3)^2 \bZ_3$ & 0\\
\hline
$\bZ_3$ & 0\\
\hline
\end{tabular} & &
\begin{tabular}{|c|c|}
\hline
 $\SU(3)^3$ & $\frac{3}{2} \tilde{N}(\tilde{N}+3) $ \\
\hline
 $\SU(3)^2 \times \U(1)_R$ & $-\frac{1}{2}  \tilde{N}(\tilde{N}+3)-6$ \\
\hline
 $\U(1)_R^3$ & $\frac{4}{3}\tilde{N}(\tilde{N}+3)-33$ \\
\hline
 $\U(1)_R$ & $-9$ \\
\hline
$\SU(3)^2 \bZ_3$ & 0\\
\hline
$\bZ_3$ & 0\\
\hline
\end{tabular}
\end{tabular}
\vspace{0.5\baselineskip}
\end{center}
where we write only those discrete anomalies which must match in
comparing two dual theories~\cite{Csaki:1997aw}. The anomalies of the
two models above match for $\tilde{N} =N -3$.

The two theories given above can also be derived by applying Seiberg duality to the $\SO$ or $\Sp$ node of the orientifolds of phase I and integrating out the massive matter.  In the remainder of this section we derive these two quiver
theories explicitly using string theory methods and show that they are
related by an orientifold transition.

\subsection{String theory}

Ordinary Seiberg duality can be understood in the context of the
derived category as a tilting of the category
\cite{Berenstein:2002fi,Braun:2002sb,Herzog:2004qw,Aspinwall:2004vm}. In
the particular case of the original
collection~\eqref{eq:usual-collection-projective} the tilting object
giving rise to the Seiberg dual theory can be easily constructed,
following the procedure in \cite{Herzog:2004qw,Aspinwall:2004vm}, as:
\begin{align}
  \label{eq:dual-collection-projective}
  \begin{split}
    \sP_0' & = \cO\\
    \sP_1' & = \underline{\cO(1)^{\oplus 3}} \to \cO(2)\\
    \sP_2' & = \cO(1)
  \end{split}\, ,
\end{align}
where the underline denotes position zero in the complex.

\begin{figure}
  \begin{center}
    \includegraphics[width=0.3\textwidth]{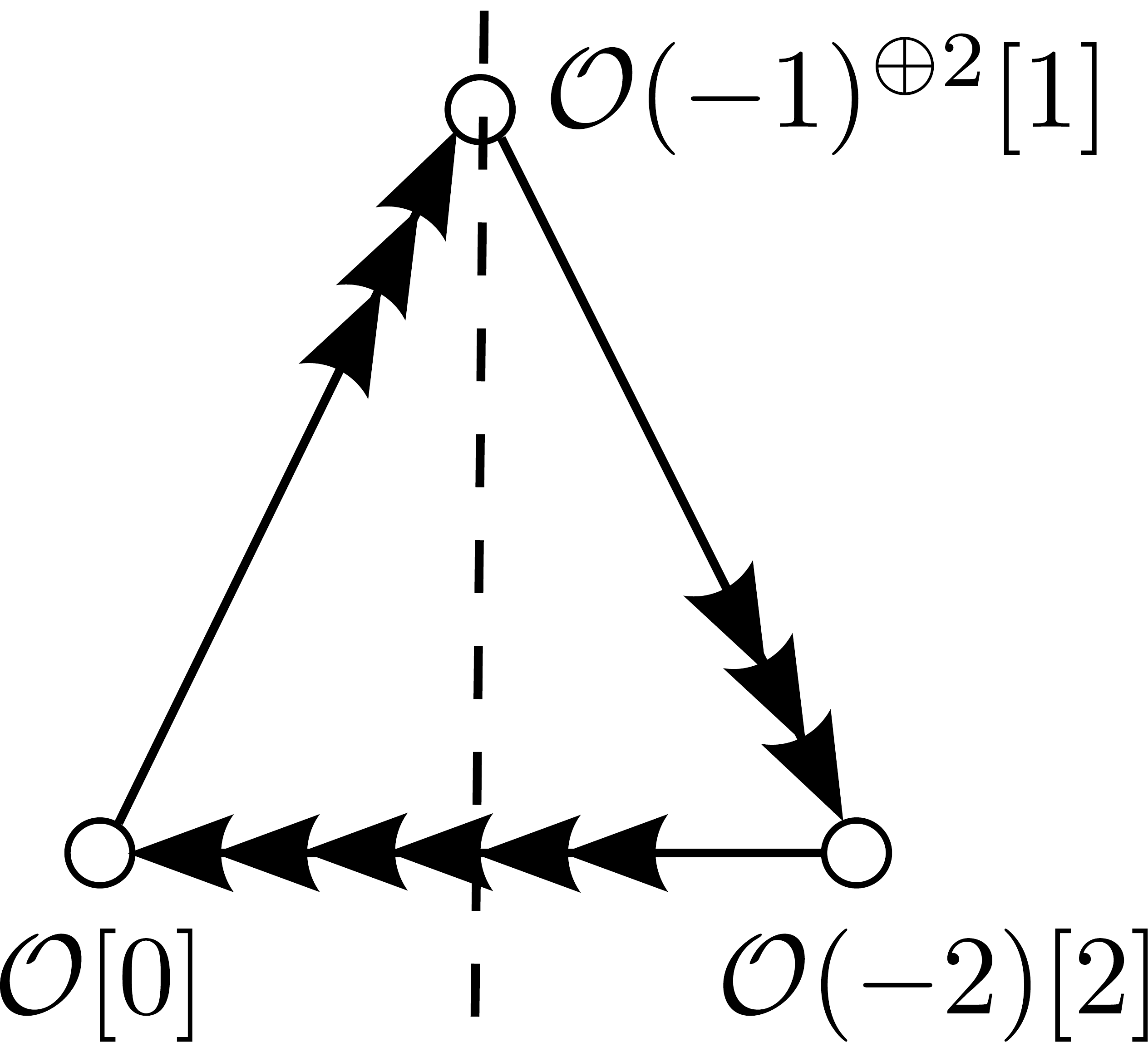}
  \end{center}
  \caption{Basis of branes for the Seiberg dual of the $\bC^3/\bZ_3$
    orbifold theory.}
  \label{fig:dp0phase2exc}
\end{figure}

We can construct the basis of fractional branes by mutating the
collection~$\{\sP_i'\}$ as usual \cite{Herzog:2003dj}, with the result:
\begin{align}
  \cC' = \bigl\{\cO, \cO(-1)^{\oplus 2}[1], \cO(-2)[2]\bigr\}\, .
\end{align}
This is the same basis of fractional branes given in
\cite{Cachazo:2001sg}, with the refinement of having the grading in
$\bZ$, rather than $\bZ_2$. We have taken two copies of $\cO(-1)[1]$
in order to cancel tadpoles. It is easy to compute the spectrum of
bifundamentals for this set of fractional branes. We show the
resulting quiver in figure~\ref{fig:dp0phase2exc}, which agrees with
the quiver in figure~\ref{fig:dp0phase2}, as it should.

On the other hand, it is clear from~\eqref{eq:orientifold-action} that
the ordinary orientifold involution does not act on the fractional
branes in the way that we expect, for example:
\begin{align}
  i_*\cO(-1)[1] \longrightarrow i_*(\cO(1)\otimes \cO(-3))[1] =
  i_*\cO(-2)[1]\, ,
\end{align}
so this action does not map the $\U(N)$ stack to itself. The solution,
as advanced in section~\ref{sec:DC-preliminaries}, is to introduce a
non-vanishing $B_2$ field, in this way modifying the orientifold action
to~\eqref{eq:orientifold-action-B}. In particular, we will choose
$B_2=\frac{1}{2}$, or equivalently $\cL_{2B_2}=\cO(1)$. The resulting
orientifold then acts as we expect:
\begin{align}
  \begin{split}
    i_*\cO(-1)[1] & \longrightarrow i_*(\cO(1)\otimes \cO(-3)\otimes \cO(1))[1] =
    i_*\cO(-1)[1]\\
    i_*\cO(-2)[2] & \longrightarrow i_*(\cO(2)\otimes \cO(-3)\otimes \cO(1))[0] =
    i_*\cO\\
    i_*\cO & \longrightarrow i_*(\cO\otimes \cO(-3)\otimes \cO(1))[2] =
    i_*\cO(-2)[2]\, .
  \end{split}
\end{align}

The matter content of the orientifolded theory can be derived using
the rules in section~\ref{sec:DC-preliminaries}. Assume that we
introduce an $O7^+$ plane. In order to cancel $D7$ tadpoles we need to
introduce 8 $\cO(-1)[1]$ planes. We will determine the projection on
the invariant branes momentarily; for now let us denote the group on
the stack of $\cO(-1)[1]$ branes by $G$, which can be either $\SO$ or
$\Sp$. Adding $N$ regular $D3$ branes, we obtain a gauge group
$G(2(N+4))\times \U(N)$. The chiral multiplet spectrum can be easily
obtained using table~\ref{table:orientifold-spectrum} and the charge
vectors
\begin{align}
  \label{eq:phaseII-charge-vectors}
  \begin{split}
    e^{-B_2}\Gamma(\cO) & = 1 + \ell + \frac{5}{8}\ell^2\\
    e^{-B_2}\Gamma(\cO(-1)[1]) & = -1 - \frac{1}{8}\ell^2\\
    e^{-B_2}\Gamma(\cO(-2)[2]) & = 1 - \ell + \frac{5}{8}\ell^2\\
    \Gamma(O7^+) & = 8 - \frac{1}{2}\ell^2\, ,
  \end{split}
\end{align}
and it is given by:
\begin{align}
  \label{eq:P2'-O7+-quiver}
  \begin{array}{cc|c}
    SO(2(N+4)) & \U(N) & \SU(3)\\
    \hline
    \ov\fund & \fund & \fund\\
    1 & \ov\symm & \ov\symm
  \end{array}
\end{align}
where we have set $G=\SO$ by comparing with the expectation from field
theory. (As in the theory before Seiberg duality, a derivation from
first principles using the techniques in
\cite{Brunner:2008bi,Gao:2010ava} should be possible, but we will not
attempt to do so here.) Notice also the flavor structure, which can be
derived as follows. The modes in the $\fund$ of $\U(N)$ come from:
\begin{align}
  \Ext^1_X(i_*\cO(-2)[2], i_*\cO(-1)[1]) =
  \Ext^0_\cS(\cO(-2),\cO(-1)) = \bC^3\, ,
\end{align}
where we have used~\eqref{eq:ext-restriction}, and the fact that
$\Ext^\bullet_\cS(\cO(-1), \cO(-2))=0$. Furthermore, we can identify
geometrically the $\SU(3)$ flavor group as $\SU(3)$ rotations on the
$(z_1,z_2,z_3)$ homogeneous coordinates on $\bP^2$. We have that
$\Ext^0_\cS(\cO(-2),\cO(-1))=\Gamma(\cO(1))$, i.e. the group of
sections of $\cO(1)$, which are described by polynomials of the form
$\sum a_iz_i$. Thus we can immediately see that the elements transforming
in the fundamental of $\U(N)$ also transform in the fundamental of
$\SU(3)$. Similarly, the fields transforming in the $\ov\symm$ of
$\U(N)$ come from:
\begin{align}
  \Ext^1_X(i_*\cO[0],i_*\cO(-2)[2]) = \Ext^0_\cS(\cO(-2), \cO)^\vee =
  \bC^6\, .
\end{align}
One has that $\Ext^0_\cS(\cO(-2), \cO)=\Gamma(\cO(2))$. These are
polynomials in the homogeneous coordinates of the form
$\sum_{ij}c_{ij}z_i z_j$, which clearly transform in the symmetric
representation of the flavor group. Notice, though, that Serre duality
gives us the \emph{dual} of $\Gamma(\cO(2))$, which accordingly
transforms in the conjugate representation.

\paragraph{Seiberg duality as motion in moduli space}

Before going into details of the orientifold transition in this
system, we would like to clarify a couple of points in the discussion
above. Notice that in the process of Seiberg dualizing we had to
introduce half a unit of $B_2$ field. It can also be easily seen that if
we start with an $O7^+$ its Seiberg dual should be an $O7^-$. We will
now argue that both statements are compatible with (and in the case of
the $B_2$ field, required from) the usual picture in string theory of
Seiberg duality as a motion in K\"ahler moduli space.

\begin{figure}
  \begin{center}
    \includegraphics[width=0.7\textwidth]{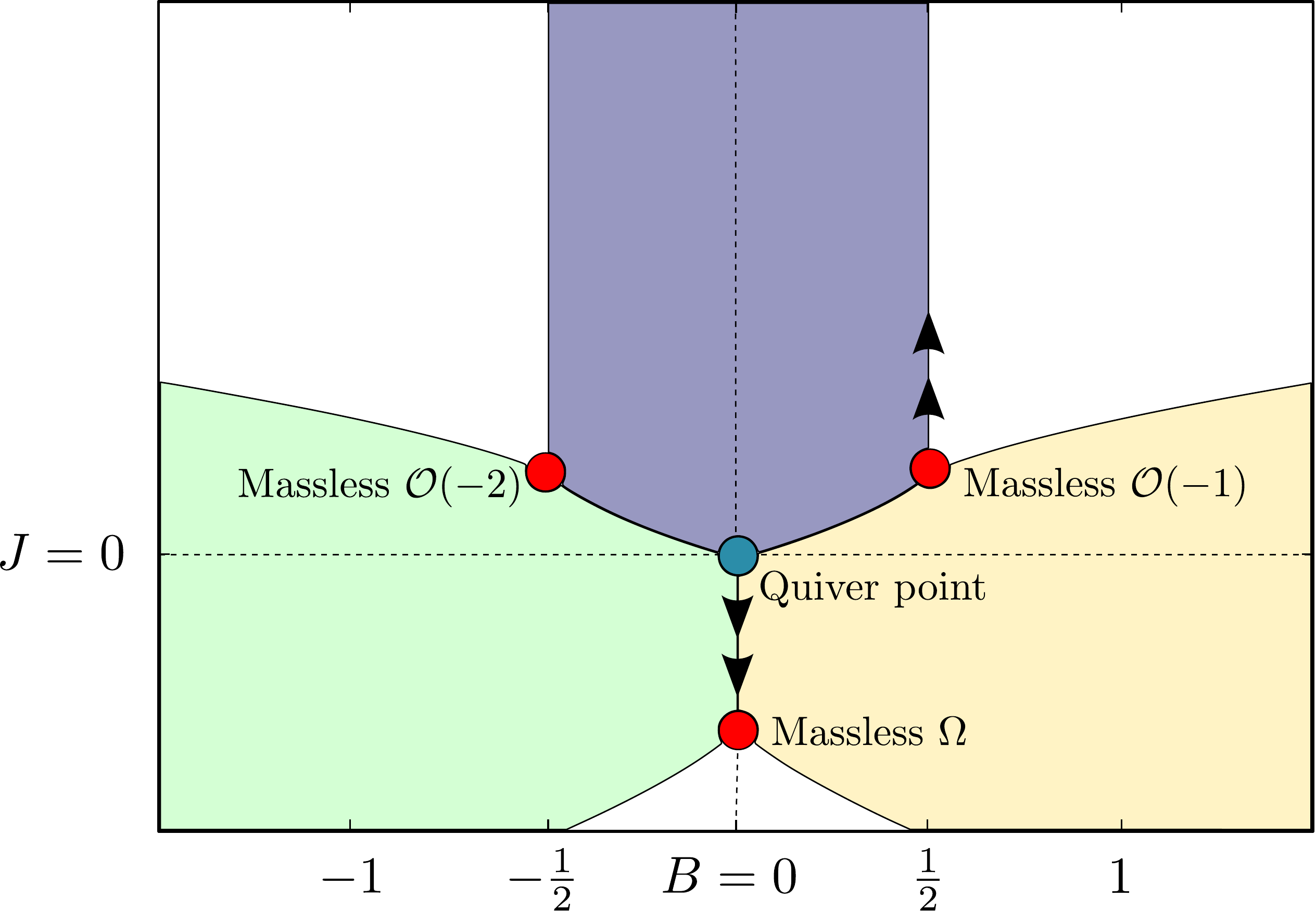}
  \end{center}
  \caption{Unfolded moduli space of $\bC^3/\bZ_3$
    \cite{Aspinwall:2004jr}. The three colored regions are $\bZ_3$
    images of each other, each region being a copy of the fundamental
    $\bP^1$ moduli space. We have denoted the points where the
    fractional branes become massless by red dots. The path followed
    in doing Seiberg duality is marked by arrows.\label{fig:dP0stringySD}}
\end{figure}

Recall that the quantum moduli space of the $\bC^3/\bZ_3$ geometry can
be seen as a $\bP^1$ with three marked points: the quiver point, the
large volume point, and a ``conifold'' point in which a certain
D-brane becomes massless. In order to visualize this structure it is
convenient, as done in \cite{Aspinwall:2004jr}, to unfold this sphere
into three copies in such a way that the quantum $\bZ_3$ symmetry of
the configuration is manifest. We present the resulting moduli space
in figure~\ref{fig:dP0stringySD}. Due to the orientifold projection
this moduli space is restricted to integer and half-integer values of
the $B_2$ field.

In brane constructions Seiberg duality can often be understood as
continuation beyond infinite coupling (see \cite{Giveon:1998sr} for a
nice review of a number of examples). In our configuration this can be
achieved as follows. We start from the quiver point at $B_2+iJ=0$, and
continue towards negative values of $J$ (we depict the motion by
arrows in figure~\ref{fig:dP0stringySD}). At a particular point along
this line the invariant brane $\Omega$ becomes massless.\footnote{In
  terms of the coordinates for the moduli space introduced below this
  is the point $\psi=1$.} Note that due to the $\bZ_3$ symmetry this
point is identified with the point at which $\cO(-1)$ becomes
massless, so we can continue beyond the singular CFT point along the
$B_2=\frac{1}{2}$ line. In this picture $\cO(-1)$ is the invariant brane
which can become massless, and we have a non-vanishing background
value for the $B_2$ field, perfectly consistent with the description of
Seiberg duality above. It is also easy to verify that the collection
of branes that we found by tilting is that given by Picard-Lefschetz
monodromy around the point where the invariant brane becomes massless,
precisely as advocated in the mirror context in
\cite{Cachazo:2001sg}. (We discuss further the mirror picture in
appendix~\ref{sec:SD-mirror}.)

There are a couple of complementary perspectives that could be
illuminating. First, notice that in figure~\ref{fig:dP0stringySD}
there are three branches coming out of the point where $\Omega$
becomes massless. One of the branches is associated with the ordinary
large volume orientifold at $B_2=0$. The other two branches are
{\ae}rientifolds of the type that we have discussed in
section~\ref{sec:aerientifolds}. So Seiberg duality in this context
involves a change in the orientifold type as we cross a conifold
point, a process quite reminiscent of the processes analyzed in
\cite{Hori:2005bk}. In our case the orientifold changes between an
ordinary orientifold with $B_2=0$ and an {\ae}rientifold, which by
composition with the quantum symmetry can be turned into an ordinary
orientifold of opposite type with $B_2=\frac{1}{2}$, as we implicitly
did above.

\begin{figure}
  \begin{center}
    \includegraphics[width=0.4\textwidth]{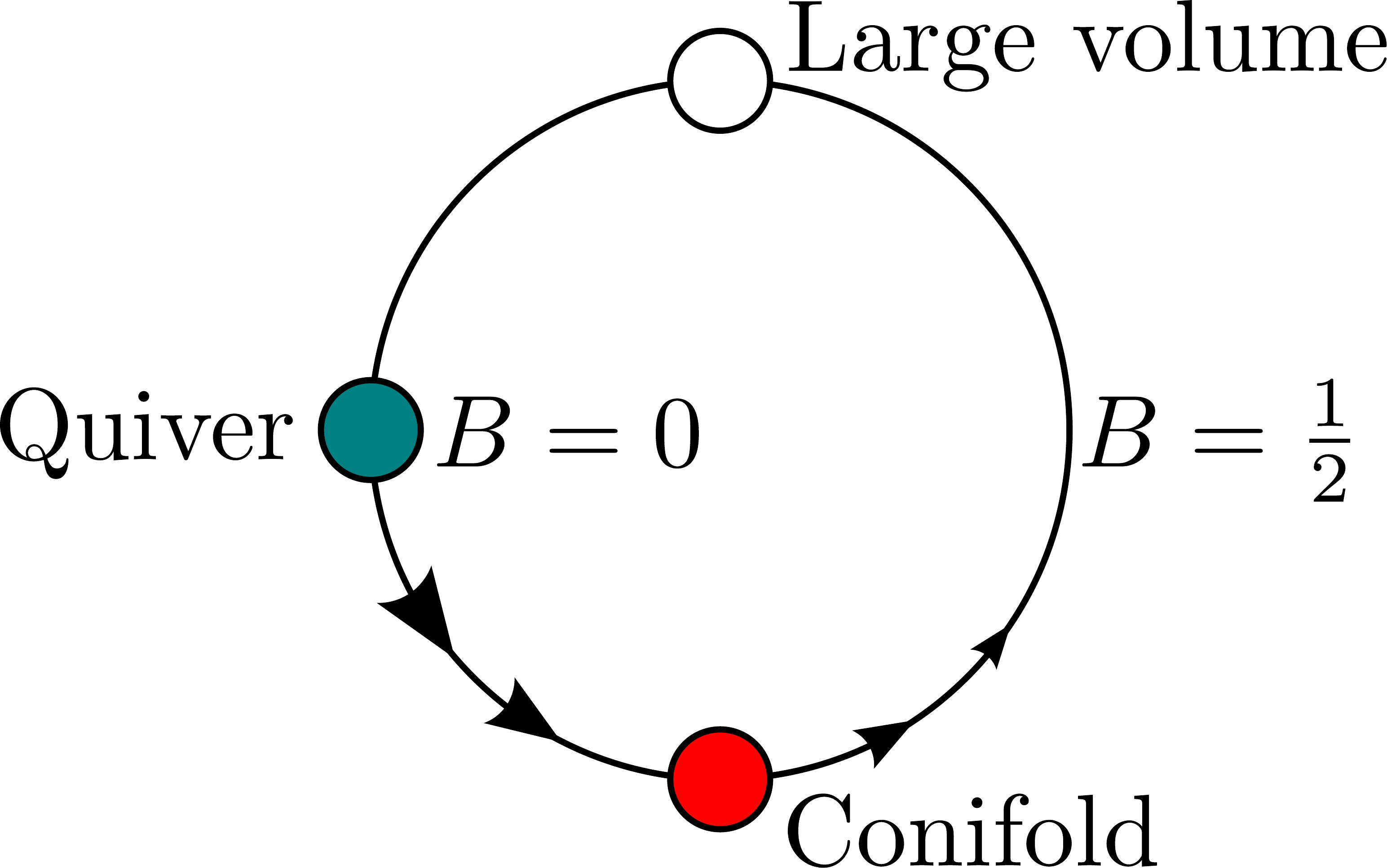}
  \end{center}
  \caption{Real moduli space for the orientifolded $\bC^3/\bZ_3$. The
    two possible values for the $B_2$ field connect at the singular CFT
    and the large volume points. Seiberg duality comes from crossing
    from the quiver side, with $B_2=0$, to the side with
    $B_2=\frac{1}{2}$, via the conifold point.\label{fig:P2-SD-circle}}
\end{figure}

Finally, in the picture of the moduli space as a $\bP^1$, we have that
the orientifold action constrains us to move along the equator. All
three special points are located along the equator, and in particular
the large volume and conifold points naturally divide the equator into
two halves, which we identify with $B_2=0$ and $B_2=\frac{1}{2}$. Seiberg
duality corresponds to crossing from one branch to the other through
the conifold point. We illustrate this structure in
figure~\ref{fig:P2-SD-circle}.

\begin{figure}
  \begin{center}
    \includegraphics[width=0.5\textwidth]{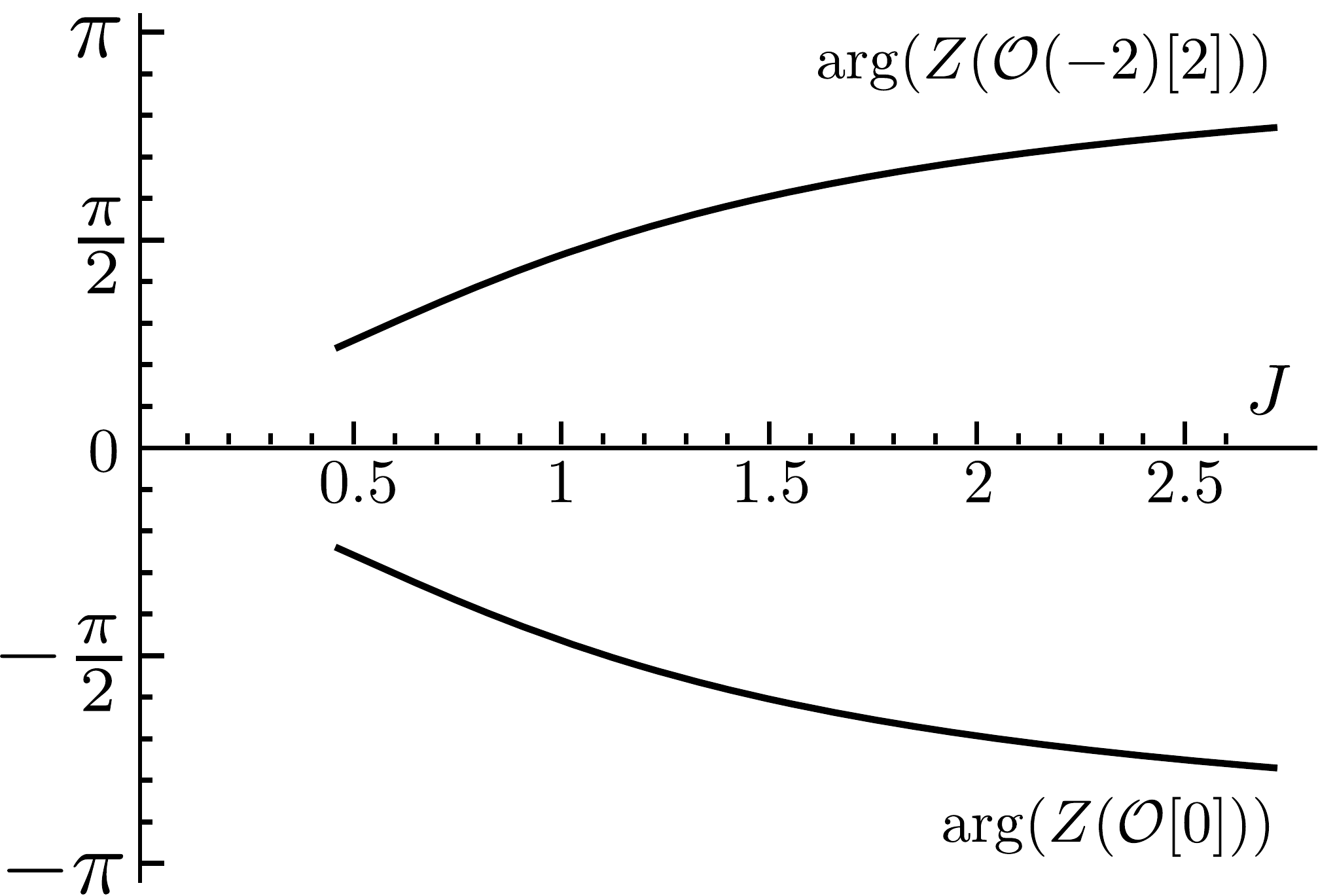}
  \end{center}

  \caption{Behavior of the phase of the central charge for the
    fractional branes of the Seiberg dual phases as we go towards
    large volume along the $B_2=\frac{1}{2}$ line, starting from the
    conifold point at $\psi=e^{2\pi i/3}$ (equivalently
    $B_2=\frac{1}{2}$, $J\approx 0.46$). The central charge of the
    $\cO(-1)[1]$ brane stays real and positive, so we have omitted it
    from the diagram. The BPS phases asymptote to $\pm \pi$.}

  \label{fig:phasesSD}
\end{figure}

A point which is clear in this last picture is that, as least at the
level of motion in K\"ahler moduli space, the Seiberg dual brane
configuration is not supersymmetric, since the quiver point lays in
the $B_2=0$ half of the real moduli space. It is not difficult to see
this explicitly by a direct computation of the periods, which satisfy
the Picard-Fuchs equation
\cite{Candelas:1990rm,Morrison:1991cd,BatyrevVariations,Aspinwall:1993xz}
\begin{align}
  \left[\left(z\frac{d}{dz}\right)^3 +
  27z\left(z\frac{d}{dz}\right)\left(z\frac{d}{dz} +
    \frac{1}{3}\right) \left(z\frac{d}{dz} +
    \frac{2}{3}\right)\right]\Phi = 0\, .
\end{align}
A basis of solutions for this equation was found in
\cite{Aspinwall:1993xz,Greene:2000ci,Aspinwall:2004jr}. A convenient
way of presenting the general form of the solution to this class of
problems and doing the analytic continuations is in terms of Meijer
$\meijerG{}$ functions \cite{Greene:2000ci}. Choosing the same
conventions we chose in writing~\eqref{eq:dp0-general-Z}, and
introducing a variable $\psi$ given by $-27z=\psi^{-3}$, we obtain:
\begin{align}
  \label{eq:C3/Z3-periods}
  \begin{split}
    \Phi_1 &= -\frac{\sqrt{3}}{4\pi^2 i}\meijerG\left[\begin{array}{cc|c}
      \frac{1}{3} & \frac{2}{3} & 1\\
      0 & 0 & 0
    \end{array}\right]\!\!\left(-\psi^{-3}\right)\, ,\\
  \Phi_2 &= -\frac{\sqrt{3}}{4\pi^3}\meijerG\left[\begin{array}{c|c}
      \frac{1}{3} \,\, \frac{2}{3} & 1\\
      0 \, 0 \, 0 & -
    \end{array}\right]\!\!\left(\psi^{-3}\right)\, .
  \end{split}
\end{align}
Plugging these values in the expression for the central
charge~\eqref{eq:dp0-general-Z} and going towards large volume along
the $B_2=\frac{1}{2}$ line, we obtain the BPS phases shown in
figure~\ref{fig:phasesSD}, which clearly show that the Seiberg dual
system of branes is not supersymmetric.

We will give further evidence for this statement by carefully
analyzing the BPS structure of the mirror in
appendix~\ref{sec:SD-mirror}.

This lack of supersymmetry is clearly something that makes the brane
construction of the Seiberg dual somewhat less appealing, but notice
that the problem is independent of the presence of the
orientifold. Since the mismatch is just at the level of D-terms, we
will just assume that the information about the field theory that we
get from the brane construction is still reliable, and proceed with
the construction. Notice that taking the discussion in this section at
face value would then imply a strong/weak duality between a pair of
non-supersymmetric theories, different from the example considered in
\cite{Uranga:1999ib,Sugimoto:2012rt}.

\paragraph{The orientifold transition.}

In the previous discussion we have considered the case in which we add
an $O7^+$ plane. The other possibility consists of adding an $O7^-$
plane, which gives the following theory:
\begin{align}
  \label{eq:P2'-O7--quiver}
  \begin{array}{cc|c}
    \Sp(2(\tilde{N}-4)) & \U(\tilde{N}) & \SU(3)\\
    \hline
    \ov\fund & \fund & \fund\\
    1 & \ov\asymm & \ov\symm
  \end{array}
\end{align}

As before, let us assume that the two configurations are dynamically
connected via a strongly coupled orientifold transition. We again expect a process of the form:
\begin{align}
  O7^+ + 4 \cO(-1)^{\oplus 2}[1] \longleftrightarrow O7^- + 4(\cO +
  \cO(-2)[2]) + n\, D3s\, .
\end{align}
Conservation of $D3$ charge then requires:
\begin{align}
   8 - \frac{1}{2}\ell^2 + (-8 - \ell^2) = (-8 + \frac{1}{2}\ell^2) +
   (8 + 5\ell^2) + n\ell^2\, ,
\end{align}
which implies $n=-7$. Adding $N$ regular branes to the $O7^+$ side one
obtains the theory~\eqref{eq:P2'-O7+-quiver}. After the transition we
thus expect the spectrum:
\begin{align}
  O7^+ + 4 \cO(-1)^{\oplus 2}[1] + N\, D3s\longleftrightarrow O7^- + 4(\cO +
  \cO(-2)[2]) + (N-7) \, D3s\, ,
\end{align}
i.e. the theory \eqref{eq:P2'-O7--quiver} with $\tilde{N}=N-3$, in
perfect agreement with the expectations from field theory.

\section{Conclusions}
\label{sec:conclusions}

In this paper we have argued that the field theory duality presented
in \cite{gauge} admits a very natural embedding in string theory as
the action of type IIB S-duality on branes at singularities.

Building on the field theory checks performed in~\cite{gauge}, in this paper we argued that
the brane configurations corresponding to the dual theories of~\cite{gauge} source
discrete torsions for the NSNS and RR two-forms related by
S-duality. Furthermore, we found that the collections of fractional branes constructing the dual theories
are in fact S-dual once the $O7^+$ is resolved into its $(p,q)$ seven-brane components.

Taken together, these arguments give very strong support to the idea
that the $\cN=1$ theories we have been discussing are indeed related
by strong/weak dualities, and illuminate the physical origin of some
of its main features, such as the change in rank and the change between
$\SO$/$\Sp$ groups and symmetric/antisymmetric projections.

There are a number of interesting directions for future work, some of which we now discuss.
First, it would be very interesting to extend
the ideas in this work to theories without supersymmetry. It was
realized in \cite{Uranga:1999ib,Sugimoto:2012rt} that the study of a
non-supersymmetric version of the brane configuration engineering
$\cN=4$ SYM would give interesting insight into the strong dynamics of
the corresponding non-supersymmetric version of $\cN=4$. The same idea
should generalize to the much larger class of $\cN=1$ duals we have
introduced in this paper (and the ones to appear in
\cite{strings-all}), potentially giving a window into
the strongly coupled dynamics of a large class of non-supersymmetric
theories.

There are also several formal problems that we have not addressed
in this work, but which would be interesting to understand. One such issue is that of K-theory tadpoles.
Typically, in the presence of orientifolds, in
addition to the usual conditions for cancellation of RR and NSNS
tadpoles one should also make sure that certain $\bZ_2$ valued
K-theory tadpoles are canceled
\cite{Witten:1998cd,Moore:1999gb,Diaconescu:2000wy,Diaconescu:2000wz,Uranga:2000xp,LoaizaBrito:2001ux,GarciaEtxebarria:2005qc}. It
would be very interesting to have a systematic understanding of such
K-theory tadpoles in the configurations we study in this series of
papers.\footnote{For our main $\bC^3/\bZ_3$ example and its orbifold
  brethren the topological structure is very similar to the $\cN=4$
  case, so there is most likely no issue with K-theory tadpoles. However, more
  involved non-orbifold examples \cite{strings-all} may exhibit
  more interesting structures.}

Finally, our discussion of the strongly coupled behavior of the $O7^+$
plane was mostly kinematical, focusing on monodromy and charge
conservation. It would be very interesting (but probably quite
involved) to study the dynamics in more detail, and see that
orientifold transitions such as~\eqref{eq:P2-orientifold-transition}
are also dynamically preferred, in the sense that the brane
recombinations proceed as we have described.

\acknowledgments

We would like to thank P.~Aspinwall, M.~Berg, P.~Berglund, M.~Buican,
P.~C\'amara, C.~Csaki, A.~Dabholkar, N.~Halmagyi, S.~Kachru,
O.~Loaiza-Brito, F.~Marchesano, L.~McAllister, N.~Mekareeya,
B.~Pioline, A.~Sagnotti, S.~Sch\"afer-Nameki, E.~Silverstein,
G.~Torroba and A.~Uranga for illuminating discussions. We particularly
thank L.~McAllister and G.~Torroba for initial collaboration and
extensive discussions.  The work of B.H.  was supported by the Alfred
P. Sloan Foundation and by the NSF under grant PHY-0757868 and in part by a John and David Boochever Prize
Fellowship in Fundamental Theoretical Physics. The work of T.W. was
supported by a Research Fellowship (Grant number WR 166/1-1) of the
German Research Foundation (DFG). B.H. gratefully acknowledges support
for this work by the Swedish Foundation for International Cooperation
in Research and Higher Education and by a Lucent Travel Award. We
would like to thank the organizers of String Phenomenology 2011,
Strings 2011, and String Phenomenology 2012 for providing stimulating
environments in which part of this work was carried out. B.H. would
also like to thank the organizers of the Nordita string phenomenology
workshop, the 2012 Carg\`{e}se summer school on Gauge Theory and
String Theory, and the Simons Center Graduate Summer School on String
Phenomenology for likewise providing a stimulating and productive
environment. I.G.-E. thanks N.~Hasegawa for kind encouragement and
constant support.

\appendix

\addtocontents{toc}{\protect\setcounter{tocdepth}{1}}

\section{Mirror picture for Seiberg duality on \alt{$\bC^3/\bZ_3$}{C3/Z3}}
\label{sec:SD-mirror}

We would like to describe the mirror geometry to the $\bC^3/\bZ_3$
orbifold. As discussed in detail in \cite{Hori:2000kt,Hori:2000ck}
(see also \cite{Feng:2005gw}), for the purposes of studying BPS
objects the mirror can be taken to be a fibration over the complex
plane with fiber $\bC^*\times T^2$. Parameterizing the base of the
fibration by $z$, the total geometry is given by:
\begin{align}
  uv & =  z - 3\psi\label{eq:C*-fibration}\\
  y^2 & = x^3 + \left(\frac{z}{2}\right)^2 x^2 +
  \left(\frac{z}{2}\right)x + \frac{1}{4}\, ,
\end{align}
with $u,v\in \bC$ and $x,y\in\bC^*$. The dependence on the K\"ahler
moduli of the original geometry is encoded in the variable $\psi$
introduced above~\eqref{eq:C3/Z3-periods}. We can recast the elliptic
part of the fibration in the usual Weierstrass form by a linear change
in variables, getting:
\begin{align}
  y^2 = x^3 + f(z) x + g(z)
\end{align}
with
\begin{align}
  \label{eq:mirror-fibration-fg}
  \begin{split}
    f(z) & = -\frac{z}{2}\left(\frac{z^3}{24} - 1\right)\\
    g(z) & = \frac{z^6}{864} - \frac{z^3}{24} + \frac{1}{4}\, .
  \end{split}
\end{align}
The resulting discriminant is then given by:
\begin{align}
  \Delta(z) = 4f(z)^3 + 27g(z)^2 = \frac{1}{16}\left(27 - z^3\right)\, ,
\end{align}
and from here we see that the elliptic fiber becomes singular at the
three points $z_* = 3\{1,\, e^{2i\pi/3},\, e^{4i \pi/3}\}$ (associated
to the conifold points $\psi=\{1,\, e^{2i\pi/3},\, e^{4i \pi/3}\}$ in
the original $\bC^3/\bZ_3$ geometry). Finally, the holomorphic 3-form
of the geometry can be taken to be:
\begin{align}
  \label{eq:Omega-embedding}
  \Omega = \frac{dx}{y}\wedge dz \wedge \frac{du}{u}\, .
\end{align}

\subsection{Lattices and elliptic fibrations}

In order to compute the structure of special Lagrangian cycles, it is
convenient to take the torus from its Weierstrass expression in terms
of an equation in $\bC^2$ to a flat description of the form $\bC/\bL$,
where $\bL$ is a lattice generated by the two vectors $2\omega_1$ and
$2\omega_3$ ($\omega_1$ and $\omega_3$ are called the
\emph{half-periods}, and they are defined up to $SL(2,\bZ)$
transformations). Notice that in the following we will be working with
$\bL$, and not just the complex structure $\tau=\omega_3/\omega_1$ of
the torus. This is important in order to explicitly see the special
Lagrangian structure of our branes. The technology we will be using in
this section is well developed, but we will quickly review it here for
the convenience of the reader. We will follow the conventions in
\cite{DLMFModularFuncs}, to which we also refer for further
explanations.

The basic map uses the Weierstrass $\wp$ function, defined by:
\begin{align}
  \wp\left(\zeta|\bL\right)=\frac{1}{\zeta^{2}}+\sum _{{\omega\in\bL\setminus\{
      0\}}}\left[\frac{1}{(\zeta-\omega)^{2}}-\frac{1}{\omega^{2}}\right]\, .
\end{align}
This is a function of one complex coordinate $\zeta$, and the chosen
lattice $\bL$. The interest of this function is that it satisfies the
differential equation
\begin{align}
  \label{eq:de-Weierstrass}
  \left(\frac{d\wp}{d\zeta}\right)^2 = 4\wp^3 - g_2 \wp - g_3\, ,
\end{align}
which is clearly the equation of Weierstrass form. The $g_2$ and $g_3$
coefficients in this equation are determined in terms of $\bL$:
\begin{align}
  \label{eq:g2g3}
  \begin{split}
    g_2 & = 60 \sum_{\omega\in\bL\setminus 0} \frac{1}{\omega^4}\\
    g_3 & = 140 \sum_{\omega\in\bL\setminus 0} \frac{1}{\omega^6}\, .
  \end{split}
\end{align}
We can thus map the $x,y$ coordinates to a flat $\zeta$ coordinate if we
set $y=\wp'$, $x=\sqrt[3]{4}\, \wp$ and choose $\bL$
appropriately.

The lattice half-periods can be determined as follows. Consider a
Weierstrass fibration rewritten in the following way:
\begin{align}
  \label{eq:Weierstrass-math-conventions}
  y^2 = 4x'^3 - \widetilde{g_2}(z) x' - \widetilde{g_3}(z)\, .
\end{align}
This is just a change in conventions, with $\widetilde{g_2}(z) \equiv
-\sqrt[3]{4}f(z)$, $\widetilde{g_3}(z)\equiv -g(z)$. The right hand
side can be written as $4(x'-e_1)(x'-e_2)(x'-e_3)$. The $e_i$ are
called the \emph{lattice roots}. The half-periods are then given by:
\begin{align}
  \label{eq:half-periods}
  \begin{split}
    \omega_1 & = \frac{\pi}{2\sqrt{e_1-e_3}}\,
    _2F_1\left(\frac{1}{2},\frac{1}{2};1;\frac{e_2-e_3}{e_1-e_3}\right)\,
    ,\\
    \omega_3 & = i\frac{\pi}{2\sqrt{e_1-e_3}}\,
    _2F_1\left(\frac{1}{2},\frac{1}{2};1;\frac{e_1-e_2}{e_1-e_3}\right)\,
    ,
  \end{split}
\end{align}
where $\,_2F_1$ denotes the ordinary or Gaussian hypergeometric
function. These half-periods define the lattice $\bL=\{2\omega_1m +
2\omega_3n,\,\, m,n\in\bZ\}$. Choosing this lattice, we have that
$\wp(\zeta|\bL)$ satisfies~\eqref{eq:de-Weierstrass} with
$g_2=\widetilde{g_2}(z)$ and $g_3=\widetilde{g_3}(z)$. We thus have an
embedding into a flat torus, as claimed.

\subsection{Special Lagrangian branes at the quiver point}

Using the embedding we have just discussed, the holomorphic 3-form
simplifies to
\begin{align}
  \label{eq:Omega-flat}
  \Omega = d\zeta \wedge dz \wedge \frac{du}{u}\, .
\end{align}
Branes wrapping a 3-cycle $\cS$ are BPS if they are special
Lagrangian, which means that $\re(e^{i\theta}\Omega|_{\cS})=0$, for
some constant $\theta$. Two sLag branes $\cS_1$, $\cS_2$ are mutually
supersymmetric if $\theta_1=\theta_2$. In our case we will be
constructing supersymmetric branes wrapping a 1-cycle in the $\bC$
base, and a 1-cycle in each of the components of the fiber. We want to
construct a supersymmetric system of branes at the quiver point, which
is located at $\psi=0$. Equation~\eqref{eq:C*-fibration} then
reduces to $uv=z$.

\begin{figure}
  \begin{center}
    \includegraphics[width=0.3\textwidth]{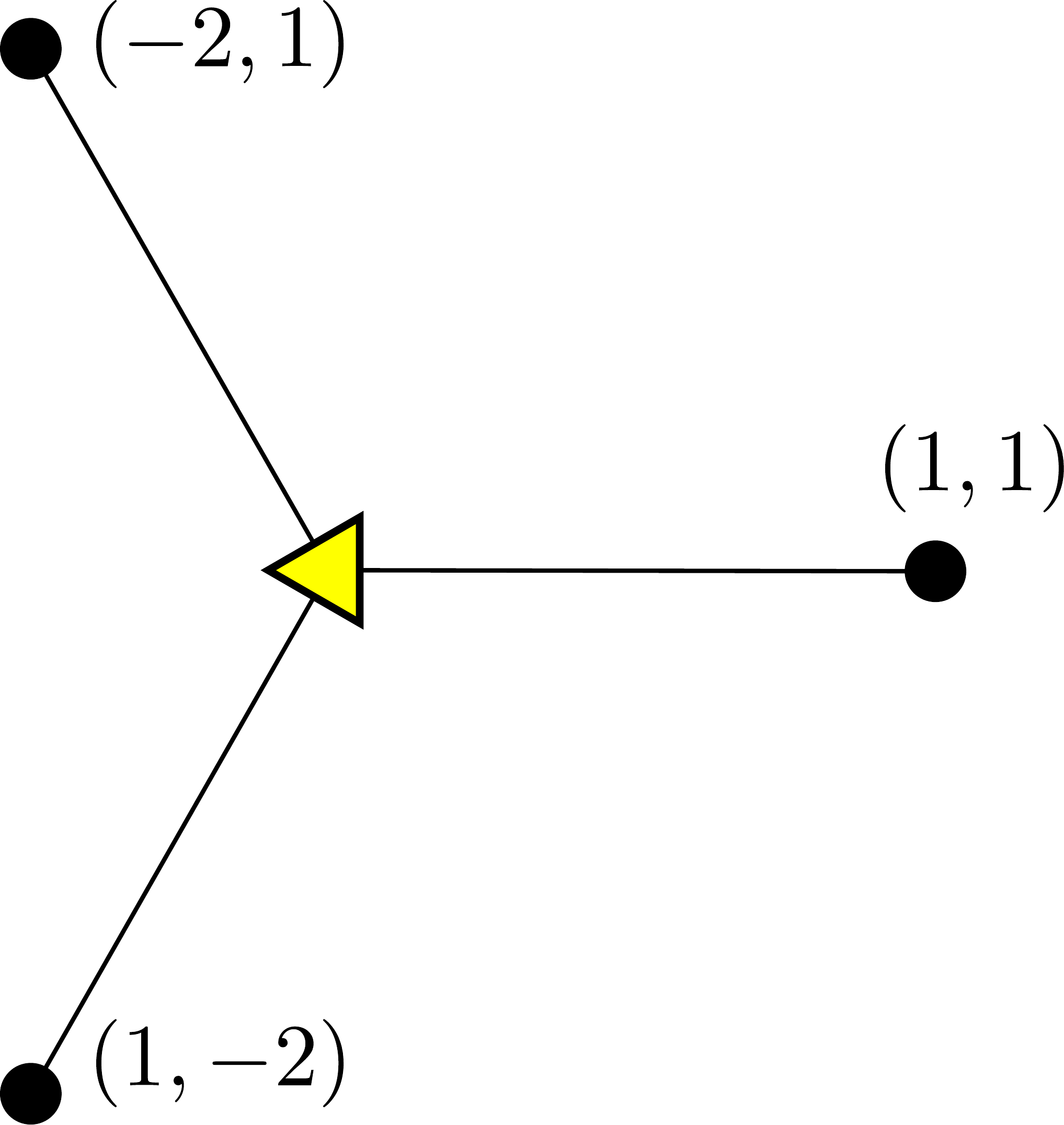}
  \end{center}
  \caption{Structure of mirror in the $z$ plane. The central yellow
    triangle is where the $\bC^*$ fiber degenerates, and the three
    black dots outside denote the zeroes of the discriminant. We have
    also indicated which $(p,q)$ cycle degenerates on which zero, in
    the particular $SL(2,\bZ)$ convention used in the text. Finally,
    the straight lines denote the segments in the base where the sLag
    branes are to be wrapped.\label{fig:mirror-fibration}}
\end{figure}

The $\bC^*$ direction is the easiest, we take a 1-cycle parameterized
by $u=e^{i\alpha}$, $v=ze^{-i\alpha}$, with $\alpha\in [0,2\pi]$. In
this case $du/u=i\, d\alpha$, so the phase of this component of
$\Omega$ is constant. In the base we will take the three straight
segments that connect $z=0$ with the three points where $\Delta$
vanishes. If the cycle degenerating at $z=z_*$ is of type $(p,q)$, we
will take a straight (in the $\zeta$-plane) line in the torus fiber, with
winding number $(p,q)$. We show the resulting geometry in
figure~\ref{fig:mirror-fibration}.
The total space is thus a
$S^1\times S^1$ fibration over a segment, with each of the fibers
degeneration at one of the ends of the segment. It is easy to convince
oneself that the topology of the resulting space is $S^3$. We will now
show that in addition this $S^3$ is sLag.

\begin{figure}
  \begin{center}
    \includegraphics[width=0.4\textwidth]{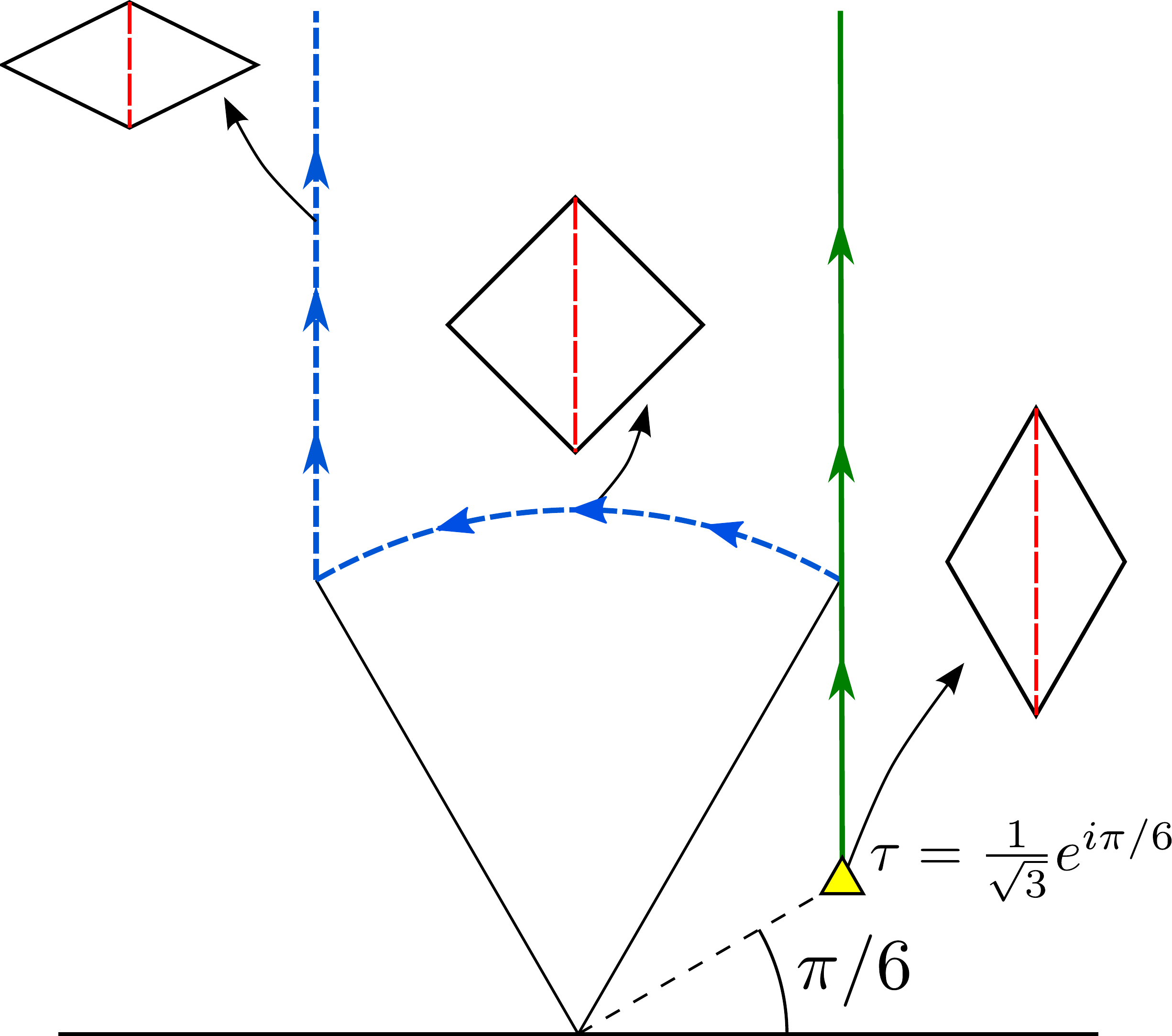}
  \end{center}
  \caption{Behavior of $\tau$ as we move along the base. We have
    depicted the corresponding rhombic lattice cell for various values
    of $\tau$. The red dashed line denotes the $(1,1)$ cycle wrapped
    by the brane. The solid green line on the right shows the path of
    $\tau$ in the natural conventions from the point of view of the
    rhombic lattice, while the dashed line on the left is an
    equivalent path along the edge of the fundamental domain, obtained
    by $\tau\to \tau/(1-\tau)$ for the component with $|\tau|\leq 1$,
    and $\tau\to\tau-1$ for $|\tau|\geq1$.\label{fig:tau-fundamental-region}}
\end{figure}

We will start with the horizontal cycle in
figure~\ref{fig:mirror-fibration}, going to the locus where the
$(1,1)$ cycle degenerates. The phase of the $dz$ term in $\Omega$ is
thus constant, and equal to 0. The structure in the elliptic fiber
is more involved, and it is here where the discussion in terms of the
lattice $\bL$ in the previous section pays off. It is useful to start
by looking to the behavior of the complex structure $\tau$ as we go
from $z=0$ to the $(1,1)$ degeneration. This can be obtained by
computing the inverse of the $j$ function. Introducing Klein's
invariant $J(\tau)=j(\tau)/1728$, one has:
\begin{align}
  \label{eq:J-inverse}
  J^{-1}(\lambda) = \frac{i(r(\lambda)-s(\lambda))}{r(\lambda)+s(\lambda)}
\end{align}
with
\begin{align}
  r(\lambda) & = \Gamma\left(\frac{5}{12}\right)^2
  \,_2F_1\left(\frac{1}{12},\frac{1}{12};\frac{1}{2};1-\lambda\right)\\
  s(\lambda) & = 2(\sqrt{3}-2)\Gamma\left(\frac{11}{12}\right)^2
  \sqrt{\lambda -1 }
  \,_2F_1\left(\frac{7}{12},\frac{7}{12};\frac{3}{2};1-\lambda\right)\, .
\end{align}
Plugging in the explicit fibration data~\eqref{eq:mirror-fibration-fg}
for our example, one can easily see that $\tau$ varies as shown in
figure~\ref{fig:tau-fundamental-region}.
 For our purposes the most
natural domain for $\tau$ comes from considering the rhombic lattice,
and it is shown as the green vertical path in
figure~\ref{fig:tau-fundamental-region}. Recall, from
\cite{DLMFModularFuncs}, that rhombic lattices are defined by
$\omega_1$ real and positive, $\im(\omega_3)\geq 0$ and
$\re(\omega_3)=\frac{1}{2}\omega_1$. Since conventionally
$\tau=\omega_3/\omega_1$, one automatically has
$\re(\tau)=\frac{1}{2}$.

In these conventions the path is straight, starting from the
equianharmonic\footnote{The equianharmonic lattice is a rhombic
  lattice where the angles are $\pi/6$ and $\pi/3$, see
  \cite{DLMFModularFuncs}.}  lattice at $z=0$, with
$\tau=e^{i\pi/6}/\sqrt{3}$, and going vertically to infinity, reached
at $z=z_*$, where the $(1,1)$ cycle degenerates. As we go along the
path, the unit cell becomes flatter and flatter in the vertical
direction. One can easily verify these statements using the
formulas~\eqref{eq:half-periods}.

Of crucial importance for us is that not only is $\tau$ that of a
rhombic lattice, but $\omega_1$ is real. This means that, no matter
our position along the segment in the $z$ plane, the $(1,1)$ cycle in
the torus always wraps real values for the flat coordinate $\zeta$, so
$d\zeta$ has constant phase, and thus $\Omega$ has constant phase, as we
claimed.

Notice that this construction strongly requires working with the whole
lattice $\bL$, and not just its complex structure $\tau$. If we had
defined the torus in the conventional way $\zeta\sim \zeta+1 \sim
\zeta+\tau$ the $(1,1)$ cycle would have had slope
$\arctan(\im(\tau)/(1+\re(\tau)))$, which can easily be seen to vary
non-trivially along the segment in the base.

Let us consider the other two $D6$ branes, going from the origin to
the discriminant points $z_*=3\{e^{2i\pi/3},\, e^{4 i \pi/3}\}$. For
simplicity let us consider $z_*=3e^{2i\pi/3}$ only, the other point
works similarly. The first observation is that the segment in the base
can be obtained simply by multiplying the segment we just considered
by $\beta \equiv e^{2i\pi/3}$ (i.e. it is just a rotation by
$120^\circ$). From the expression for $f$ and $g$
in~\eqref{eq:mirror-fibration-fg} it is clear that upon doing this
rotation $f\to\beta f$, $g\to g$. In addition, since $j(\tau)\sim
f^3/\Delta$, one has that the $j$ function is invariant upon doing
this rotation, and this implies (by taking the inverse
via~\eqref{eq:J-inverse}) that $\tau$ is unaffected by the
rotation. Since now we want to consider $(p,q)$ cycles different from
$(1,1)$ this would immediately imply (if we just look to $\tau$) that
the phase of $\Omega$ would vary as we move along the base, making the
cycle non-calibrated.

\begin{figure}
  \begin{center}
    \includegraphics[width=0.3\textwidth]{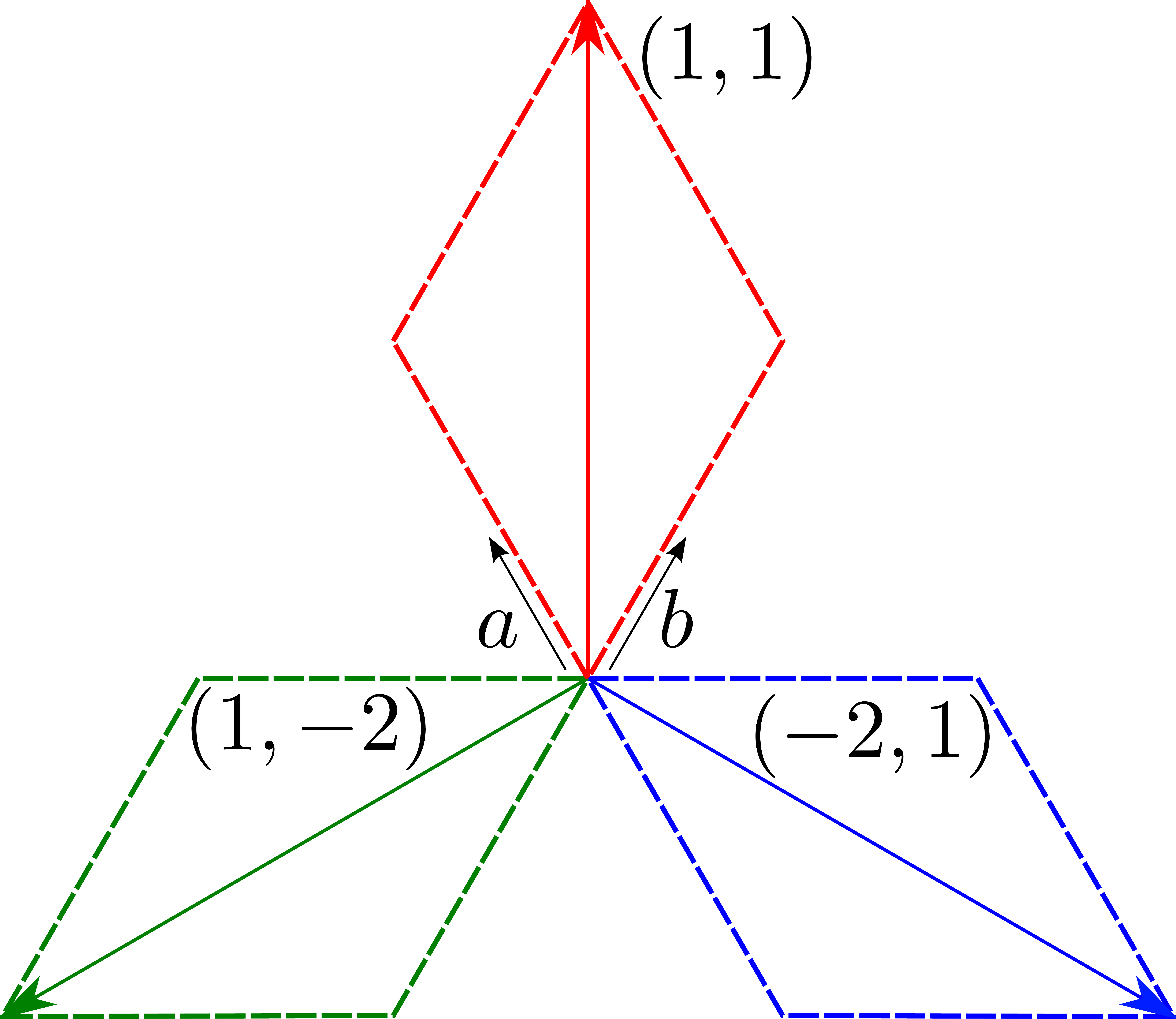}
  \end{center}
  \caption{Three fundamental cells of the equianharmonic lattice,
    rotated by powers of $\beta$. We have drawn the cycles wrapped by
    the $D6$ branes.\label{fig:EquianharmonicLattice}}
\end{figure}

The resolution is as before looking to $\bL$, instead of just
$\tau$. In particular, from~\eqref{eq:g2g3} one easily sees that
$\omega_i\to \beta^{-1}\omega_i$, so the lattice is rotated in the
direction opposite to the rotation in the base. This also implies
that, as we move towards $z_*$, the equianharmonic lattice will start
deforming, in a way that is a $\beta^{-1}$ rotation of the deformation
we saw before. In particular the brane related by a $\beta^{-1}$
rotation to the $(1,1)$ brane will be the one with constant slope. As
depicted in figure~\ref{fig:EquianharmonicLattice}, in this particular
case the brane of interest is the $(-2,1)$ brane.
This argument also
shows that in addition to each brane being individually sLag, the
three branes are also mutually supersymmetric, since the rotation in
the base and in the fiber are precisely opposite.

\subsection{The orientifold configuration and Seiberg duality}

Let us now discuss the mirror to the discussion in section
\ref{sec:Z3-phaseII}, once we introduce the orientifold and move in
moduli space. Looking to figure~\ref{fig:mirror-fibration}, or by
recalling that in the $\bC^3/\bZ_3$ orbifold the orientifold acted as
$\psi\to\ov{\psi}$, it is clear that the mirror orientifold acts on the
$z$ plane by conjugation: $z\to\ov{z}$. Similarly, it acts by complex
conjugation on the flat torus $\bC/\bL$, which in terms of cycles can
be easily seen to correspond to a $(a,b)\leftrightarrow(b,a)$ exchange
for the basis of cycles that we have chosen for the rhombic cell. The
net effect is that the $(1,-2)$ and $(-2,1)$ branes are exchanged,
while the $(1,1)$ brane stays invariant.

In the current picture Seiberg duality comes from changing the value
of $\psi$ in~\eqref{eq:C*-fibration} \cite{Cachazo:2001sg}. A possible
choice is to move in moduli space to a point where the $\bC^*$ fiber
degenerates at some real value of $z$ larger than 3. The moduli space
of the configuration is now restricted to moving $z$ along the real
axis, so exactly as in the IIB picture, we necessarily pass through
the singular $z=3$ point when doing this.

\begin{figure}
  \begin{center}
    \includegraphics[width=0.6\textwidth]{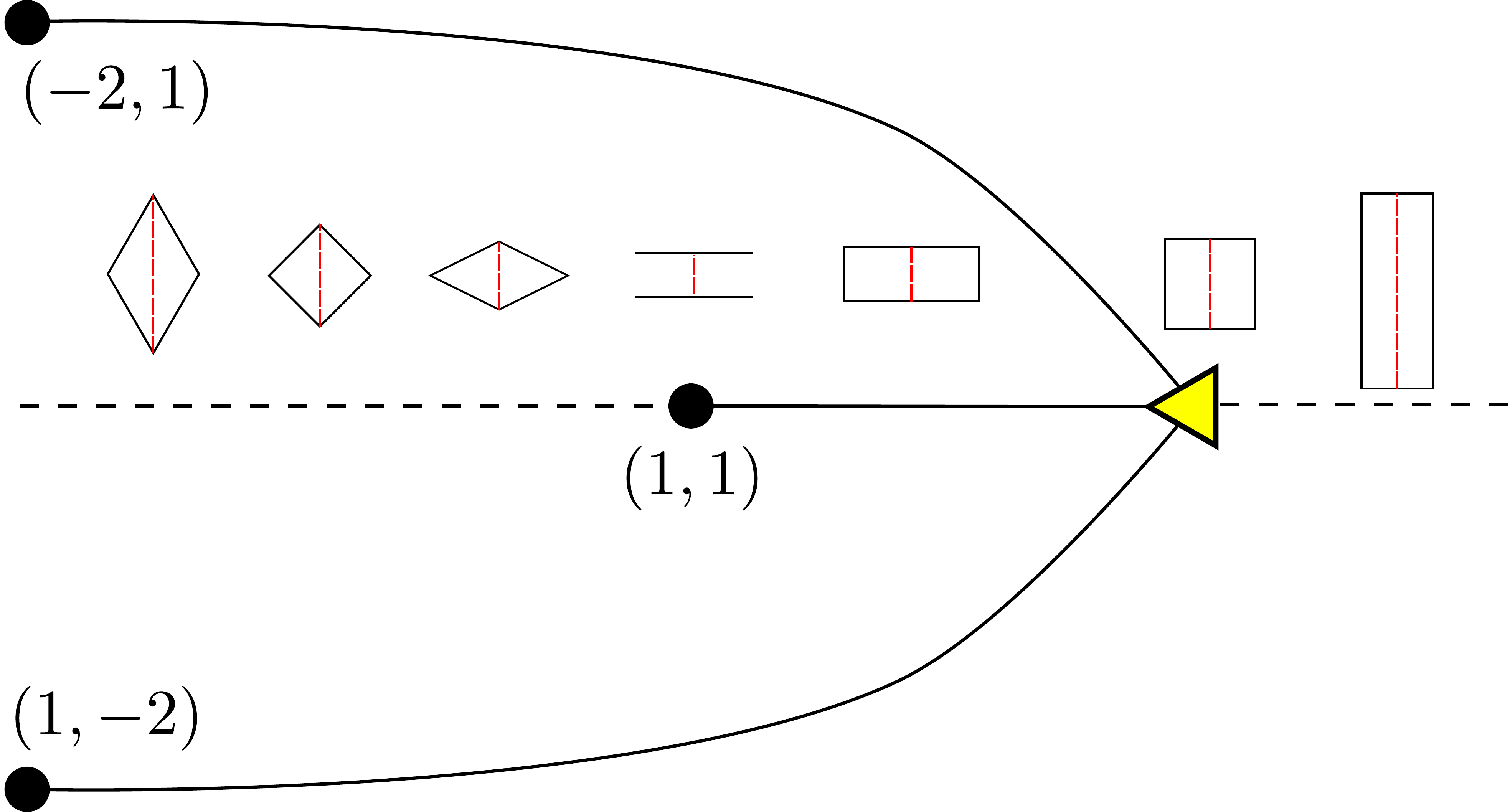}
  \end{center}
  \caption{The Seiberg dual configuration. We have depicted
    schematically the structure of the special Lagrangian branes and
    the behavior of the elliptic fiber.\label{fig:mirror-duality}}
\end{figure}

It is interesting to look to the detailed behavior of the torus
lattice as we move along the real line, we depict this in
figure~\ref{fig:mirror-duality}.
On the side of the quiver point the
lattice cell has rhombic structure. This cell becomes singular at the
conifold point, and on the opposite side of the real line a
rectangular lattice emerges. This is presumably the mirror
manifestation of the $B_2=0$ to $B_2=\frac{1}{2}$ transition on the IIB
side. In a natural basis for the rectangular lattice the invariant
brane wraps the $(1,0)$ cycle, while the $(-1,2)$ and $(-2,1)$ branes
wrap the $(1,3)$ and $(1,-3)$ cycles respectively. The orientifold
acts on the $z$ plane as before, and acts on the rectangular lattice
by sending $(p,q)$ cycles to $(p,-q)$ cycles (i.e. inverting the sign
of the second coordinate).

\bibliographystyle{JHEP}
\bibliography{refs}

\end{document}